\documentclass[%
preprint,
 amsmath,amssymb,
aip,
longbibliography
]{revtex4-2}

\usepackage{graphicx}% Include figure files
\usepackage{bm}% bold math
\usepackage{natbib}
\usepackage{appendix}
\usepackage{amsbsy}
\usepackage{psfrag}
\usepackage{amsmath,amssymb}
\usepackage{cases}
\usepackage{soul}
\usepackage{comment}
\usepackage{enumitem}
\usepackage{verbatim}
\usepackage[dvipsnames]{xcolor}
\usepackage{fdsymbol}
\usepackage{multirow}
\usepackage{booktabs}
\usepackage[colorlinks=true,
            linkcolor=Maroon,
            urlcolor=blue,
            citecolor=blue]{hyperref}
\usepackage{lineno}

\begin{document}

% \preprint{}

\title{A scaling improved inner-outer decomposition of near-wall turbulent motions}% Force line breaks with \\
% \thanks{A footnote to the article title}

\author{Limin Wang}
% \email{wanglm2015@lzu.edu.cn}
\author{Ruifeng Hu}%
\email{hurf@lzu.edu.cn}
\author{Xiaojing Zheng}%
 \email{Author to whom the correspondence should be addressed: xjzheng@lzu.edu.cn}
\affiliation{%
 Center for Particle-Laden Turbulence, Key Laboratory of Mechanics on Disaster and Environment in Western China, Ministry of Education and College of Civil Engineering and Mechanics, Lanzhou University, Lanzhou 730000, PR China
}%

\date{\today}% It is always \today, today,

\begin{abstract}
Near-wall turbulent velocities in turbulent channel flows are decomposed into small-scale and large-scale components at $y^+<100$ by improving the predictive inner-outer model of Baars \emph{et al.} [Phys. Rev. Fluids 1, 054406 (2016)], where $y^+$ is the viscous-normalized wall-normal height. The small-scale one is obtained by reducing the outer reference height (a parameter in the model) from the center of the logarithmic layer to $y^+=100$, which can fully remove outer influences. On the other hand, the large-scale one represents the near-wall footprints of outer energy-containing motions. We present plenty of evidences that demonstrate that the small-scale motions are Reynolds-number invariant with the viscous scaling, at friction Reynolds numbers between 1000 and 5200. At lower Reynolds numbers from 180 to 600, the small scales can not be scaled by the viscous units, and the vortical structures are progressively strengthened as Reynolds number increases, which is proposed as a possible mechanism responsible for the anomalous scaling behavior. Finally, it is found that a small-scale part of the outer large-scale footprint can be well scaled by the viscous units. 

%\begin{description}
%use the optional argument of the \verb+\item+ command to give the category of each %item. 
%\end{description}
\end{abstract}

\maketitle

%\tableofcontents

\section{\label{sec:level1}Introduction}

Reynolds-number dependence and scaling laws for mean and fluctuating values of flow quantities have always been one of the most fundamental topics in turbulence research. For wall-bounded turbulent flows, the celebrated law of the wall for mean streamwise velocity is well known, albeit the debate on the logarithmic and the power laws \cite[]{Marusic2010b}. 
% In channel flow, the mean velocity in the inner layer follows a linear relationship, while the overlap region yields fit the logarithmic law of the wall. 
For turbulence quantities, Townsend \cite{Townsend1976} proposed the attached eddy hypothesis (AEH) and predicted scaling relationships of fluctuating velocity variances at high-Reynolds-number condition, where it was postulated that the logarithmic layer of  turbulent flow can be modelled as an ensemble of self-similar energy-containing eddies. The size and population density of these eddies are presumed to be proportional and inversely proportional to their wall normal height $y$ \cite[]{Perry1982,Hwang2018,Marusic2019attached}, respectively. There also exists a number of other theories in the literature that aim to predict the Reynolds-number effect and scaling laws of wall turbulence quantities \cite[]{Monkewitz2015,Chen2018Quantifying,Chen2019Nonuniversal}.
% The random superposition of attached eddies of different sizes embodies the multi-scale characteristics in wall-bounded turbulence and extracts the statistics of all three velocity fluctuation components. 
In the near-wall region, it is now well recognized that the peak of streamwise turbulence intensity has a weak Reynolds number dependence when it is scaled by the friction velocity $u_\tau$ \cite[]{De2000Reynolds,marusic2017scaling}, where $u_\tau=\sqrt{\overline{\tau}_w/\rho}$ ($\overline{\tau}_w$ is the mean wall-shear stress and $\rho$ is the fluid density). Although the mean flow is well accepted to follow the law of the wall, there is less consensus about the scaling of the Reynolds normal stresses, especially the Reynolds-number dependence of the near-wall peak and its physical origins. 

Reynolds-number effect on near-wall turbulence statistics has been explored in many investigations over the past several decades. The classical view of wall-bounded turbulence considers an inner region near the wall where the viscous effect dominates, so that all velocity statistics should be universally scaled by the friction velocity and the kinematic viscosity of the fluid \cite[]{Tennekes1972}. This is referred as the inner or viscous scaling that leads to the classical von K\'arm\'an's law of the wall \cite[]{Karman1931}. Although some studies seem to support this hypothesis \cite[]{Perry1975Scaling,Mochizuki1996reynolds,Tachie2003Low,Hultmark2012,Vallikivi2015b}, much more evidence definitely shows an increasing trend of the streamwise (inner) peak turbulence intensity with Reynolds number. The existence of Reynolds-number effect on near-wall turbulence intensities have been observed from numerous simulations and experiments in various types of canonical wall-bounded flows, including boundary layers, channels and pipes, which have provided strong evidence that near-wall turbulence statistics of fluctuating quantities do not follow the inner scaling \cite[]{Purtell1981turbulent,Spalart1988direct,Wei1989Reynolds,Erm1991,Antonia1992Low,Antonia1994Low,Ching1995Low,De2000Reynolds,Metzger2001,Morrison2004scaling,hoyas2006scaling,Hutchins2007a,Schultz2013reynolds,Vincenti2013,Bernardini2014velocity,Lee2015,willert2017near,Samie2018fully}. There have been some excellent reviews on the Reynolds-number scaling issue which provide much more historical details \cite[]{Gad1994reynolds,Fernholz1996incompressible,Klewicki2010reynolds,marusic2017scaling}. 
% Besides, several works attempted to reveal the physical mechanism responsible for the Reynolds-number dependence of near-wall turbulence statistics. 
% For example, \cite{Wei1989Reynolds} measured the streamwise and wall-normal velocity components $u$ and $v$ with a laser-Doppler anemometer system in a turbulent channel flow and showed a systematic dependence of turbulence statistics on Reynolds number. They hypothesized that it may be a result of modifications of coherent structures in the vicinity of the wall, and suggested two possible mechanisms. The first one is an increase in near-wall vortex stretching with Reynolds number. The second one  raises possibility of effect from the opposite wall, especially at low Reynolds number. 
% \cite{Antonia1992Low} investigated the mechanism by introducing heating at one channel wall, and no strong evidence was found for direct interaction between inner regions of the opposite walls. It was further suggested that the Reynolds number effects of various turbulence quantities were likely to be associated with the increased intensities and stretching of quasi-streamwise vortices in the near-wall region \cite[]{Antonia1994Low}.

In recent years, a prominent view is that the augmentation of near-wall turbulence intensities with Reynolds number can be attributed to the increasing influence of outer energetic motions in the inner region \cite[]{Metzger2001,Del2003Spectra,Abe2004,Hutchins2007}. 
As Reynolds number increases, very long and energy containing motions prevail in the logarithmic layer of wall-bounded turbulent flows, and they are conventionally termed as large-scale motions (LSMs), very-large-scale motions (VLSMs), superstructures or global modes \cite[]{Kovasznay1970,Brown1977,Kim1999,Del2004scaling,Hutchins2007a,Lee2011}. It was found that the turbulence kinetic energies carried by these structures increase with Reynolds number \cite[]{hoyas2006scaling,Balakumar2007,Hutchins2007a,Vallikivi2015}. 
% Recent studies conducted at higher values of Reynolds-number have also noted the presence of very-large scale motions (VLSMs, also referred to as “superstructures”) in the logarithmic layer of turbulent boundary layers. 
Several studies \cite[]{Del2003Spectra,Abe2004,Hutchins2007} among others have clearly demonstrated that the aforementioned large outer energetic motions can penetrate deep down to the wall, playing as strong imprints or footprints, which is also consistent with Townsend's AEH. As a consequence, the presence of large-scale footprints in the near-wall region can be associated with the failure of inner scaling of near-wall turbulence intensities \cite[]{marusic2017scaling}. 

On the other hand, another major progress recently in wall turbulence research is the discovery of a self-sustaining near-wall regeneration cycle, which comprises of quasi-cyclic regeneration of streaks and quasi-streamwise vortices \cite[]{Hamilton1995,Waleffe1997,Kawahara2001periodic,Schoppa2002}. In this process, streaks can be profoundly amplified by quasi-streamwise vortices through transferring energy of mean shear to streamwise velocity fluctuations, i.e., the so-called lift-up effect \cite[]{Ellingsen1975stability,Landahl1990sublayer,Butler1993optimal,Brandt2014lift}. Then the amplified streaks rapidly oscillate and break down due to instability or transient growth, which in turn leads to the generation of new quasi-streamwise vortices \cite[]{Hamilton1995,Schoppa2002}. This cycle can also be uncovered via nonlinearly equilibrium or temporally periodic invariant solutions of the incompressible Navier-Stokes equations, which are termed by the exact coherent states as well \cite[]{Waleffe1997,Kawahara2001periodic,Waleffe2001exact}. In addition, the near-wall cycle is found to be autonomous in that it could be well self-sustained by artificially removing outer turbulent fluctuations \cite[]{Jimenez1999}. Therefore, it should be reasonable to hypothesize that the statistics of the near-wall cycle can be completely scaled with the viscous units and thus Reynolds-number invariant \cite[]{Mathis2011}, since it is now increasingly recognized that the Reynolds-number dependence of near-wall turbulence is solely introduced by outer footprints.

Based on the current understanding of near-wall turbulence, Marusic and co-workers proposed an algebraic predictive model for near-wall turbulence statistics with outer inputs \cite[]{Marusic2010,Mathis2011,Baars2016}, by incorporating the effects of superposition (footprints) and amplitude modulation of outer large scales on inner small-scale turbulent motions (near-wall cycle). Hence it suggests that outer footprints and near-wall autonomous cycle co-exist and interact in the near-wall region. In their model, a Reynolds-number-independent small-scale velocity component, namely $u^*$, is supposed to be a surrogate of the near-wall cycle and should be determined \emph{a priori} in a calibration measurement. 
% The input large-scale velocity signal is measured at the centre of the logarithmic layer where the outer fluctuation is the strongest. 
% The details of the model will be given in \S 3. 
The model has been demonstrated to work well in turbulent boundary layers at $Re_\tau=2800\sim19000$ \cite[]{Mathis2011,Baars2016}. 
% Similar results were obtained by \cite{Baars2016} with a refined predictive model using a spectral linear coherence estimation instead of a definite spectral cut off. 
% For instance, the universal velocity signal $u^*$ was acquired in turbulent boundary layer measurement at $Re_\tau=7300$ in \cite{Mathis2011}, and predictions at $Re_\tau=2800\sim19000$ show reasonable agreements with the direct laboratory measurement data.
Here the friction Reynolds number is defined by $Re_\tau=u_\tau \delta/\nu$, $\delta$ is the outer length scale (boundary layer thickness, channel half height or pipe radius), and $\nu$ is the fluid kinematic viscosity. 
% In fact, the predictive model of \cite{Marusic2010} provides a natural inner-outer decomposition of near-wall streamwise fluctuating velocity, at least in the statistical sense.

However, there still exists several issues that should be addressed and clarified. The first and the most important one is, as a key component and assumption, explicit assessment of Reynolds-number invariance of $u^*$ was never put out, which is vital for the correctness of the model. Moreover, the predictive model is somewhat inconsistent with the attached eddy model \cite[]{Perry1982,Marusic2019attached}. In the predictive model, the near-wall footprint is evaluated with an input large-scale velocity signal at $y_O^+=3.9\sqrt{Re_\tau}$ where the outer fluctuation is the strongest, \textcolor{black}{and $y_O^+$ is the viscous-scaled wall-normal height of the input large-scale velocity signal, and $y^+$ is the viscous-scaled wall-normal height}. However, the attached eddy model of Perry \& Chong \cite{Perry1982} admits the smallest self-similar wall-attached eddies of height $l_y^+ \sim O(100)$ in viscous units, which is smaller than $3.9\sqrt{Re_\tau}$ at $Re_\tau>660$ and can also impose footprints in the near wall region. In other words, if $y_O^+=3.9\sqrt{Re_\tau}$ is used, the contributions of outer eddies with sizes of $l_y^+<3.9\sqrt{Re_\tau}$ are not accounted for footprints, thus the extracted $u^*$ is expected to be Reynolds number dependent, violating its elemental assumption. 
Furthermore, in practical implementation of the model, Mathis \emph{et al.} \cite[]{Mathis2011} claimed that the Fourier phases of the large-scale signal in the calibration measurement need to be retained and replace the large-scale velocity phases measured under the prediction condition. \textcolor{black}{Without this procedure, one-point moments, especially high order ones, would be erroneously predicted.} This indicates that the extracted universal signal $u^*$ may still contain a fraction of outer large-scale footprints \cite[]{Mathis2011}. The above mentioned inconsistency or nonphysical manipulation, may be closely related to the Reynolds number dependence of $u^*$.

Moreover, there are some other relevant studies that tried to extract universal near-wall turbulence. Hwang \cite[]{Hwang2013near} designed numerical experiments that the near-wall turbulent motions with $\lambda_z^+ >100$ at 
$Re_\tau$ up to 660 were removed using the spanwise minimum flow unit (MFU) \cite[]{Jimenez1991minimal}, \textcolor{black}{$\lambda_z^+$ is the viscous-scaled spanwise wavelength}. It was found that the streamwise velocity fluctuations at $y^+<40$ are well scaled by the viscous units, whereas the wall-normal and spanwise velocity fluctuations are not. Yin \emph{et al.} \cite{Yin2017near} extended the work of Hwang \cite{Hwang2013near} to higher Reynolds numbers, i.e, $Re_\tau=1000\sim4000$, confirming similar findings. Then Yin \emph{et al.} \cite[]{Yin2018} modified the predictive model of Marusic and co-workers by replacing experimentally calibrated $u^*$ with three-dimensional turbulent velocity fields obtained from MFU simulation. Yin \emph{et al.} \cite{Yin2018} also compared the intensities of the extracted $(u^*, v^*, w^*)$ from the predictive model and MFU, and found good agreements, while the comparison was taken only at a single not a wide range of Reynolds number.
Agostini, Leschziner and others \cite[]{Agostini2014,Agostini2016c} employed the "Empirical Mode Decomposition" method to extract the small-scale motions, however they also did not explicitly demonstrate its $Re_\tau$ independence. 
Hearst \emph{et al.} \cite{Hearst2018robust} utilized a windowing technique to extract the universal inner small-scale spectrum measured from turbulent boundary layers subjected to high-intensity freestream turbulence, but they did not apply this method to instantaneous flow field. Carney \emph{et al.} \cite{Carney2020near} proposed an interesting near-wall patch approach, while which needs explicit high-pass filtering to obtain Reynolds-number invariant solutions.

In the present study, we firstly check whether the $u^*$ (as well as $v^*$ and $w^*$) extracted by adopting the refined predictive model of Baars \emph{et al.} \cite{Baars2016} is statistically Reynolds number independent in the Reynolds number range of $180 \le Re_\tau \le 5200$. And it is indeed found that the statistics of the extracted $u^*$, $v^*$ and $w^*$ are definitely Reynolds number dependent, due to the fact that the outer eddies of $100<l_y^+<3.9\sqrt{Re_\tau}$ should be further included to calculate the footprints. Therefore, we simply let $y_O^+=100$ and extract $u^*$, $v^*$ and $w^*$ with negligible Reynolds number dependence in turbulent channel flows at $Re_\tau=1000\sim5200$ using high-fidelity DNS data, which may help to further improve the predictive model for near-wall turbulence, or shed light on the physics of interactions of turbulent motions with different scales.

The paper is organised as follows. In \S 2, the data sets used in this work are described. In \S 3, we outline the decomposition scheme all three velocity components. Section \S 4 shows the evidence for Reynolds-number-independent small-scale motions in the Reynolds number range of $Re_\tau=1000\sim5200$. The low-Reynolds-number effect at $Re_\tau=180\sim600$ is given in \S 5, as well as the characteristics and scalings of large-scale outer footprints in \S 6. The final conclusion of the paper is drawn in \S 7. In this paper, the streamwise ($x$), wall-normal ($y$), and spanwise ($z$) velocity fluctuations are denoted as $u$, $v$ and $w$, respectively. The superscript '$+$' indicates the viscous scaling, i.e. the normalization by the friction velocity $u_\tau$ and the kinematic viscosity $\nu$. The angle brackets represent the spatio-temporal averaging in each of the homogeneous directions and in time.

\section{Data sets} \label{sec:data}

The main data sets used in this study are from DNS (direct numerical simulation) of fully developed turbulent channel flows. 
% which are one of the most common canonical wall-bounded turbulent flows, sharing many similarities with boundary layer and pipe flows especially in the near-wall region, although differences exist more apparently in the outer region. 
The friction Reynolds numbers are $Re_\tau=180$, 310, 600, 1000, 2000 and 5200, covering a wide range over at least one order of magnitude, which could help to display the Reynolds number effect on the near-wall turbulence clearly.  

\begin{table}
\centering
\caption{Summary of the DNS data sets. The $Re_\tau=180\sim600$ data are from our own DNS, the $Re_\tau=1000$ and $Re_\tau=5200$ data are from Lee \& Moser \cite{Lee2015} indicated by LM15, and the $Re_\tau=2000$ data are from Hoyas \& Jim\'enez \cite{hoyas2006scaling} indicated by HJ06. $L_x$ and $L_z$ are computation domain sizes in streamwise and spanwise, respectively. The outer length scale, i.e., boundary-layer thickness, channel half height or pipe radius, is denoted by $\delta$. $\Delta x^+$ and $\Delta z^+$ are the streamwise and spanwise viscous-scaled grid size. $\Delta y^+_w$ and $\Delta y^+_c$ are the viscous-scaled wall-normal grid spacings at the wall and the channel centre, respectively. FD denotes finite difference scheme, and SP denotes spectral method.} 
\begin{tabular}{cccccccccc} \toprule
 $Re_\tau$ & Reference & Method & $L_x/\delta$ & $L_z/\delta$ & $\Delta x^+$ & $\Delta z^+$  & $\Delta y^+_w$ & $\Delta y^+_c$ &  Line and Symbol \\ 
 180 & Present & FD & 8$\pi$ & 3$\pi$ & 10 & 4.99 & 0.196 & 7.08  &
 {\color{blue}$\minus$}{\color{blue}$\medblackcircle$}{\color{blue}$\minus$} \\  
 310 & Present & FD & 6$\pi$ & 2$\pi$ & 10 & 5.00 & 0.270 & 9.84  &  \textcolor{red}{$\minus$}\textcolor{red}{$\medblackdiamond$}\textcolor{red}{$\minus$} \\ 
 600 & Present & FD & 4$\pi$ & 2$\pi$ & 10 & 5.00 & 0.325 & 11.97  &  \textcolor{green}{$\minus$}\textcolor{green}{$\medblacktriangleup$}\textcolor{green}{$\minus$} \\
 1000 & LM15 \cite[]{Lee2015} & SP & 8$\pi$ & 3$\pi$ & 12.3 & 6.14 & 0.017 & 6.16  &  \textcolor{orange}{$\minus$}\textcolor{orange}{$\medblacktriangledown$}\textcolor{orange}{$\minus$} \\
 2000 & HJ06 \cite[]{hoyas2006scaling} & SP & 8$\pi$ & 3$\pi$ & 8.2 & 4.09 & 0.323 & 8.89  &  \textcolor{black}{$\minus$}\textcolor{black}{$\medblacktriangleright$}\textcolor{black}{$\minus$} \\
 5200 & LM15 \cite[]{Lee2015} & SP & 8$\pi$ & 3$\pi$ & 12.8 & 6.38 & 0.498 & 10.3 &  \textcolor{cyan}{$\minus$}\textcolor{cyan}{$\medblackstar$}\textcolor{cyan}{$\minus$} \\ \bottomrule
\end{tabular}
\label{tab:tab1}
\end{table}

The turbulent channel data sets at $Re_\tau$=180, 310 and 600 are obtained from the DNS by ourselves. The DNS code adopts a fourth-order accurate compact difference scheme in the homogeneous directions and a second-order accurate central difference scheme in the wall-normal direction for the descretization of the incompressible Navier-Stokes equations on a staggered grid \cite[]{Hu2018application}. 
% The exact projection method is used for velocity-pressure decoupling. For temporal advancement, the Adams-Bashforth scheme is implemented for the convection terms, and Crank-Nicolson scheme for the viscous terms is employed to permit large time steps. The pressure Poisson equation is solved efficiently by Fast Fourier Transformation (FFT). 
In our previous work, a series of low-Reynolds-number channel DNS (up to $Re_\tau=180$) were conducted using the code \cite[]{Hu2018energy} and the results were well validated against Lee \& Moser \cite{Lee2015}. In this study, we perform simulations at two higher Reynolds numbers, i.e. $Re_\tau=310$ and 600, following the same standard at the lower Reynolds numbers, which is validated with Lee \& Moser \cite{Lee2015} at similar Reynolds numbers in Appendix \ref{sec:appenA}.
The DNS data sets of channel flows at $Re_\tau$ = 1000 and 5200 were computed by the group at The University of Texas at Austin (UTA) \cite[]{Lee2015}, the raw data of which are assessed from the Johns Hopkins Turbulence Database (JHTDB) \cite[]{Graham2016web}. And the data set at $Re_\tau = 2000$ was generated and assessed from the group at Universidad Polit\'ecnica de Madrid (UPM) \cite[]{hoyas2006scaling}. All the cases of $Re_\tau = 1000$, 2000 and 5200 were solved using spectral method in the wall-parallel planes, and the UTA group adopted a 7th-order B-spline collocation method while the UPM group employed a seven-point compact finite difference scheme in the wall-normal direction. The detailed information of the data sets is listed in table~\ref{tab:tab1}.

\begin{figure}
\centering
\begin{minipage}{0.49\linewidth}
\centerline{\includegraphics[width=\textwidth]{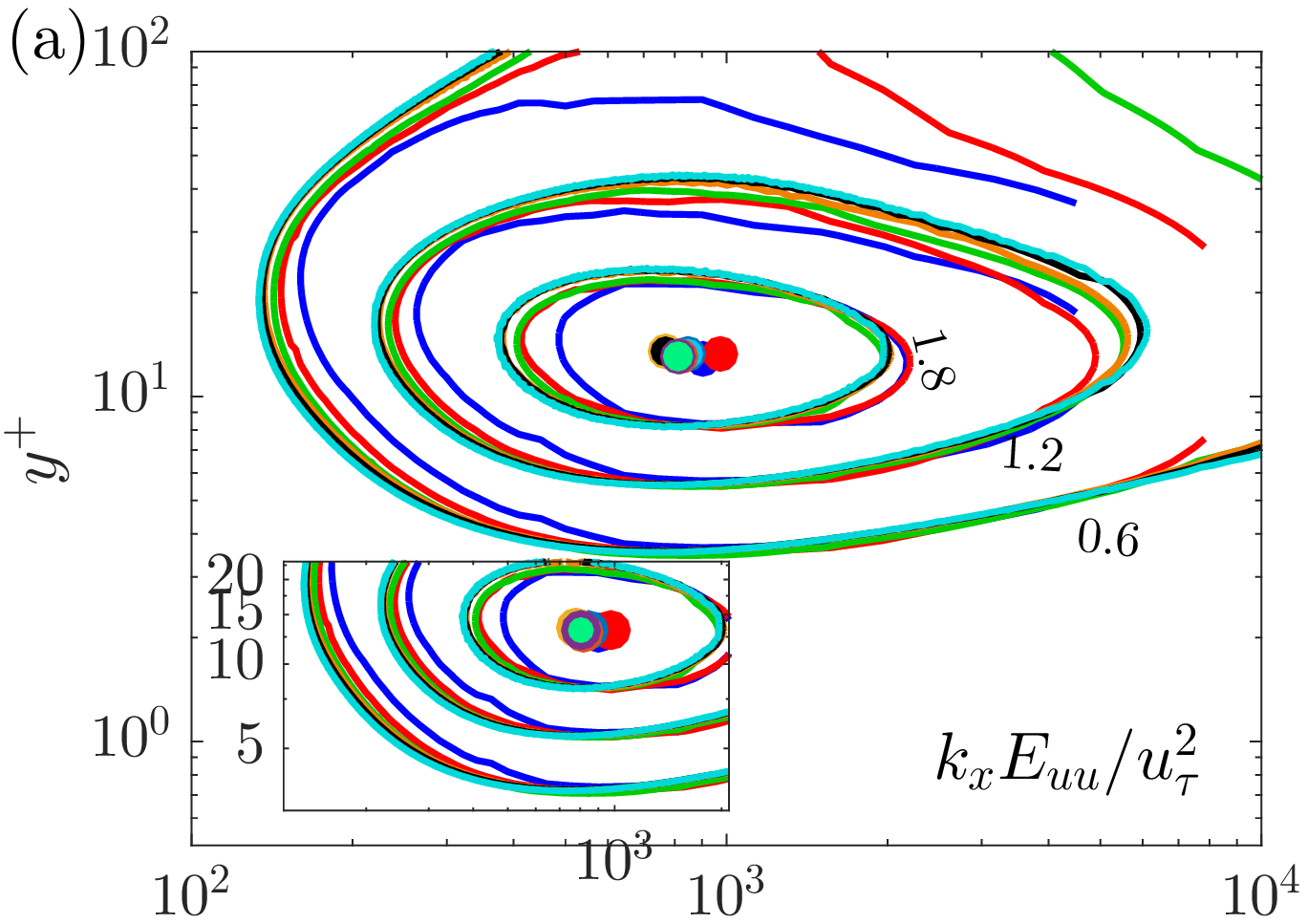}}
\end{minipage}
\hfill
\begin{minipage}{0.49\linewidth}
\centerline{\includegraphics[width=\textwidth]{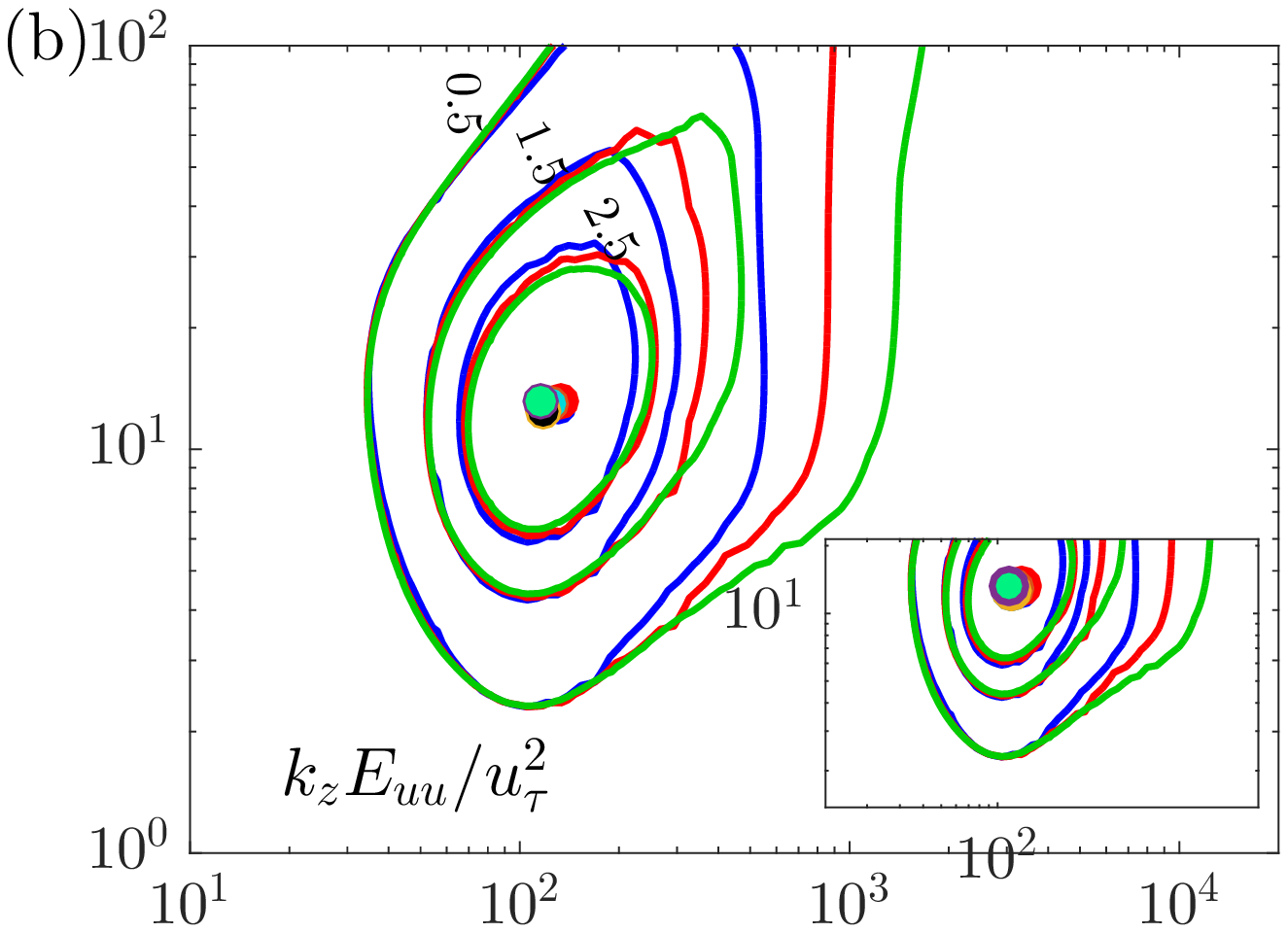}}
\end{minipage}
\vfill
\begin{minipage}{0.49\linewidth}
\centerline{\includegraphics[width=\textwidth]{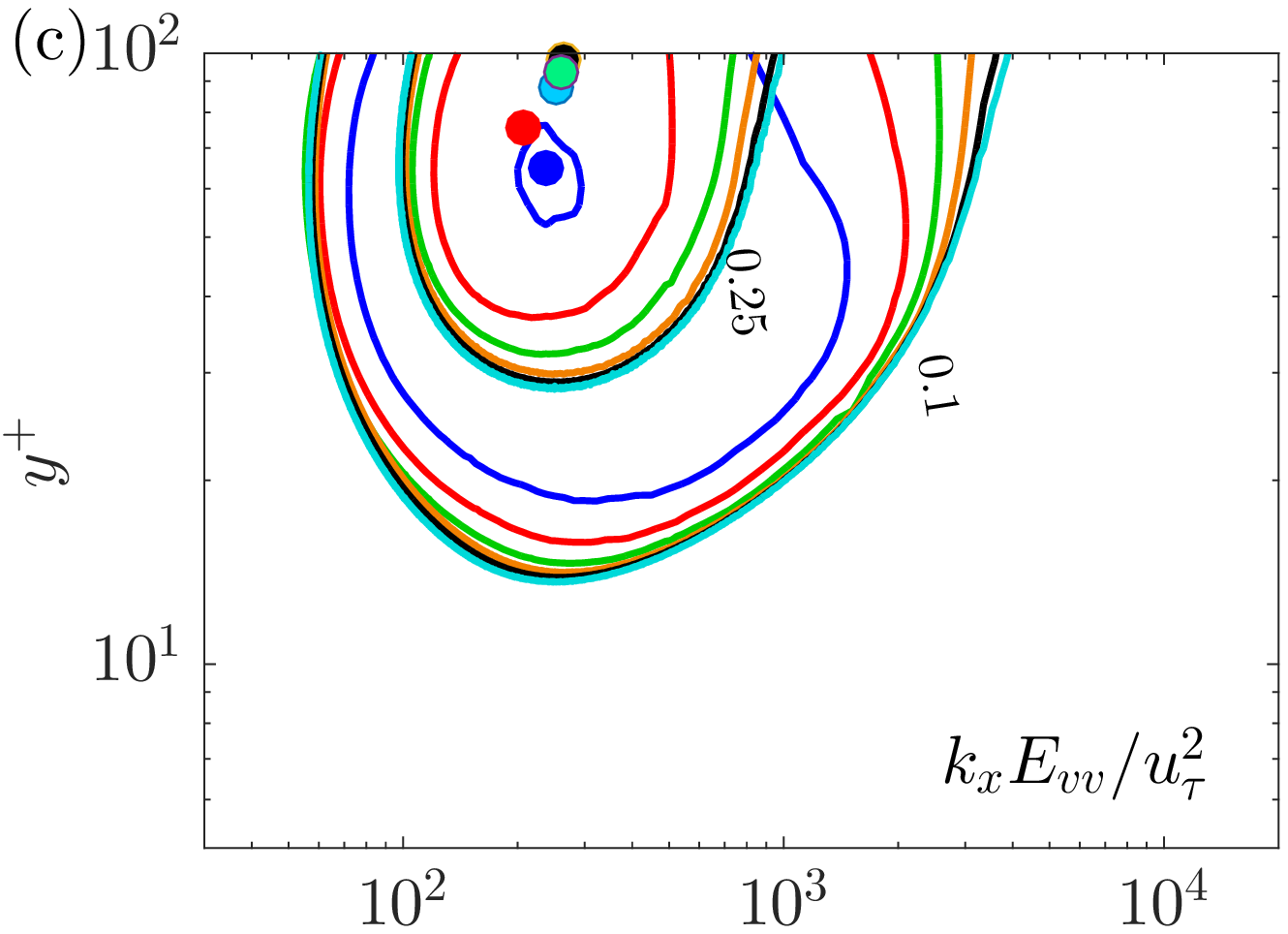}}
\end{minipage}
\hfill
\begin{minipage}{0.49\linewidth}
\centerline{\includegraphics[width=\textwidth]{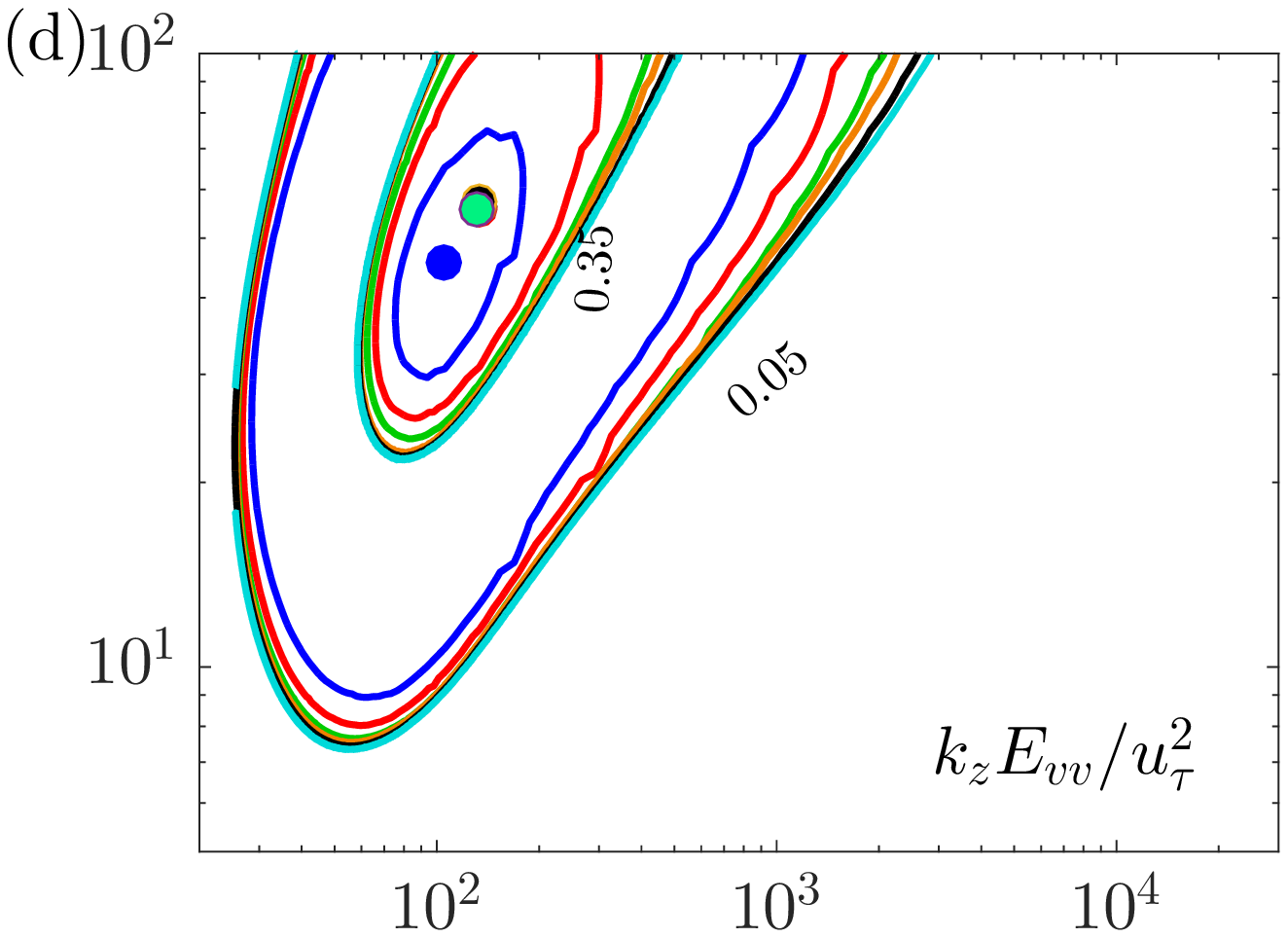}}
\end{minipage}
\begin{minipage}{0.49\linewidth}
\centerline{\includegraphics[width=\textwidth]{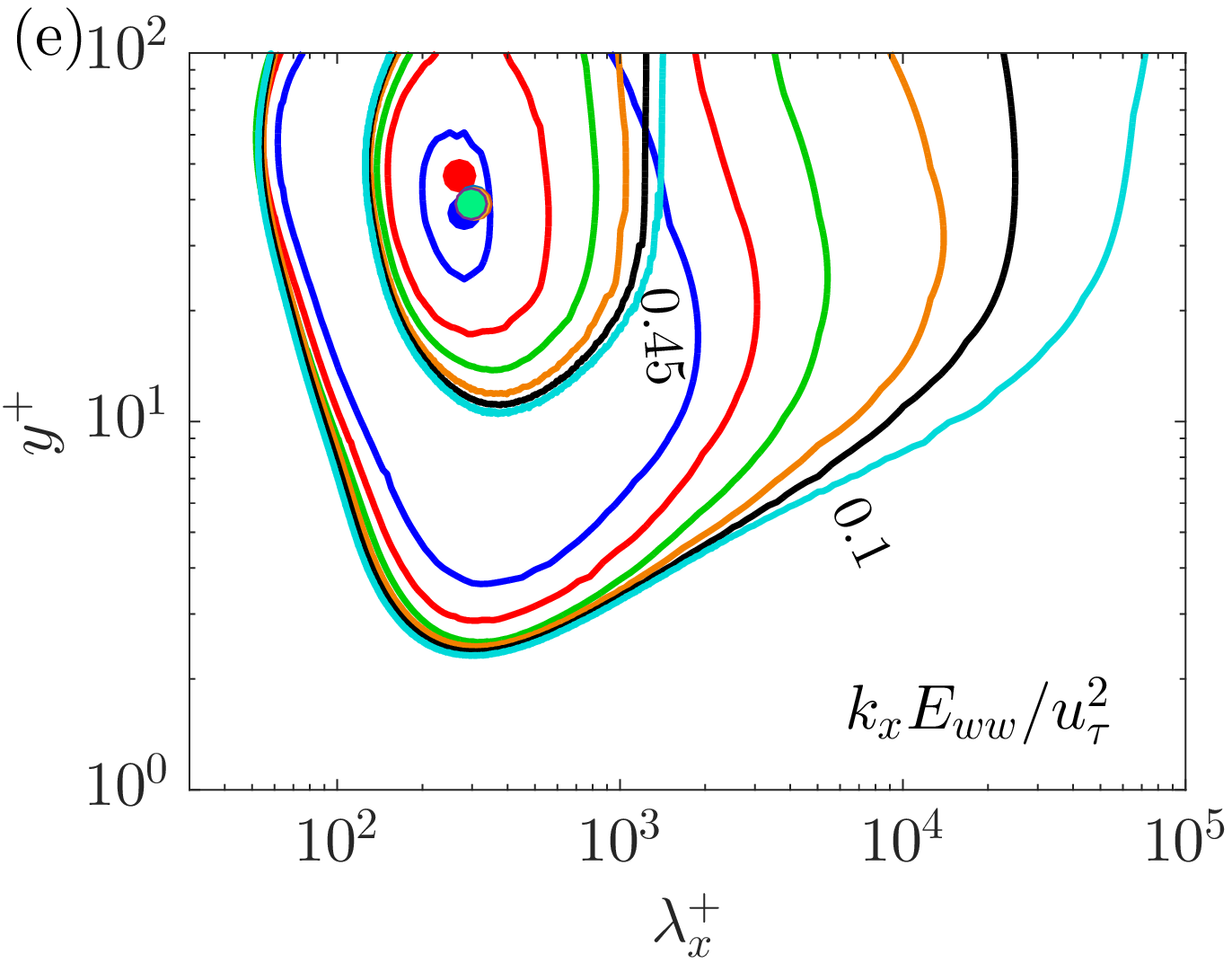}}
\end{minipage}
\hfill
\begin{minipage}{0.49\linewidth}
\centerline{\includegraphics[width=\textwidth]{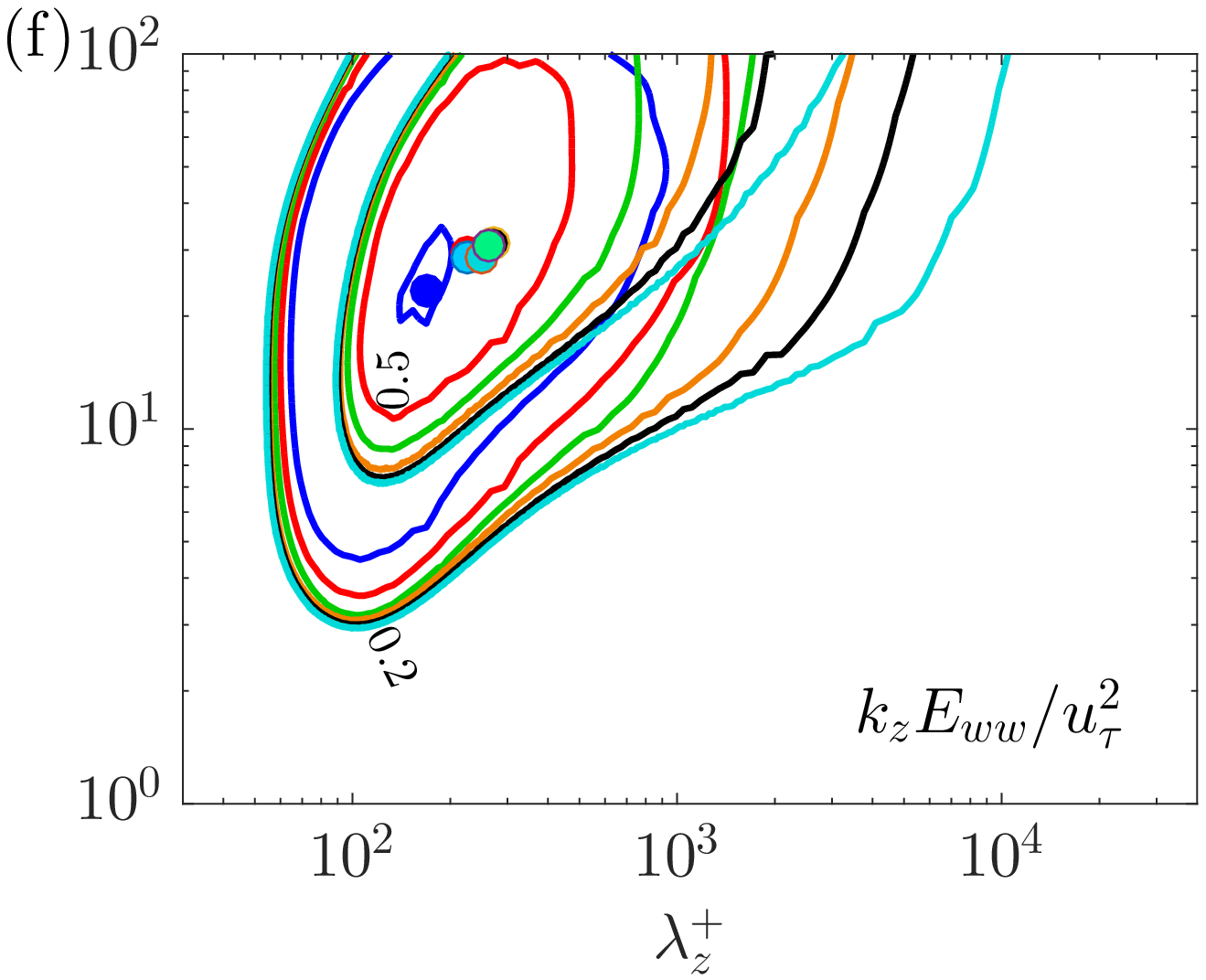}}
\end{minipage}
\caption{Pre-multiplied energy spectra of the three velocity components at different Reynolds numbers in the near-wall region ($y^+<100$). Lines with different colors correspond to different Reynolds numbers, see table~\ref{tab:tab1}. (a) $k_x E_{uu}/u_\tau^2$, (b) $k_z E_{uu}/u_\tau^2$, (c) $k_x E_{vv}/u_\tau^2$, (d) $k_z E_{vv}/u_\tau^2$, (e) $k_x E_{ww}/u_\tau^2$ and (f) $k_z E_{ww}/u_\tau^2$. \textcolor{black}{The locations of inner spectral peaks have been marked with symbols and zoomed areas for the \textcolor{black}{small-scale signal in the near wall-region}.}}
\label{fig:fig1}
\end{figure}
 
% Thirdly, we move on to the scale-to-scale measures of turbulence kinetic energies in the three directions and the corresponding Reynolds number effects. The analysis tool of pre-multiplied energy spectrum is intensely used here, which can effectively aid to reveal the spectral energy distribution of turbulence fluctuations in the logarithmic coordinate. 
% We exhibit the viscous-scaled pre-multiplied spectral energy densities of three velocity components into a contour plot against viscous-scaled wavelength and wall-normal height. 
% Here, $E_{uu}$ is the spectral energy density of $u$ at a wavenumber $k_x$ or wavelength $\lambda_x=2\pi/k_x$. The definitions for other velocity components and in the spanwise direction are similar and straightforward. This tool has been widely adopted for analyzing the scale-to-scale properties of turbulence fluctuations \cite[]{Kim1999,Guala2006,Balakumar2007,Hutchins2007a,Hwang2013near,Hwang2015statistical,Lee2015,Lee2019spectral,Vallikivi2015,Wang2016,Hu2018energy,wang2018spanwise,Cheng2019identity,Wang2019wall}.
\textcolor{black}{FIG}~\ref{fig:fig1} shows the viscous-scaled streamwise and spanwise pre-multiplied energy spectra of the three components of velocity fluctuations in the near-wall region, i.e. $y^+<100$. 
% As Reynolds number increases, the viscous-scaled wall-normal range increases accordingly, resulting in the upwards extended contour patterns. 
It is seen that distinct inner peaks can be clearly observed, which are the spectral signatures of predominant near-wall coherent structures (streaks, quasi-streamwise vortices, etc) with specific characteristic length scales. 
The streamwise and spanwise pre-multiplied energy spectra of streamwise velocity fluctuations at $Re_\tau = 180 \sim 5200$ are shown in \textcolor{black}{FIG~\ref{fig:fig1} (a) and (b). $k_xE_{u_iu_i}(\lambda_x^+, y^+) = k_x \langle \hat{u_i}^+(\lambda_x^+,y^+,z^+,t^+)\overline{\hat{u_i}^+(\lambda_x^+,y^+,z^+,t^+)}\rangle$, where $<>$ denotes the average in time and in the spanwise direction, $\hat{u_i}$ is the Fourier coefficients of $u_i$ along $x$ direction (i = 1,2,3 for $u$, $v$, $w$) and the overbar indicates complex conjugate, $k_x = 2\pi/\lambda_x$ and $\lambda_x$ is streamwise wavenumber. So did the spanwise wavenumber pre-multiplied energy spectra.}
The inner peak \textcolor{black}{(marked with symbols in FIG~\ref{fig:fig1} (a) for $Re_\tau$ = 180 and 5200)} locates at $y^+=10\sim20$, with $\lambda_x^+\sim O(10^3)$ and $\lambda_z^+\sim O(10^2)$, which is consistent with the well known characteristic streamwise length and spanwise spacing of near-wall streaks obtained from measurements or DNS \cite[]{Kline1967,Smith1983,Kim1987}. 
The spectra of wall-normal and spanwise velocity components are displayed in \textcolor{black}{FIG}~\ref{fig:fig1} (c-f). 
% In other words, $v$ and $w$ show evident Reynolds number dependence in the spectra. 
It is seen that the inner peaks of $v$- and $w$-spectra locate at $y^+=30\sim70$ (\textcolor{black}{except for $k_xE_{vv}$, where the higher Reynolds number inner peak positions are close to 100}), since $v$ and $w$ are primarily induced by quasi-streamwise vortical structures which generally ride above the near-wall streaks \cite[]{Jeong1997,Hwang2013near,Cheng2019identity}.
\textcolor{black}{For the small scales in the near-wall region} (e.g., $\lambda_x^+ <\sim 7000$ or $\lambda_z^+ <\sim 200$ for the $u$-spectra), we can see that the spectra at $Re_\tau \ge 1000$ collapse generally well, \textcolor{black}{shown in the zoomed area in {FIG}~\ref{fig:fig1} (a,b)}, indicating Reynolds-number-independent near-wall small-scale turbulent motions \cite[]{Hwang2013near,Lee2015,Hearst2018robust,wang2018spanwise}. 
%However, a certain degree of mismatch exists in the streamwise spectra, and the streamwise length of the inner peak is longer at lower Reynolds numbers (the locations of inner spectral peaks at $Re_\tau=180$ and 5200 are marked by filled circles in \textcolor{black}{FIG}~\ref{fig:fig1}(a)).
However, the spectra can not be well collapsed with viscous scaling at $Re_\tau=180 \sim 600$ \textcolor{black}{of the small-scale signal in the near wall-region}, especially for the $v$- and $w$-spectra, indicating possible low-Reynolds-number effect in this Reynolds number range.
The spectral imprints of outer large-scale components into the near-wall region are stronger and extending to longer wavelength with viscous scaling for $u$ and $w$ if $Re_\tau$ is larger, demonstrating increasing outer influences, which has been well known according to many previous works \cite[]{Metzger2001,Del2003Spectra,Abe2004,Hutchins2007,Morrison2007,Hwang2016mesolayer,Lee2019spectral}. 
% And the major objective of the present study is to remove the large-scale imprints on the near-wall region and extract truly Reynolds-number-independent universal near-wall motions. 

\begin{figure}
\centering
\begin{minipage}{0.49\linewidth}
\centerline{\includegraphics[width=\textwidth]{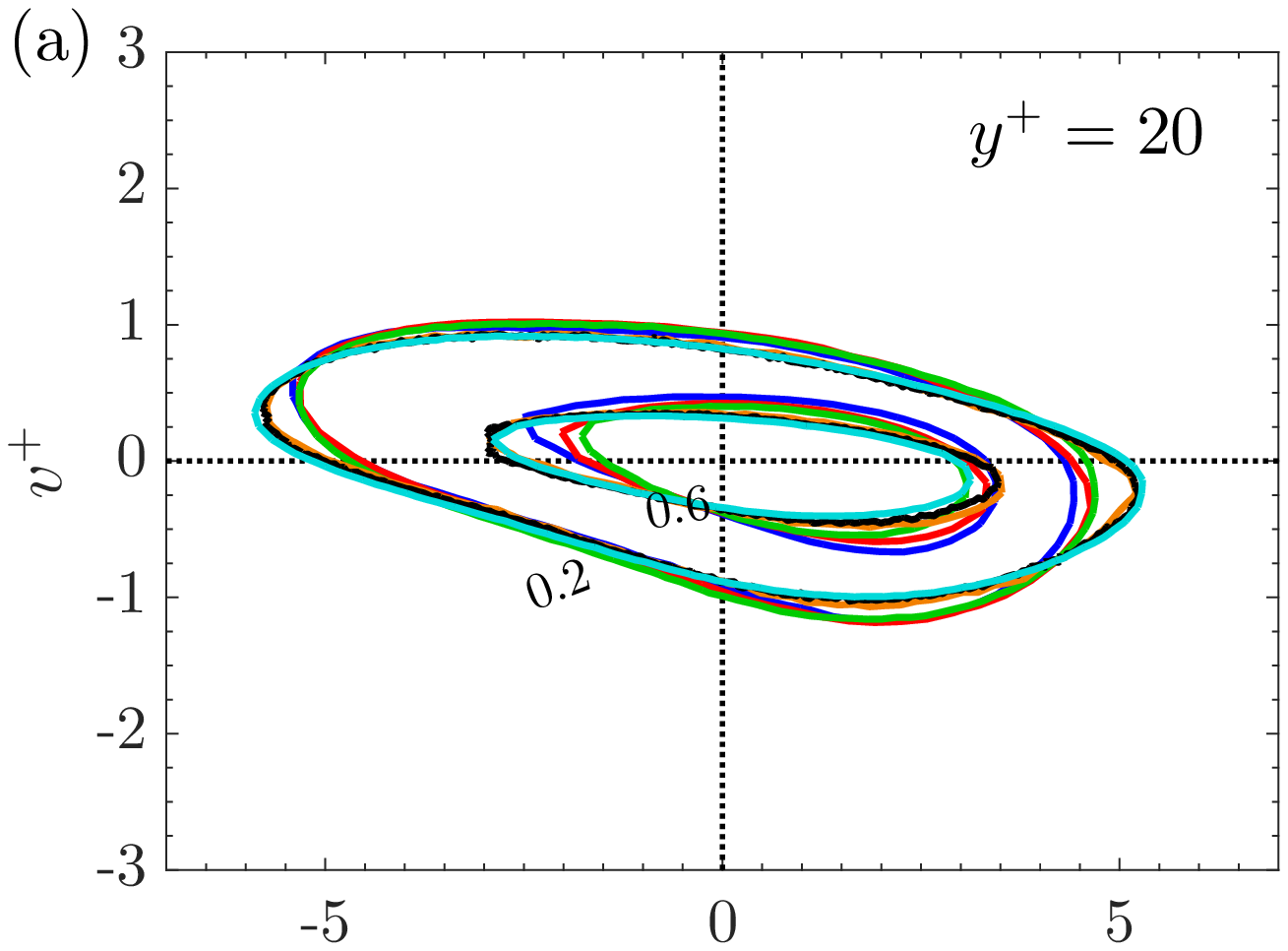}}
\end{minipage}
\hfill
\begin{minipage}{0.49\linewidth}
\centerline{\includegraphics[width=\textwidth]{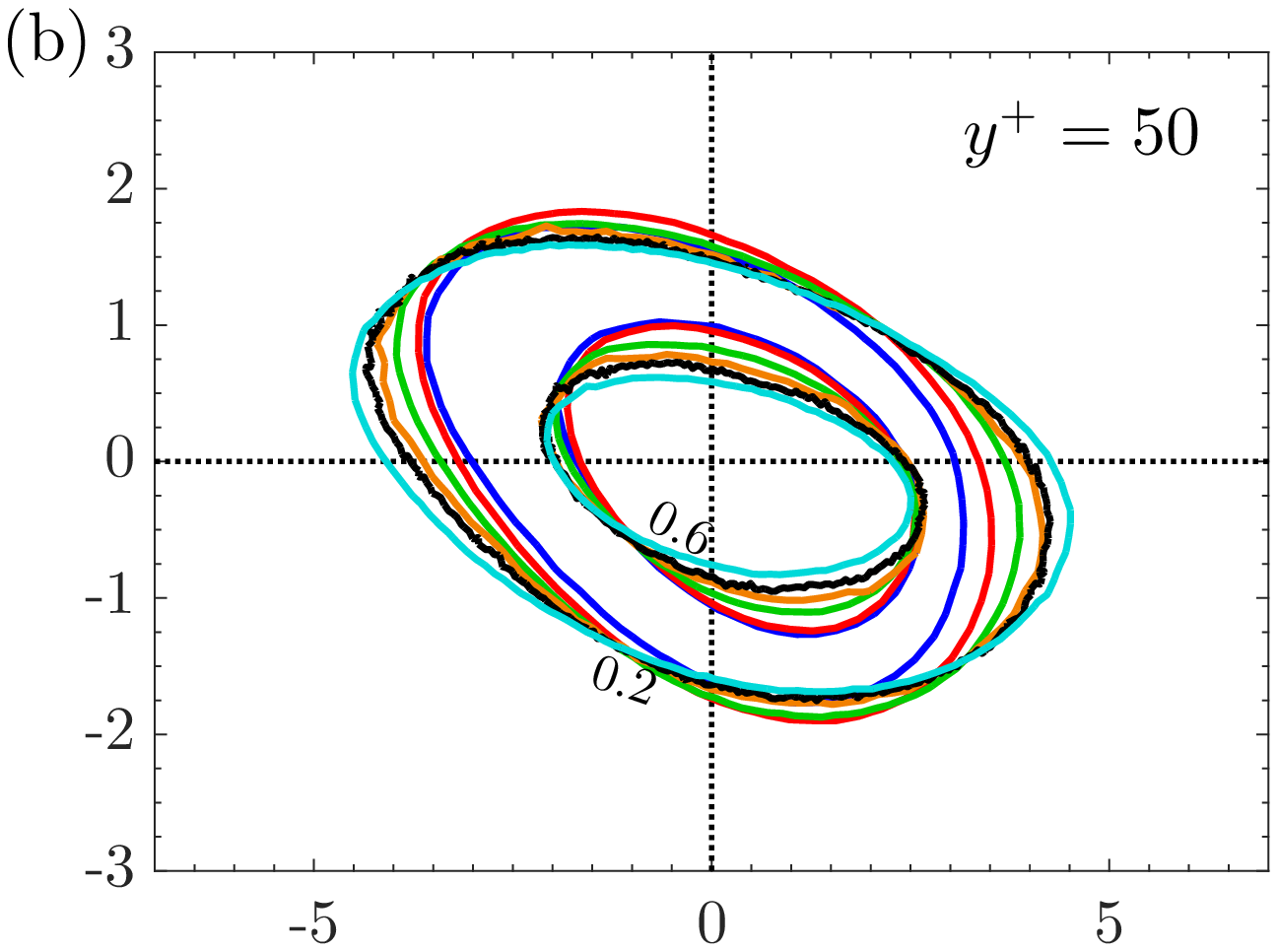}}
\end{minipage}
\vfill
\begin{minipage}{0.49\linewidth}
\centerline{\includegraphics[width=\textwidth]{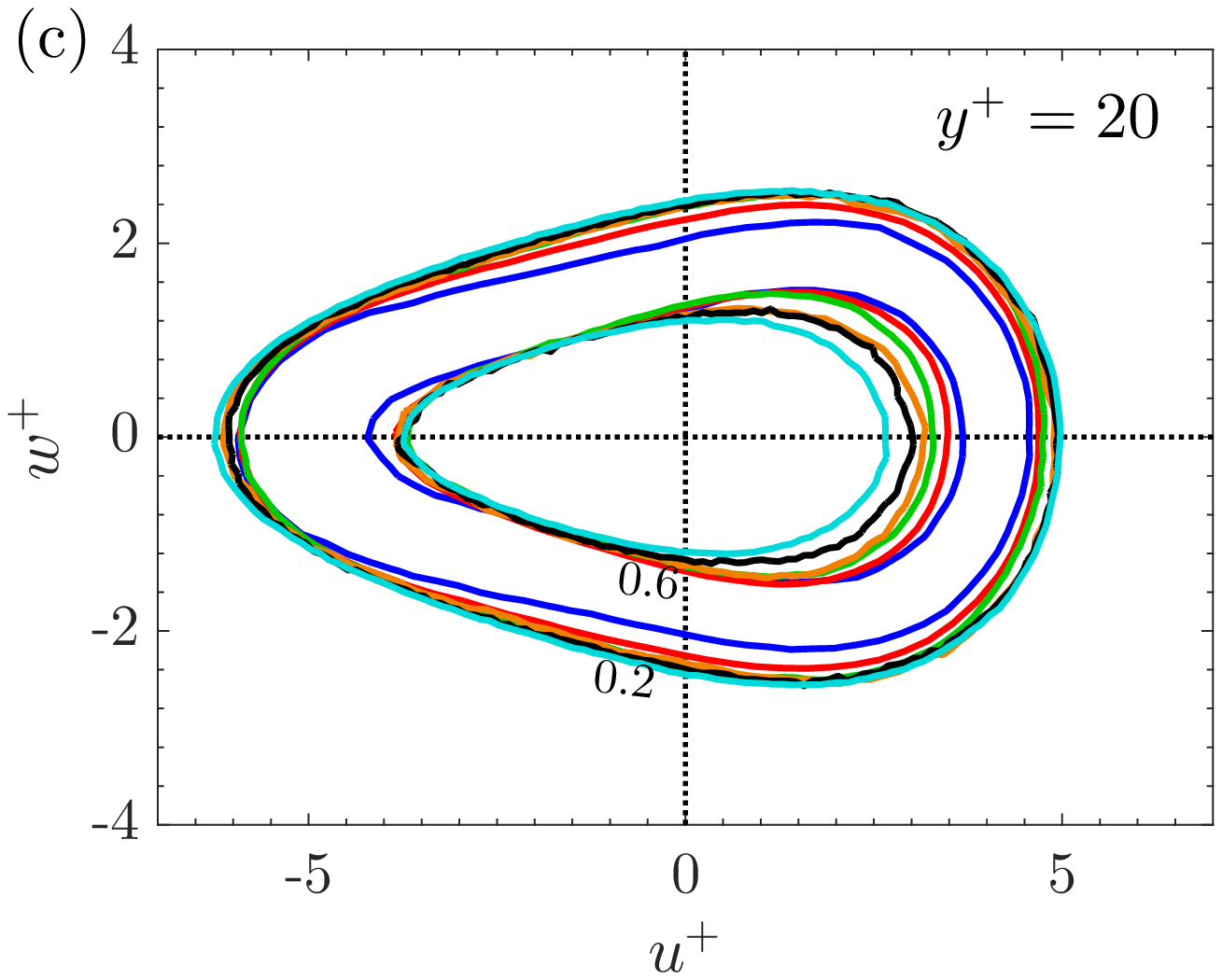}}
\end{minipage}
\hfill
\begin{minipage}{0.49\linewidth}
\centerline{\includegraphics[width=\textwidth]{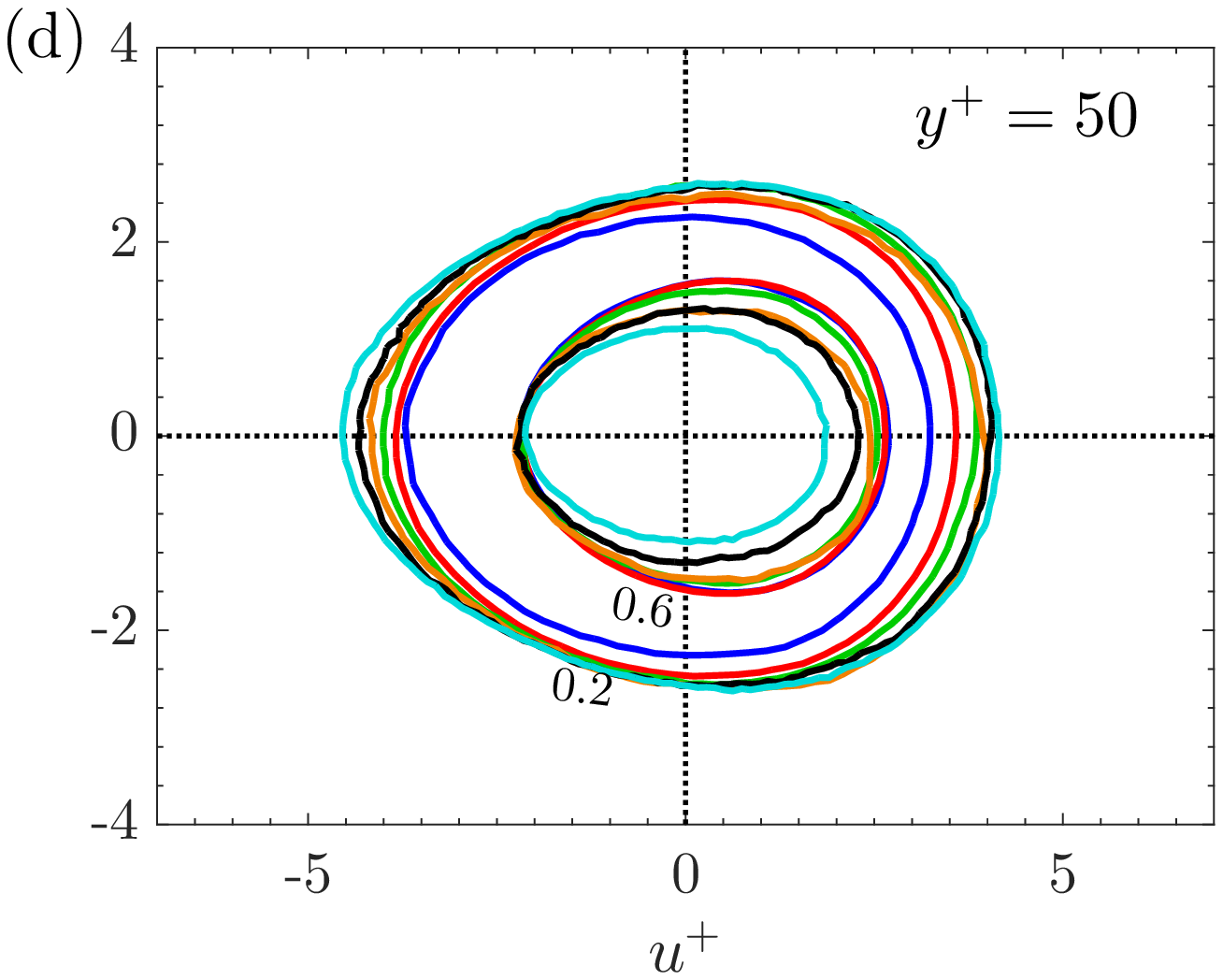}}
\end{minipage}
%\begin{minipage}{0.49\linewidth}
%\centerline{\includegraphics[width=\textwidth]{figure003e.eps}}
%\end{minipage}
%\hfill
%\begin{minipage}{0.49\linewidth}
%\centerline{\includegraphics[width=\textwidth]{figure003f.eps}}
%\end{minipage}
\caption{Joint p.d.f.s  $P(u^+, v^+)$ (a, b) and $P(u^+, w^+)$ (c, d) at $y^+$ = 20 (a, c) and $y^+$ = 50 (b, d) in the fully developed turbulent channel flows at the Reynolds numbers $Re_\tau=180\sim5200$. Lines with different colors correspond to different Reynolds numbers, see table~\ref{tab:tab1}. The contour levels are normalized by the peak value of the p.d.f.}
\label{fig:fig2}
\end{figure}

% The joint probability distribution functions (p.d.f.s) between velocity components, i.e., $P(u^+$, $v^+)$ and $P(u^+$, $w^+)$ at $y^+=20$ and $y^+=50$, are shown in figure~\ref{fig:fig2}. 
% The joint p.d.f.s are useful and can be applied to examine the most probable turbulent motions via the one-point velocity cross-correlations. 
The joint probability distribution function (p.d.f.) between velocity components is a useful tool related to the quadrant analysis, which was proposed nearly fifty years before and has been thoroughly adopted to detect the outward (Q1, $u>0$ and $v>0$), ejection (Q2, $u<0$ and $v>0$), inward (Q3, $u<0$ and $v<0$) and sweep (Q4, $u>0$ and $v<0$) events \cite[]{wallace1972wall,Willmarth1972structure,Lu1973measurements,Wallace2016quadrant}.
As displayed in \textcolor{black}{FIG}~\ref{fig:fig2} (a, b), the shape of $P(u^+, v^+)$ is inclined with the major axis in the Q2-Q4 direction, implying much higher probabilities of the ejection and sweep events. The p.d.f. contours at the two heights both show Reynolds number dependence, and the major axis tends to $v^+=0$ if $Re_\tau$ increases, which indicates $u^+$ increases more rapidly than $v^+$. If comparing the p.d.f.s at the two heights, i.e. $y^+=20$ and $y^+=50$ in \textcolor{black}{FIG}~\ref{fig:fig2} (a) and (b), it can be seen that the inclination of the major axis is steeper at $y^+=50$, that means the ejecting and sweeping angles of coherent motions are smaller towards the wall. 
\textcolor{black}{FIG}~\ref{fig:fig2} (c, d) demonstrates that the shape of $P(u^+$, $w^+)$ is symmetric about $w^+ = 0$ in all cases. However, the p.d.f.s are not symmetric about $u^+=0$, showing quite different velocity distributions at $u>0$ and $u<0$. In general, the spanwise velocity $w$ has a wider distribution at $u>0$ than $u<0$. This is probably due to the splatting effect \cite[]{Agostini2014,Agostini2016c,Pan2018extremely} or the dispersive motions of high-speed structures \cite[]{Hwang2016}, i.e., high-speed sweeping (Q4) motions (mainly $u$ component) may be partly converted into spanwise motions ($w$ component) near the wall. The splatting/dispersive effect becomes weaker away from the wall, which could be confirmed in \textcolor{black}{FIG}~\ref{fig:fig2} (d). Also, there exists visible Reynolds number dependence of $P(u^+$, $w^+)$. Also, some modern developments of the quadrant analysis have been applied to reveal the spatial organization and time evolution of sweeps and ejections \cite[]{Lozano2012three,Lozano2014Time,Dong2017coherent,Fiscaletti12018}.

% We have normalized the contour level using the peak value of the p.d.f for $Re_\tau = 1000$ and the contour level of 0.2 show that the higher $Re_\tau$ has a larger ellipse range, while the contour level of 0.6 is in contrast with the level 0.2. 
% Fig.\ref{fig:fig2} (c,d) thus demonstrates that larger ($u^+$, $w^+$) range exist at larger Reynolds number.
% Moreover, the fig.\ref{fig:fig2} (a) shows the same p.d.f.s for $Re_\tau$ of 1000, 2000 and 5200, but the low $Re_\tau$ is a little different between each other. 
% Fig.\ref{fig:fig2} (a,c) reflect joint p.d.f.s of ($u^+$, $v^+$) and ($u^+$, $w^+$) of fluctuation velocity does not yield a same signal, even at the $y^+ = 20$. The joint ($u^+$, $v^+$) and ($u^+$, $w^+$) p.d.f.s at $y^+ = 50$, however, given in fig.\ref{fig:fig2} (b,d), show a strong Re-independence.
% Furthermore, the question of what characteristic of the Reynolds-number-independent distributions of universal signals $u^*$, $v^*$, $w^*$ are discussed in next section. 

% In general, the experimental measurements get the universal signal $u^*$ by using a calibration experiment. While the universal signals targeted to the DNS data are backward extraction using predictive models. Then, the extraction of the universal signals for the three components of velocity fluctuations from the DNS data of the full-sized channel turbulence is simply described.

In summary, we have presented some statistics (pre-multiplied energy spectra and joint p.d.f.s) of the channel flow DNS data in the near-wall region, covering a wide range of Reynolds numbers at $Re_\tau=180\sim5200$. From the spectra, we found that the near-wall small-scale turbulence may be Reynolds number independent at $Re_\tau \ge 1000$, while which is Reynolds number dependent at $Re_\tau=180\sim600$. However, there are very few attempts to extract the part of the flow which should be Reynolds number independent. In the following, we will employ the above DNS data to work out the decomposition of near-wall turbulent motions and verify whether the extracted $(u^*, v^*, w^*)$ are really Reynolds number independent. 

\section{Decomposition methodology} \label{sec:meth}
% \subsection{General description}
% \subsection{Extracting the universal signals from DNS data}
% \subsection{The tool: spectral linear stochastic estimation}
The decomposition of near-wall turbulent motions is based on the framework of the predictive inner-outer (PIO) model proposed by Marusic and co-workers \cite[]{Marusic2010,Mathis2011}. The basic idea of the PIO model is that the near-wall turbulence fluctuations could be decomposed into two components, i.e., the footprints of outer large-scale fluctuations (the superposition effect), and the small-scale fluctuations with modulated amplitudes by large scales (the modulation effect). Here we resort to the refined PIO model of Baars \emph{et al.} \cite{Baars2016}, which eliminates the need for a specific spectral cut-off filter to separate small- and large-scale velocities as in the original version of this model.

The refined PIO model proposed for streamwise velocity takes the form of
\begin{equation}
   \textcolor{black}{u^+_p(x^+,y^+,z^+,t^+) = \underbrace{u^*(x^+,y^+,z^+,t^+)\left[1+ \Gamma_{uu}(y^+) u_L^+(x^+,y^+,z^+,t^+)\right]}_{\textbf{modulation}} + \underbrace{u_L^+(x^+,y^+,z^+,t^+)}_{\textbf{superposition}}.}
    \label{eqn:equ2}
\end{equation}
Here $u_p^+$ is the predicted streamwise fluctuating velocity near the wall\textcolor{red}{.} All of the variables are normalized by the viscous units. In the right-hand-side of equation (\ref{eqn:equ2}), $\Gamma_{uu}(y^+)$ is the modulation coefficient and \textcolor{black}{$u^*(x^+,y^+,z^+,t^+ )$} is the near-wall Reynolds number independent signal in the absence of outer influence, which are usually determined through a synchronously two-point calibration experiment, and assumed to be $Re_\tau$ independent \cite[]{Mathis2011,Baars2016}. \textcolor{black}{Where the second term $u_L^+(x^+,y^+,z^+,t^+)$ denotes the large-scale component of streamwise velocity fluctuation in the near wall region.}
Here we also follow Baars \emph{et al.} \cite{Baars2016} to calculate the outer footprint of streamwise velocity \textcolor{black}{$u_L^+(x^+,y^+,z^+,t^+)$} as
\begin{equation}
  \textcolor{black}{u_L^+(x^+,y^+,z^+,t^+) = F_x^{-1} \{ H_{Lu}(\lambda_x^+, y^+) F_x[u_O^+(x^+,y_O^+,z^+,t^+)] \},} 
  \label{eqn:equ3}
\end{equation}
and the large-scale footprints $v_L$ and $w_L$ are obtained similar to \textcolor{black}{$u_L^+(x^+,y^+,z^+,t^+)$} \citep{Yin2018}, i.e.,
\begin{equation}
  \textcolor{black}{v_L^+(x^+,y^+,z^+,t^+) = F_x^{-1} \{H_{Lv}(\lambda_x^+, y^+) F_x[v_O^+(x^+,y_O^+,z^+,t^+)] \}, }
  \label{eqn:equ4}
\end{equation}
\begin{equation}
  \textcolor{black}{w_L^+(x^+,y^+,z^+,t^+) = F_x^{-1} \{ H_{Lw}(\lambda_x^+, y^+) F_x[w_O^+(x^+,y_O^+,z^+,t^+)] \}, } 
  \label{eqn:equ5}
\end{equation}
in which, $u_O^+, v_O^+$ and $w_O^+$ are the input outer fluctuating velocity at $y_O^+$ \textcolor{black}{and $y_O^+ = 3.9\sqrt{Re_\tau}$ is usually used to approximate the centre of the logarithmic layer which also corresponds to the location of the outer spectral peak \cite[]{Mathis2011,Baars2016}. By checking the exact locations of the outer spectral peaks at $Re_\tau=2000$ and 5200 \cite[]{Hutchins2007a,Altintas2019}, we found the approximation $y_O^+ = 3.9\sqrt{Re_\tau}$ is quite accurate.} $F_x$ and $F_x^{-1}$ denote FFT and inverse FFT, respectively. 
\textcolor{black}{In equation (\ref{eqn:equ3}-\ref{eqn:equ5})}, $H_{Lu}$, $H_{Lv}$ and $H_{Lw}$ are scale-dependent complex-valued kernel functions for calculating footprints, representing the spectral linear stochastic estimation of outer velocity components in the near-wall region, defined by
\begin{equation}
    \textcolor{black}{H_{Lu}(\lambda_x^+, y^+) = \frac{\langle \hat{u}^+(\lambda_x^+,y^+,z^+,t^+)\overline{\hat{u}^+(\lambda_x^+,y_O^+,z^+,t^+)}\rangle}{\langle \hat{u}^+(\lambda_x^+,y_O^+,z^+,t^+)\overline{\hat{u}^+(\lambda_x^+,y_O^+,z^+,t^+)}\rangle}=|H_{Lu}|e^{j\phi_u},}
    \label{eqn:equ6}
\end{equation}
\begin{equation}
    \textcolor{black}{H_{Lv}(\lambda_x^+, y^+) = \frac{\langle \hat{v}^+(\lambda_x^+,y^+,z^+,t^+)\overline{\hat{v}^+(\lambda_x^+,y_O^+,z^+,t^+)}\rangle}{\langle \hat{v}^+(\lambda_x^+,y_O^+,z^+,t^+)\overline{\hat{v}^+(\lambda_x^+,y_O^+,z^+,t^+)}\rangle}=|H_{Lv}|e^{j\phi_v},}
    \label{eqn:equ7}
\end{equation}
\begin{equation}
    \textcolor{black}{H_{Lw}(\lambda_x^+, y^+) = \frac{\langle \hat{w}^+(\lambda_x^+,y^+,z^+,t^+)\overline{\hat{w}^+(\lambda_x^+,y_O^+,z^+,t^+)}\rangle}{\langle \hat{w}^+(\lambda_x^+,y_O^+,z^+,t^+)\overline{\hat{w}^+(\lambda_x^+,y_O^+,z^+,t^+)}\rangle}=|H_{Lw}|e^{j\phi_w}.}
    \label{eqn:equ8}
\end{equation}
\textcolor{black}{and $<>$ denotes the average in time and in the spanwise direction}. $\phi_u$, $\phi_v$ and $\phi_w$ are the phase differences of the velocities at the two heights. Following Baars \emph{et al.} \cite{Baars2016}, we also use a bandwidth moving filter of 25\% to smooth the original spectral kernel functions. 

Moreover, Talluru \emph{et al.} \cite{Talluru2014} has revealed similar amplitude modulations of the three velocity components by outer large-scale streamwise velocity. Then the universal velocity components $(u^*, v^*, w^*)$ can be obtained separately in the three directions, once the modulation coefficients $(\Gamma_{uu}, \Gamma_{uv}, \Gamma_{uw})$ are determined, i.e.
\begin{gather}
    \textcolor{black}{u^*(x^+,y^+,z^+,t^+) = \frac{u^+(x^+,y^+,z^+,t^+)-u_L^+(x^+,y^+,z^+,t^+)}{1+\Gamma_{uu}(y^+)u_L^+(x^+,y^+,z^+,t^+)},} \label{eqn:equ9} \\
    \textcolor{black}{v^*(x^+,y^+,z^+,t^+) = \frac{v^+(x^+,y^+,z^+,t^+)-v_L^+(x^+,y^+,z^+,t^+)}{1+\Gamma_{uv}(y^+)u_L^+(x^+,y^+,z^+,t^+)},}  \\
    \textcolor{black}{w^*(x^+,y^+,z^+,t^+) = \frac{w^+(x^+,y^+,z^+,t^+)-w_L^+(x^+,y^+,z^+,t^+)}{1+\Gamma_{uw}(y^+)u_L^+(x^+,y^+,z^+,t^+)}.}
    \label{eqn:equ10}
\end{gather}
The amplitude modulation coefficients $(\Gamma_{uu}, \Gamma_{uv}, \Gamma_{uw})$ are determined through iterative procedures separately, that stop when the amplitude modulation factor AM is zero \citep{Marusic2010,Mathis2011,Baars2016}, which can be written as
\begin{gather}
    AM(u^*) = \frac{\langle E_L(u^*) u_L^+ \rangle}{\sqrt{\langle E_L(u^*)^2 \rangle \langle u_L^{+2} \rangle}}, \label{eqn:equ11} \\
    AM(v^*) = \frac{\langle E_L(v^*) u_L^+ \rangle}{\sqrt{\langle E_L(v^*)^2 \rangle \langle u_L^{+2} \rangle}}, \\
    AM(w^*) = \frac{\langle E_L(w^*) u_L^+ \rangle}{\sqrt{\langle E_L(w^*)^2 \rangle \langle u_L^{+2} \rangle}},
    \label{eqn:equ12}
\end{gather}
where $E_L(u^*)$, $E_L(v^*)$ and $E_L(w^*)$ denote the envelopes of $u^*$, $v^*$ and $w^*$, respectively, which are obtained by Hilbert transform. More details about the procedure can be found in Mathis \emph{et al.} \cite{Mathis2011} and Baars \emph{et al.} \cite{Baars2016}.

According to the aforementioned works \cite[]{Marusic2010,Mathis2011,Baars2016}, the determination procedure of near-wall demodulated small-scale fluctuating velocities ($u^*$, $v^*$ and $w^*$) as well as the modulation coefficients ($\Gamma_{uu}$, $\Gamma_{uv}$ and $\Gamma_{uw}$) can be summarized as follows:

\begin{enumerate}
    \item \quad Calculate the kernel functions $H_{Lu}$, $H_{Lv}$ and $H_{Lw}$, given the outer reference height $y^+_O$, according to (\ref{eqn:equ6})-(\ref{eqn:equ8}).
    \item \quad Calculate the near-wall large-scale velocity footprints $u_L^+$, $v_L^+$ and $w_L^+$, according to (\ref{eqn:equ3})-(\ref{eqn:equ5}). 
    \item \quad Get the near-wall small-scale velocity components by subtracting the large-scale footprints from the total fluctuations, i.e. $(u_{S}^+, v_S^+, w_S^+) = (u^+, v^+, w^+) - (u_L^+, v_L^+, w_L^+)$.
    \item \quad De-modulate ($u_S^+$, $v_S^+$, $w_S^+$) to obtain ($u^*$, $v^*$, $w^*$) and ($\Gamma_{uu}$, $\Gamma_{uv}$, $\Gamma_{uw}$) through the iterative procedure, i.e., (\ref{eqn:equ9})-(\ref{eqn:equ12}). To be more specific, one firstly assumes an initial guess of ($\Gamma_{uu}$, $\Gamma_{uv}$, $\Gamma_{uw}$) at each height, substitutes into (\ref{eqn:equ9})-(\ref{eqn:equ10}) to get ($u^*$, $v^*$, $w^*$), and uses them in (\ref{eqn:equ11})-(\ref{eqn:equ12}) to check whether the AMs are zero. If not, choosing another set of ($\Gamma_{uu}$, $\Gamma_{uv}$, $\Gamma_{uw}$) to repeat the above procedure, until finding a set of ($u^*$, $v^*$, $w^*$) that leads to zero amplitude modulation coefficients at this height. 
\end{enumerate}

\section{Reynolds-number-independent near-wall motions} 
% \label{sec:res}

% \subsection{Extracting results at $Re_\tau \ge 1000$}
% \subsubsection{Turbulence statistics}

\begin{figure}
\centering
\begin{minipage}{0.325\linewidth}
\centerline{\includegraphics[width=\textwidth]{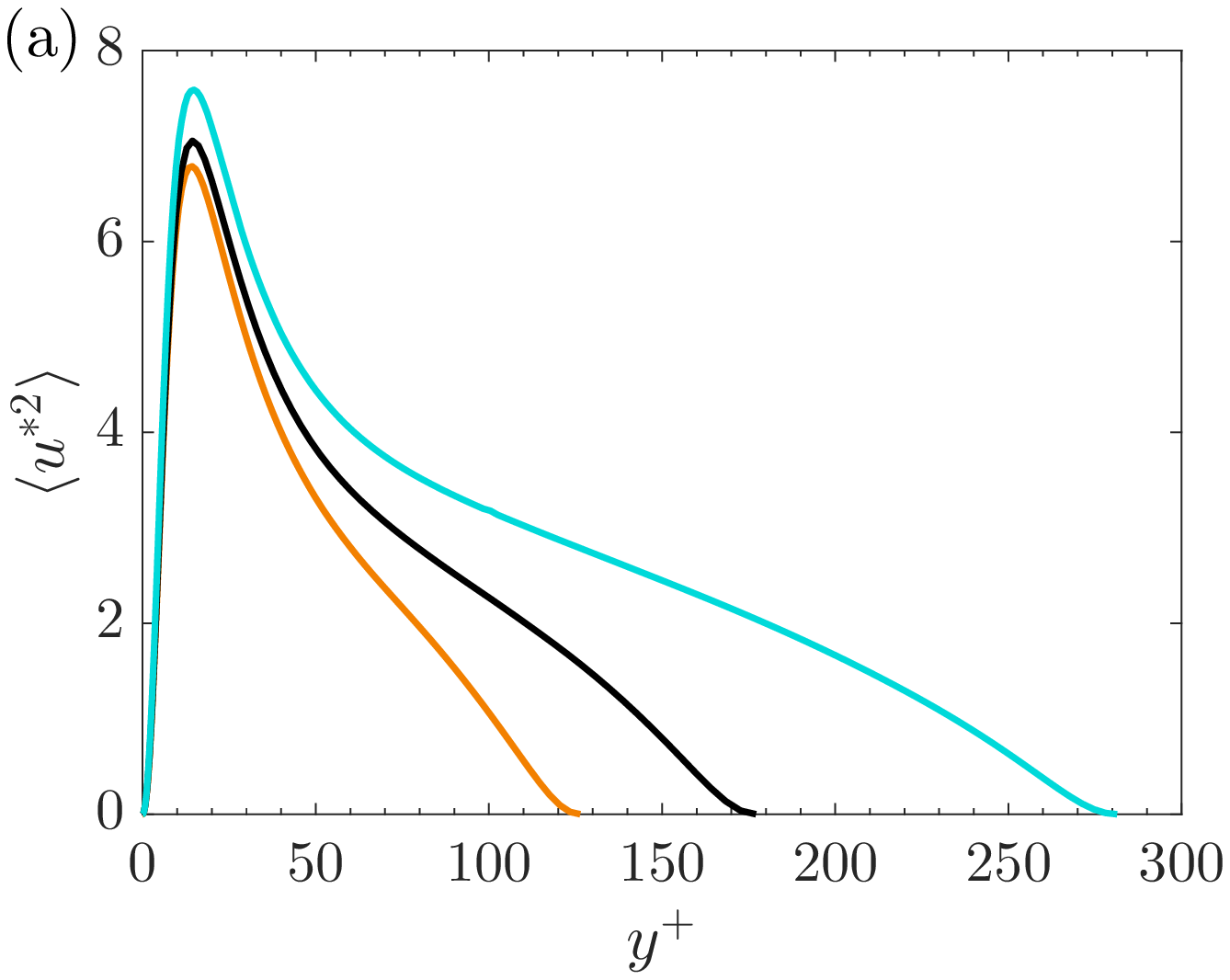}}
\end{minipage}
\hfill
\begin{minipage}{0.325\linewidth}
\centerline{\includegraphics[width=\textwidth]{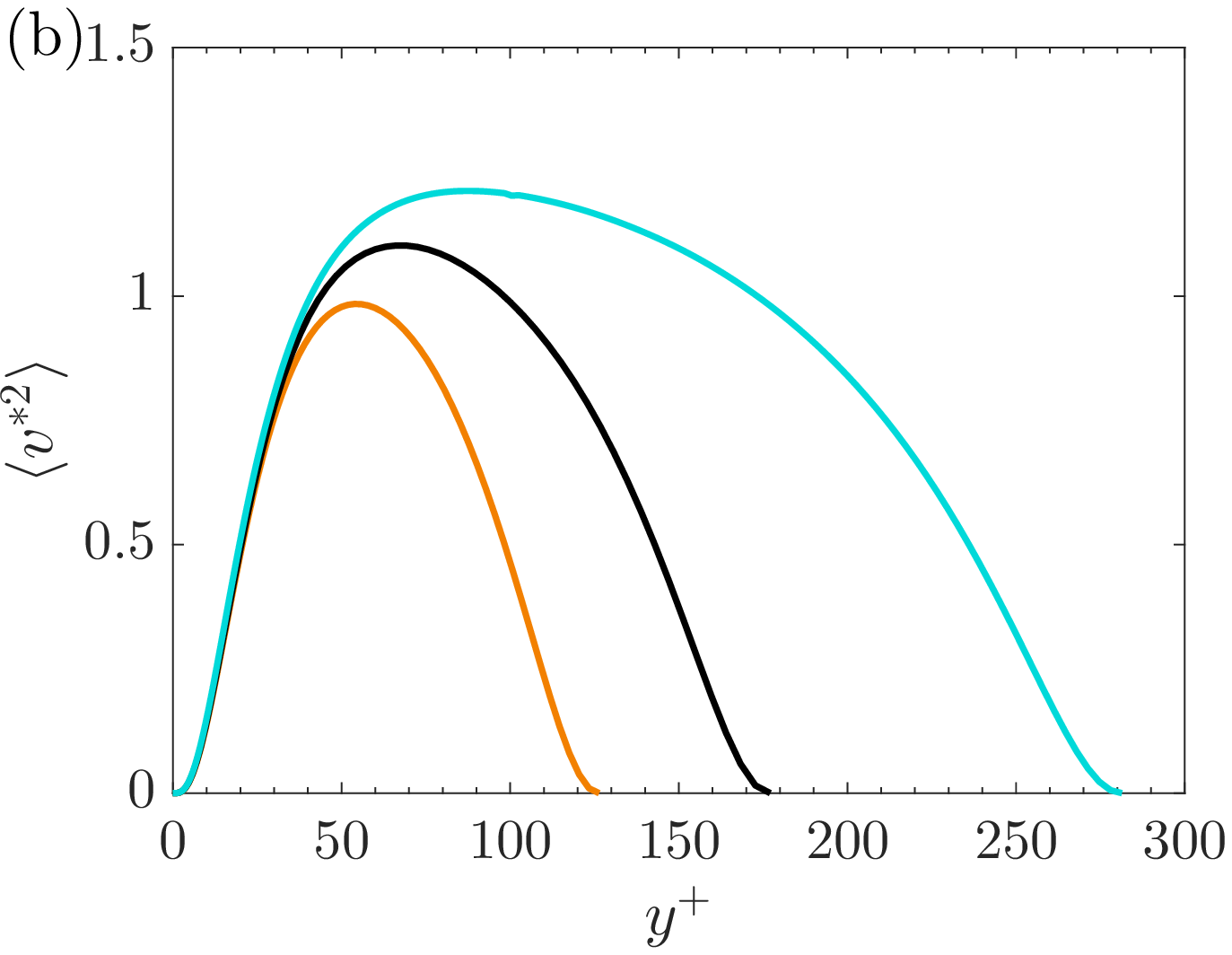}}
\end{minipage}
\hfill
\begin{minipage}{0.325\linewidth}
\centerline{\includegraphics[width=\textwidth]{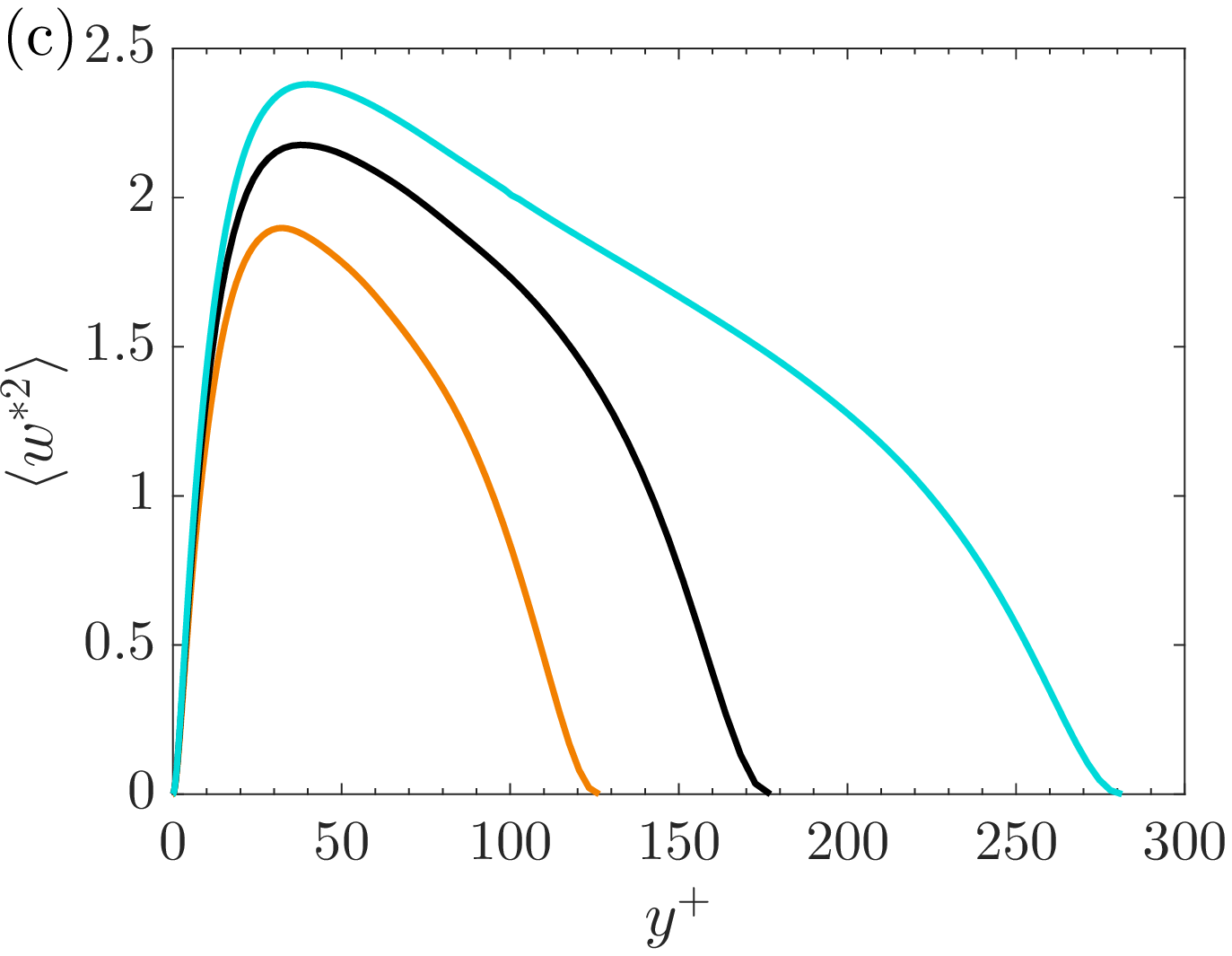}}
\end{minipage}
\caption{Turbulence intensities of the extracted small-scale near-wall fluctuating velocities at $Re_\tau=1000$, 2000 and 5200 using $y_O^+=3.9\sqrt{Re_\tau}$: (a) streamwise turbulence intensity, (b) wall-normal turbulence intensity and (c) spanwise turbulence intensity. Refer to table~\ref{tab:tab1} for the line colors.}
\label{fig:fig3}
\end{figure}

\begin{figure}
\centering
\begin{minipage}{0.49\linewidth}
\centerline{\includegraphics[width=\textwidth]{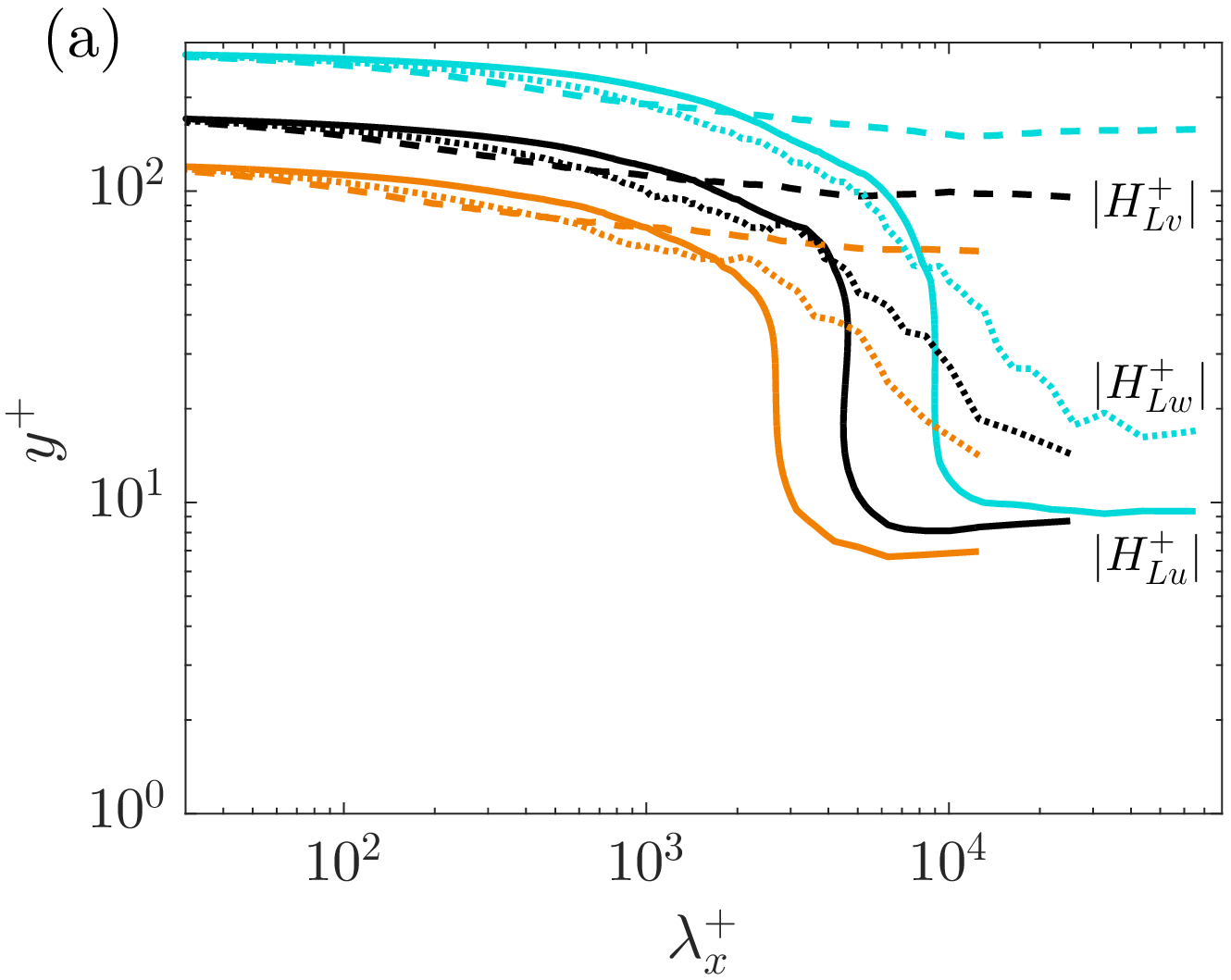}}
\end{minipage}
\hfill
\begin{minipage}{0.49\linewidth}
\centerline{\includegraphics[width=\textwidth]{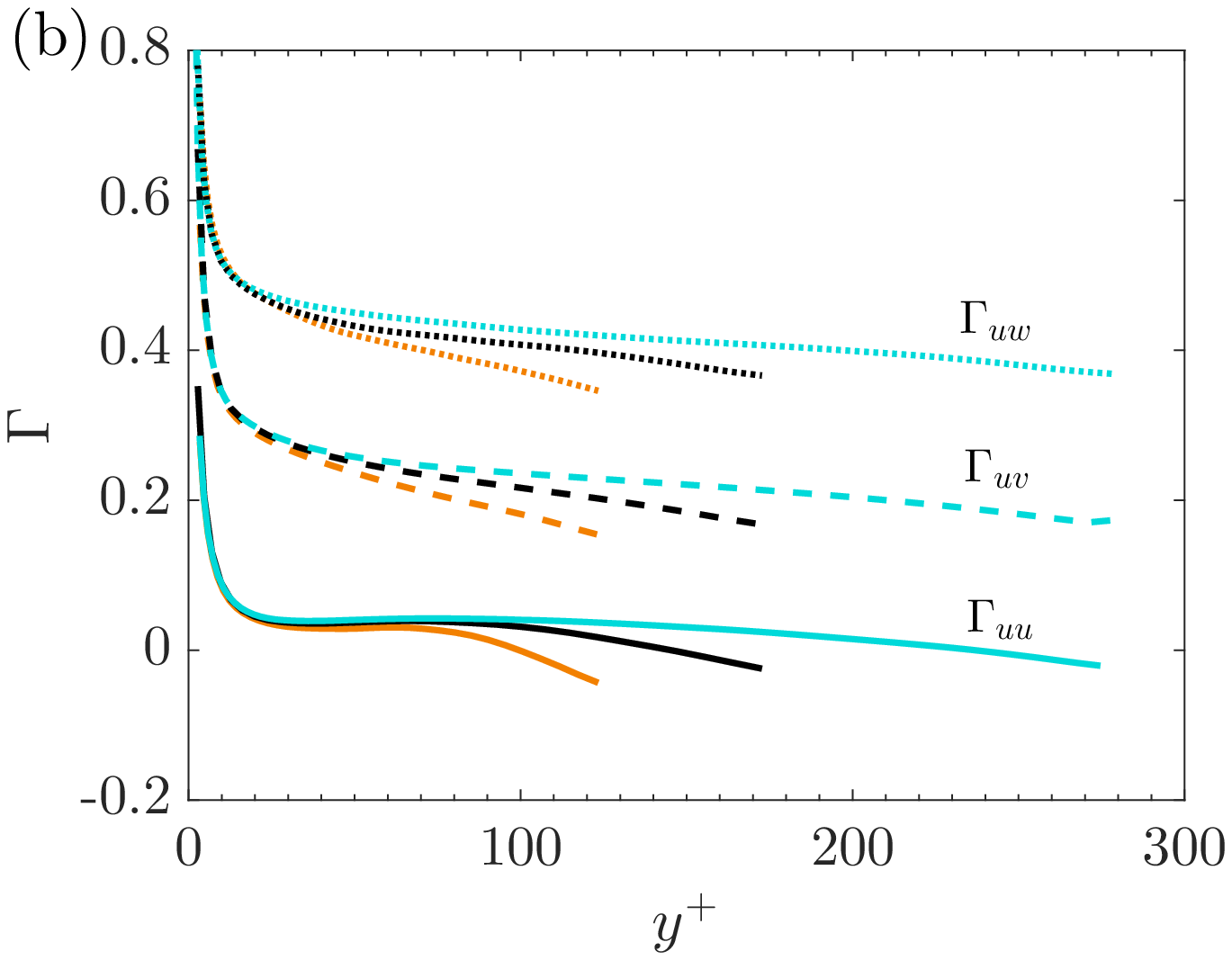}}
\end{minipage}
\caption{(a) Contour lines for the magnitudes of the scale-dependent kernel functions ($|H_{Lu}|$, $|H_{Lv}|$ and $|H_{Lw}|$) with a constant value of 0.6, and (b) profiles of the modulation coefficients ($\Gamma_{uu}$, $\Gamma_{uv}$ and $\Gamma_{uw}$) at $Re_\tau=1000, 2000$ and 5200, {where the solid lines represent $|H_{Lu}|$ and $\Gamma_{uu}$, dashed lines for $|H_{Lv}|$ and $\Gamma_{uv}$ and dotted lines for $|H_{Lw}|$ and $\Gamma_{uw}$}, using $y_O^+=3.9\sqrt{Re_\tau}$, the vertical and spanwise modulation coefficients are moved up by 0.2 and 0.4, respectively. Refer to table~\ref{tab:tab1} for the line colors.}
\label{fig:fig4}
\end{figure}

Now we present application results of the extracting scheme (\ref{eqn:equ9} and \ref{eqn:equ10}) for the near-wall Reynolds-number-independent velocity fields. The primary input is the outer reference height $y_O^+$, since the kernel functions ($H_{Lu}$, $H_{Lv}$, $H_{Lw}$) and the imprint velocities ($u_L^+$, $v_L^+$, $w_L^+$) can be directly calculated once $y_O^+$ is given. In the majority of the previous studies, $y_O^+$ is chosen at the centre of the logarithmic layer \cite[]{Mathis2011,Inoue2012,Baars2016}, i.e. $y_O^+\approx3.9\sqrt{Re_\tau}$, because the outer spectral peak of streamwise velocity fluctuations is located at this height \cite[]{Hutchins2007a}. 
% Although generally good agreement of the extracted $u^*$ statistics has been reported at high Reynolds numbers \cite[]{Baars2016}, 
As shown in \textcolor{black}{FIG}~\ref{fig:fig3}, we find that the extracted near-wall motions from the channel DNS data are actually Reynolds number dependent in the range of $Re_\tau=1000 \sim 5200$. 
Additionally, the magnitudes of the footprint kernel functions ($H_{Lu}$, $H_{Lv}$ and $H_{Lw}$) and the amplitude modulation coefficients ($\Gamma_{uu}$, $\Gamma_{uv}$ and $\Gamma_{uw}$) are also Reynolds-number dependent, which is displayed in \textcolor{black}{FIG}~\ref{fig:fig4}.
These results indicate that the near-wall influence from a portion of outer motions may be still included, if we trust the inner-outer interaction hypothesis and choose $y_O^+\approx3.9\sqrt{Re_\tau}$. This could be true as Townsend \cite{Townsend1976} already proposed that the wall-attached eddies can influence the near-wall flow, and the geometrically self-similar eddies with sizes of $l_{y}^+<3.9\sqrt{Re_\tau}$ are probably active near the wall \cite[]{Perry1982}. Therefore it suggests that the truly Reynolds-number-independent near-wall motions may be extracted by reducing the input reference height $y_O^+$.
In addition, the small scale velocities $(u^*, v^*, w^*)$ should be zero at $y^+=y^+_O$ according to the PIO model (see \S 3). Therefore, we stress that the optimal $y^+_O$ should be a constant in viscous scaling, since Reynolds number independence requires that the zero crossing height of $(u^*,v^*, w^*)$ should be the same at different Reynolds numbers.

% \textcolor{red}{I'm here!}

\begin{figure}
\centering
\begin{minipage}{0.49\linewidth}
\centerline{\includegraphics[width=\textwidth]{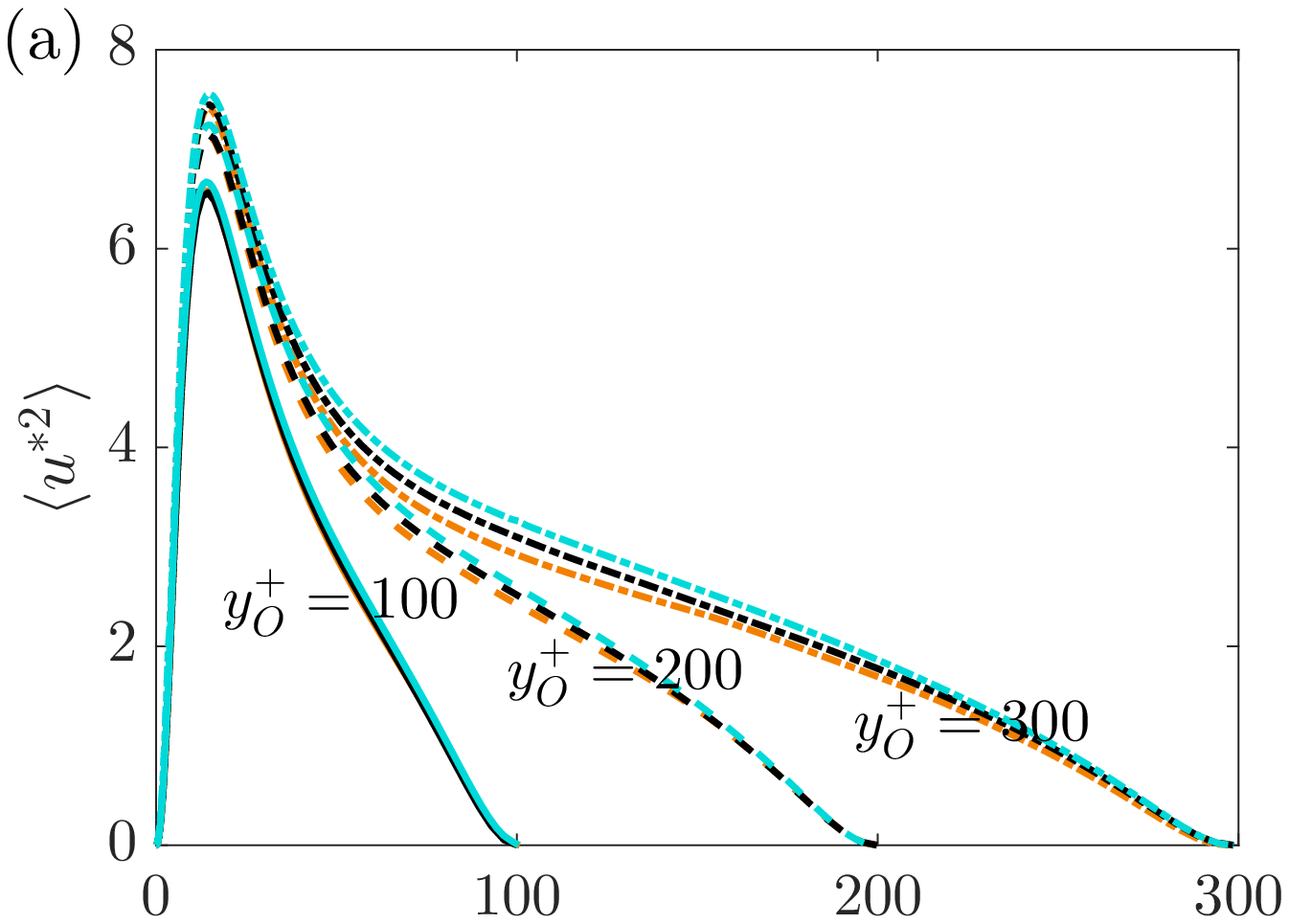}}
\end{minipage}
\hfill
\begin{minipage}{0.49\linewidth}
\centerline{\includegraphics[width=\textwidth]{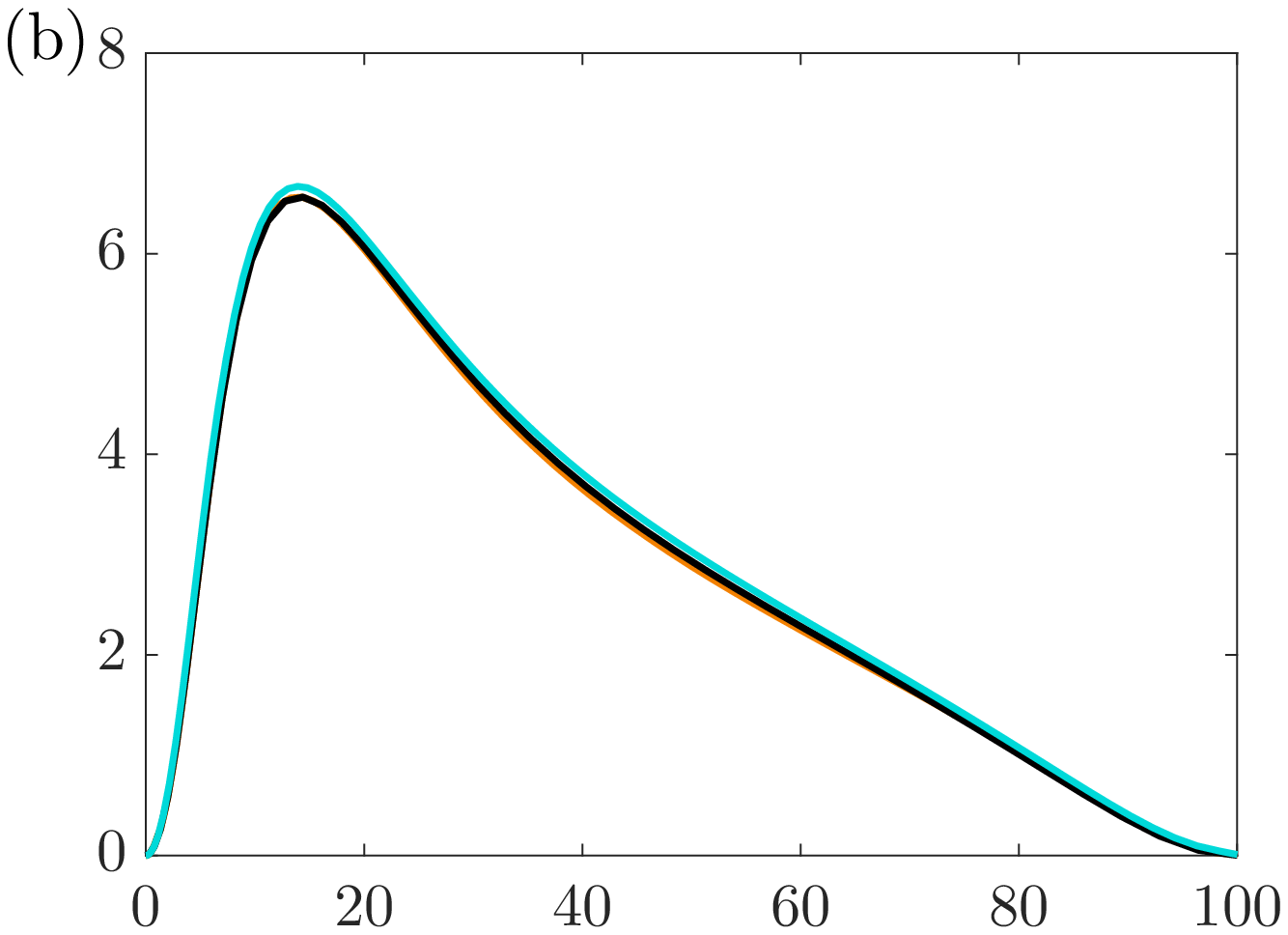}}
\end{minipage}
\vfill
\begin{minipage}{0.49\linewidth}
\centerline{\includegraphics[width=\textwidth]{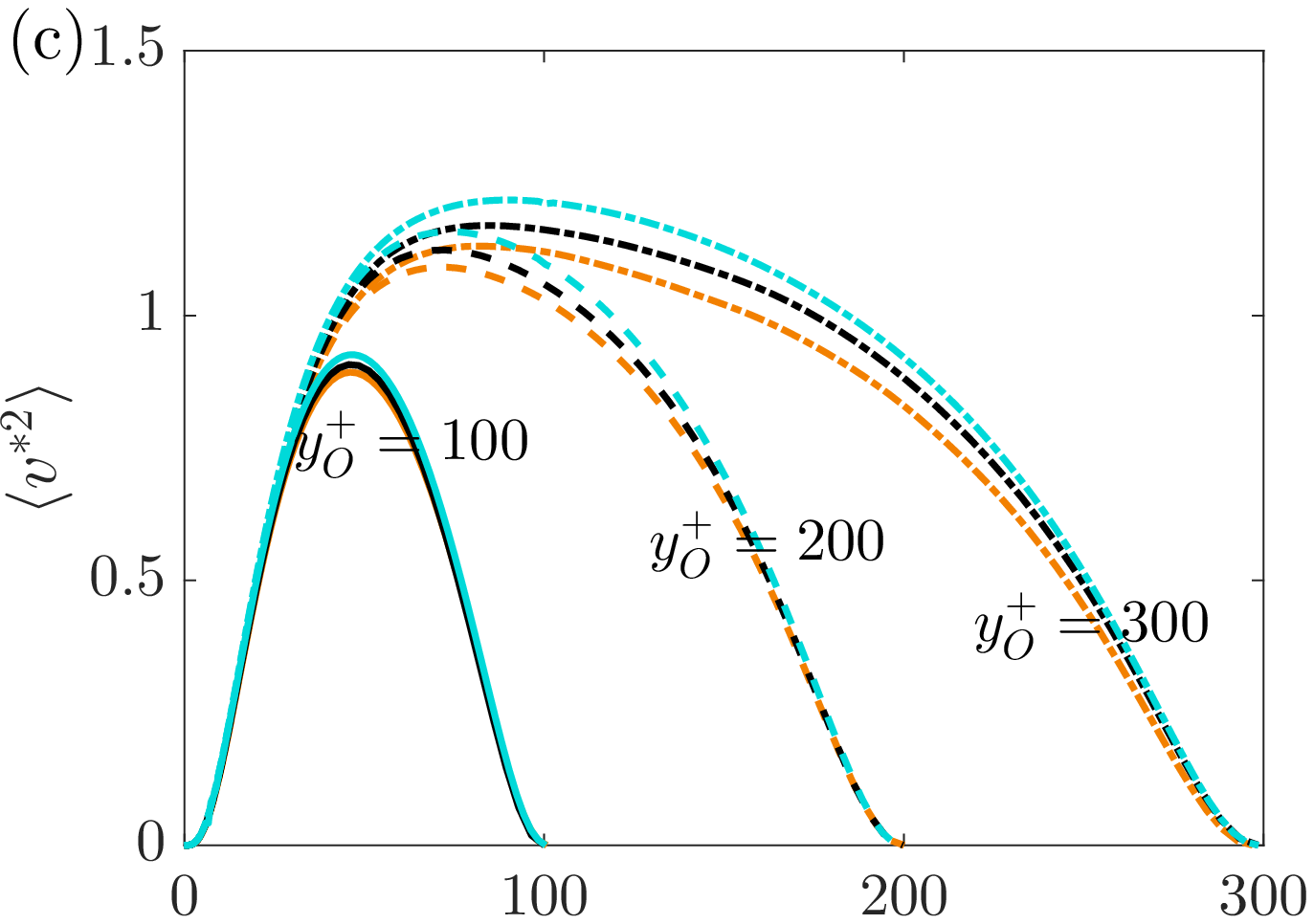}}
\end{minipage}
\hfill
\begin{minipage}{0.49\linewidth}
\centerline{\includegraphics[width=\textwidth]{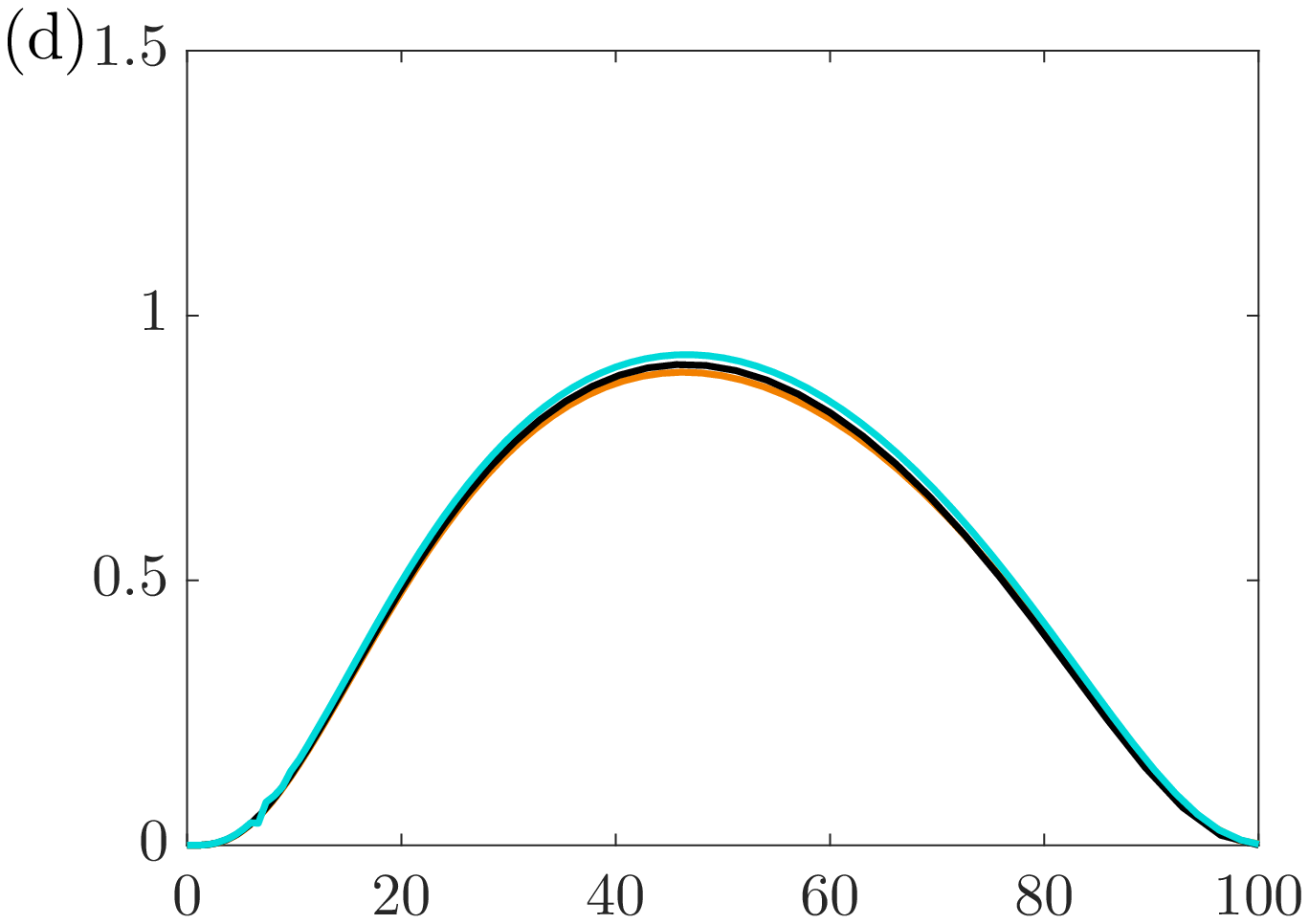}}
\end{minipage}
\vfill
\begin{minipage}{0.49\linewidth}
\centerline{\includegraphics[width=\textwidth]{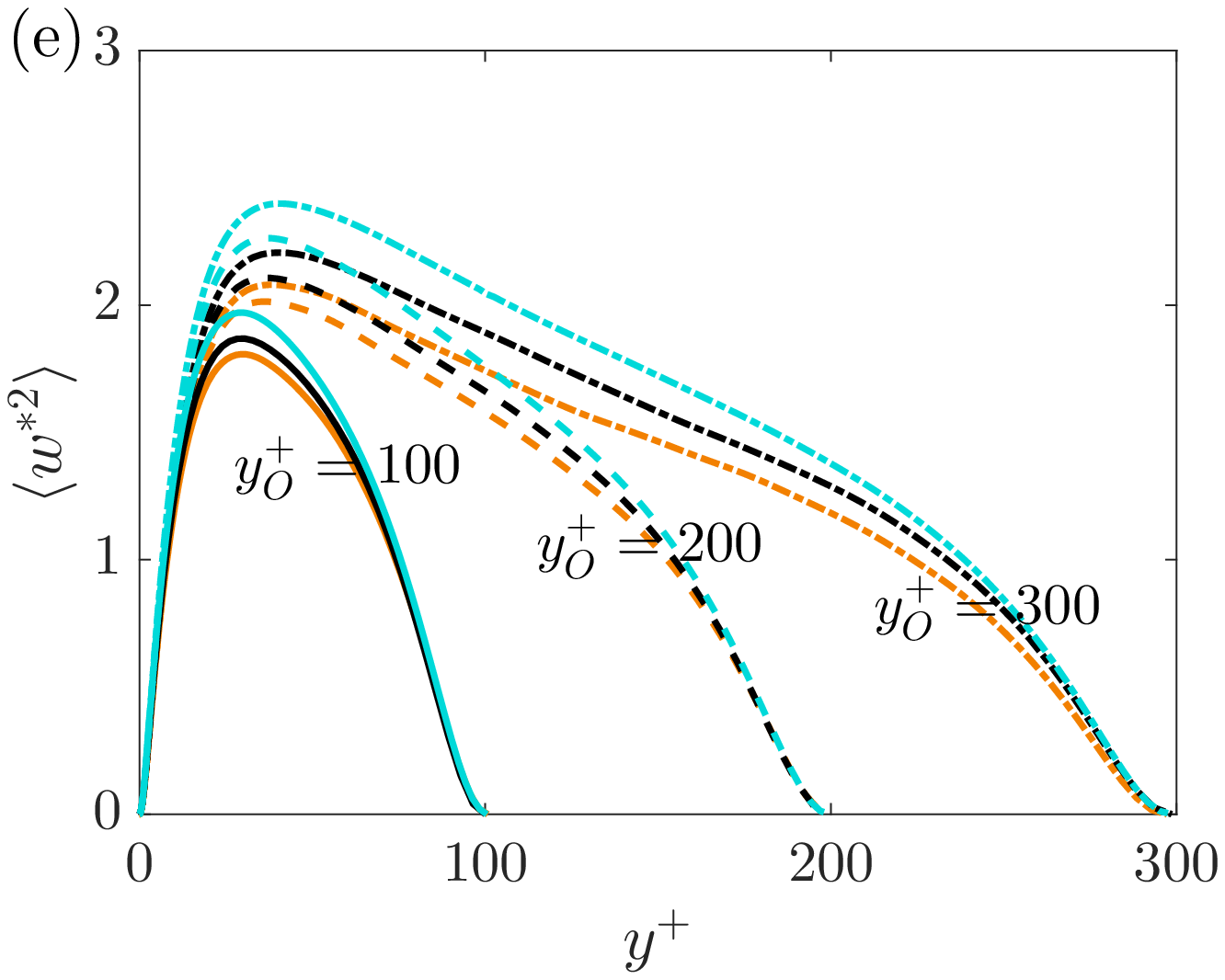}}
\end{minipage}
\hfill
\begin{minipage}{0.49\linewidth}
\centerline{\includegraphics[width=\textwidth]{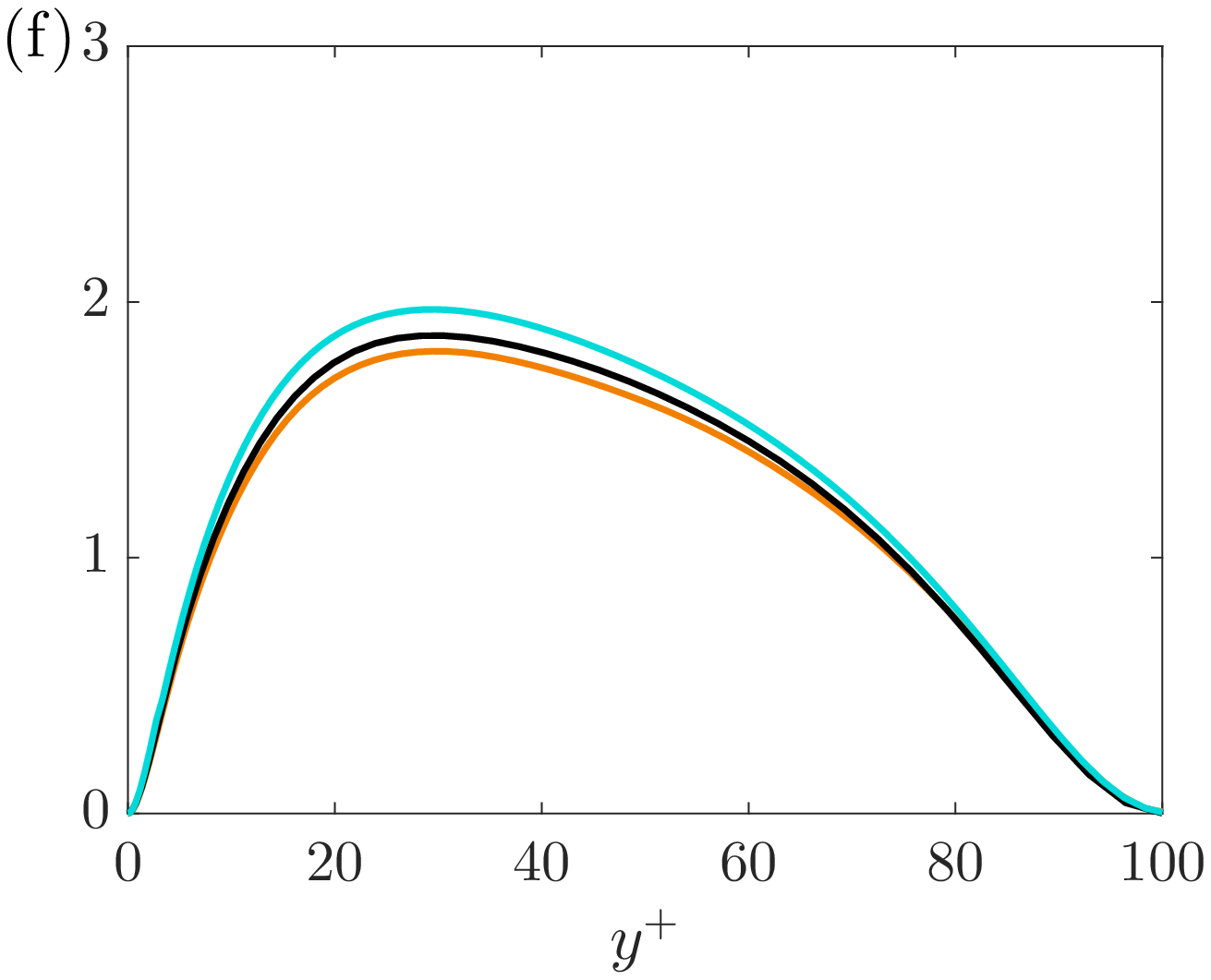}}
\end{minipage}
\caption{Turbulence intensities of the extracted Reynolds-number-independent fluctuating velocity components at $Re_\tau =$ 1000, 2000 and 5200: (a, b) streamwise velocity; (c, d) wall-normal velocity; (e, f) spanwise velocity.  In (a, c, e), the reference wall-normal position $y_O^+$ varies from 300 to \textcolor{black}{100}: $y_O^+ = 300$ (dot-dashed lines), $y_O^+ = 200$ (dashed lines) and $y_O^+ = 100$ (solid lines). In (b, d, f), only the results using the reference position $y_O^+ = 100$ are shown. Refer to table~\ref{tab:tab1} for the line colors.}
\label{fig:fig5}
\end{figure}

\begin{figure}
\centering
\begin{minipage}{0.49\linewidth}
\centerline{\includegraphics[width=\textwidth]{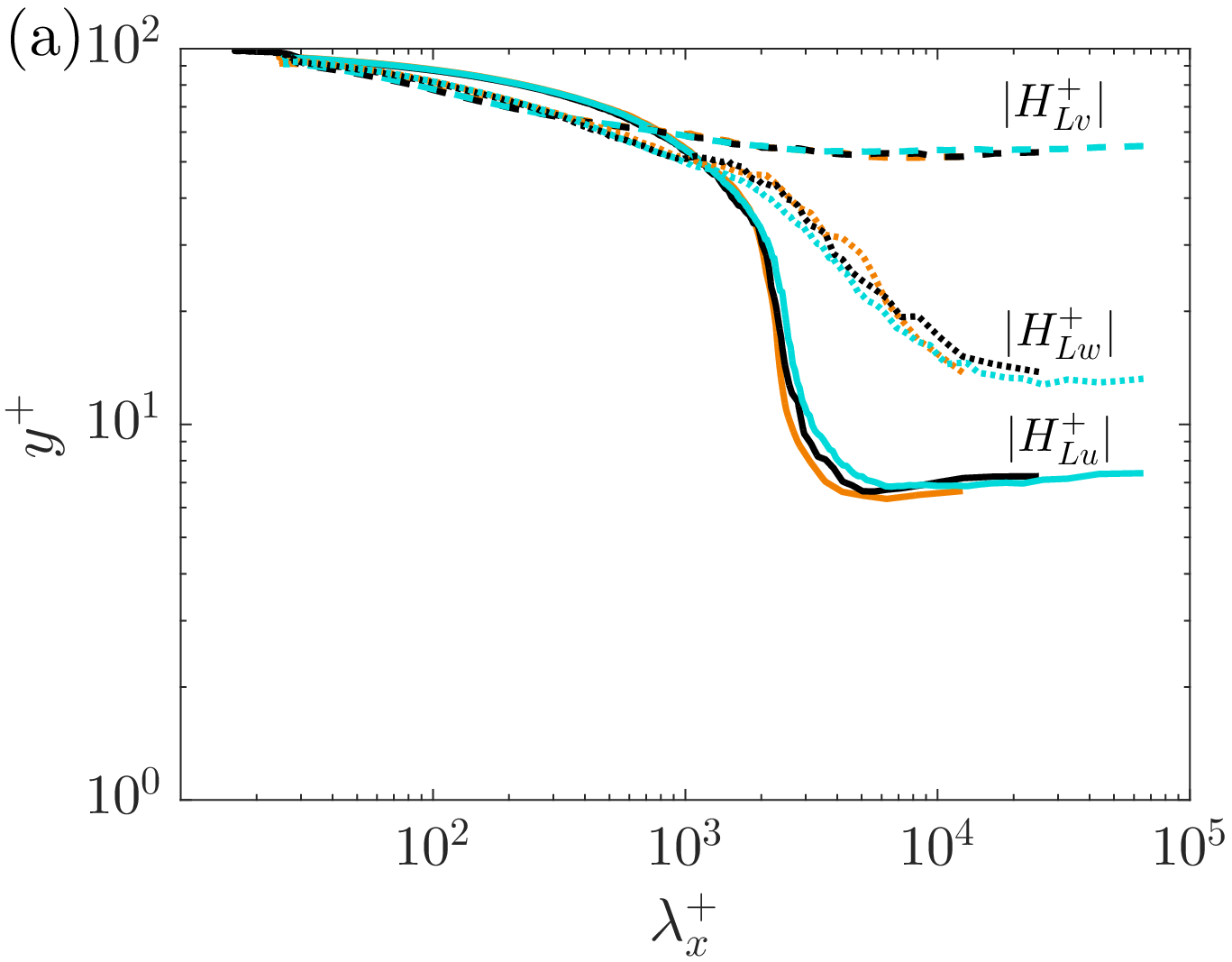}}
\end{minipage}
\hfill
\begin{minipage}{0.49\linewidth}
\centerline{\includegraphics[width=\textwidth]{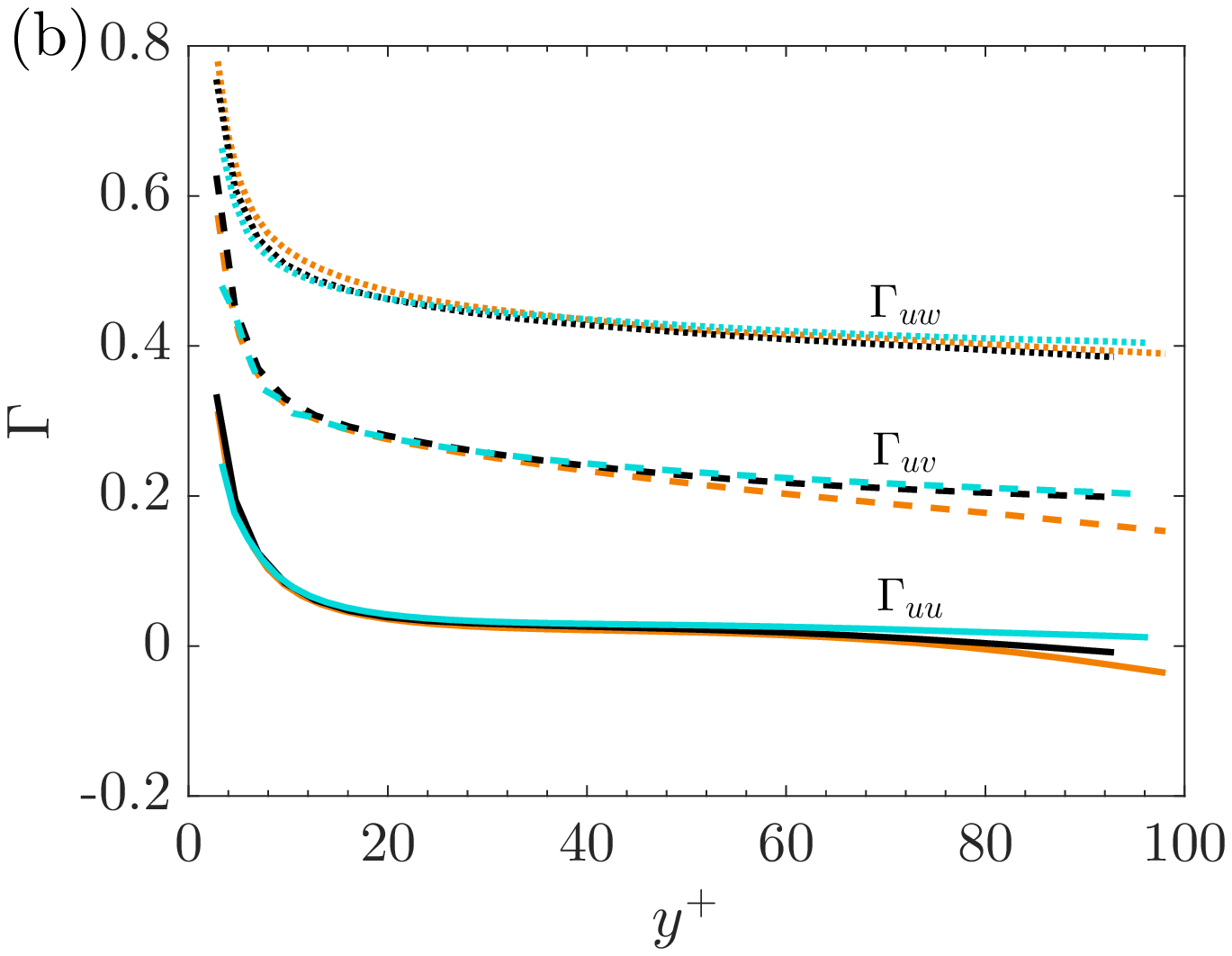}}
\end{minipage}
\caption{(a) Contour lines for the magnitudes of the scale-dependent kernel functions ($|H_{Lu}|$, $|H_{Lv}|$ and $|H_{Lw}|$) with a constant value of 0.6, and (b) profiles of the modulation coefficients ($\Gamma_{uu}$, $\Gamma_{uv}$ and $\Gamma_{uw}$) at $Re_\tau=1000, 2000$ and 5200, {where the solid lines represent $|H_{Lu}|$ and $\Gamma_{uu}$, dashed lines for $|H_{Lv}|$ and $\Gamma_{uv}$ and dotted lines for $|H_{Lw}|$ and $\Gamma_{uw}$}, using $y_O^+=100$. And the vertical and spanwise modulation coefficients are moved up by 0.2 and 0.4, respectively. Refer to table~\ref{tab:tab1} for the line colors.}
\label{fig:fig6}
\end{figure}

% In the following, the turbulence intensities of the extracted near-wall velocity fluctuations according to equations (\ref{eqn:equ5}-\ref{eqn:equ7}) are compared at the Reynolds numbers $Re_\tau=1000, 2000$ and 5200, as shown in figure~\ref{fig:fig13} (a, c, e). 
By systematically decreasing the reference height $y_O^+$ from 300 to 200 and finally 100, it clearly shows that the extracted $\langle u^{*2} \rangle$, $\langle v^{*2} \rangle$, $\langle w^{*2} \rangle$ are less dependent on Reynolds number, if $y_O^+$ is smaller, as shown in \textcolor{black}{FIG}~\ref{fig:fig5} (a, c, e). Furthermore, the Reynolds-number invariant $\langle u^{*2} \rangle$, $\langle v^{*2} \rangle$, $\langle w^{*2} \rangle$ could be well defined with $y_O^+=100$ and at $Re_\tau = 1000 \sim 5200$. In \textcolor{black}{FIG}~\ref{fig:fig5} (b, d, f), only the extracted ($\langle u^{*2} \rangle$, $\langle v^{*2} \rangle$, $\langle w^{*2} \rangle$) with $y_O^+=100$ at the three Reynolds numbers are displayed and good collapse can be found. 
In addition, excellent Reynolds-number independence of the magnitudes of the footprint kernel functions ($H_{Lu}$, $H_{Lv}$ and $H_{Lw}$) and the amplitude modulation coefficients ($\Gamma_{uu}$, $\Gamma_{uv}$ and $\Gamma_{uw}$) are also obtained and demonstrated in \textcolor{black}{FIG}~\ref{fig:fig6}. 
% It should also be noted that \citet{Perry1982} proposed that the height of the smallest attached eddies is $l_y^+ \sim O(100)$.
The contour plots of $|H_{Lu}|$, $|H_{Lv}|$ and $|H_{Lw}|$ with $y^+_O=100$, 200 and 300 at $Re_\tau = 1000 \sim 5200$ are also presented in Appendix \ref{sec:appenB}. It should be noted that although $y_O^+=200$ and 300 are constants in viscous units, they actually locate in the logarithmic layer or outer layer. This is why $y^+_O=100$ works best, as it may be regarded as a critical height dividing the inner and outer regions.

\begin{figure}
\centering
\begin{minipage}{0.49\linewidth}
\centerline{\includegraphics[width=\textwidth]{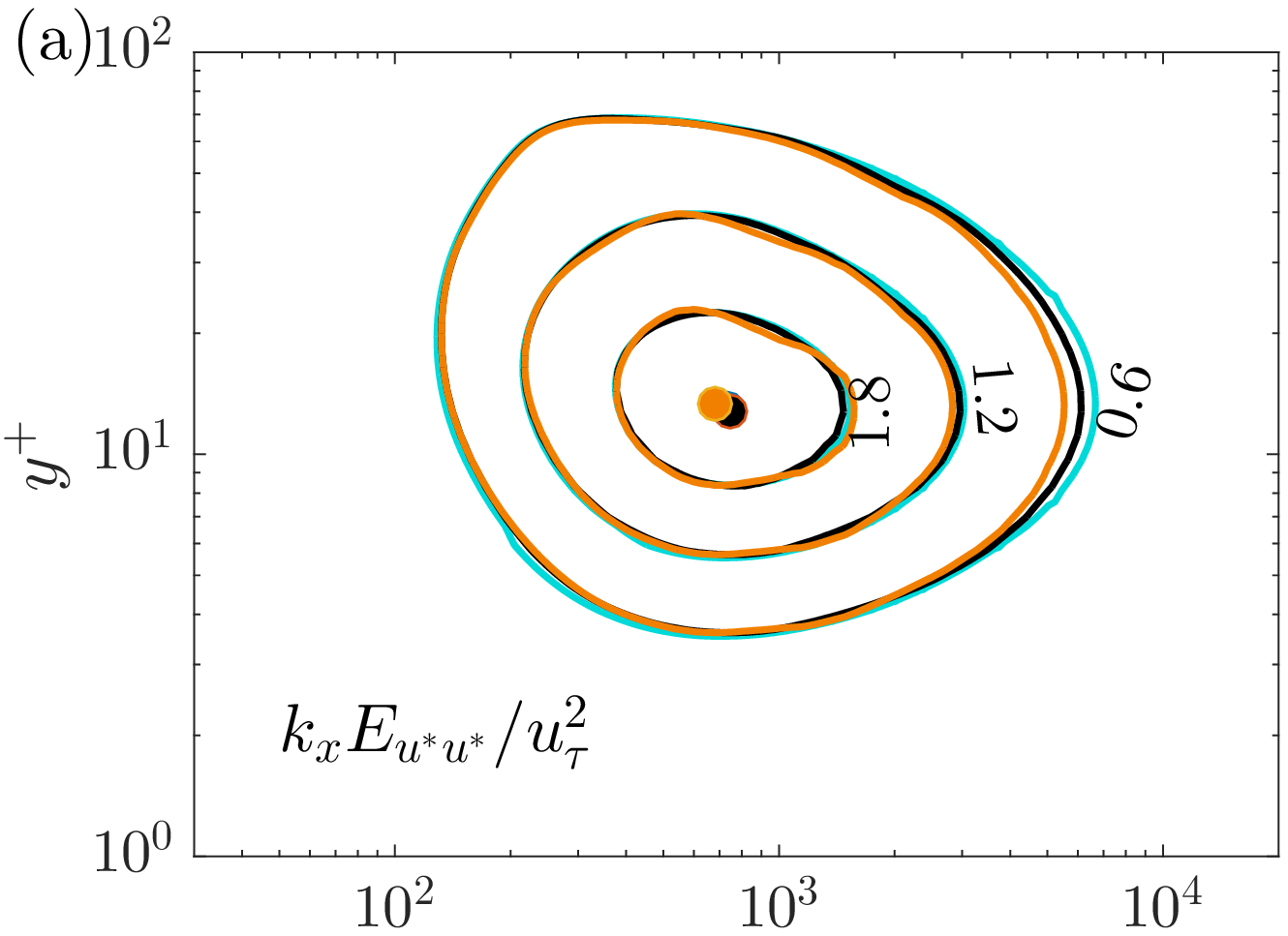}}
\end{minipage}
\hfill
\begin{minipage}{0.49\linewidth}
\centerline{\includegraphics[width=\textwidth]{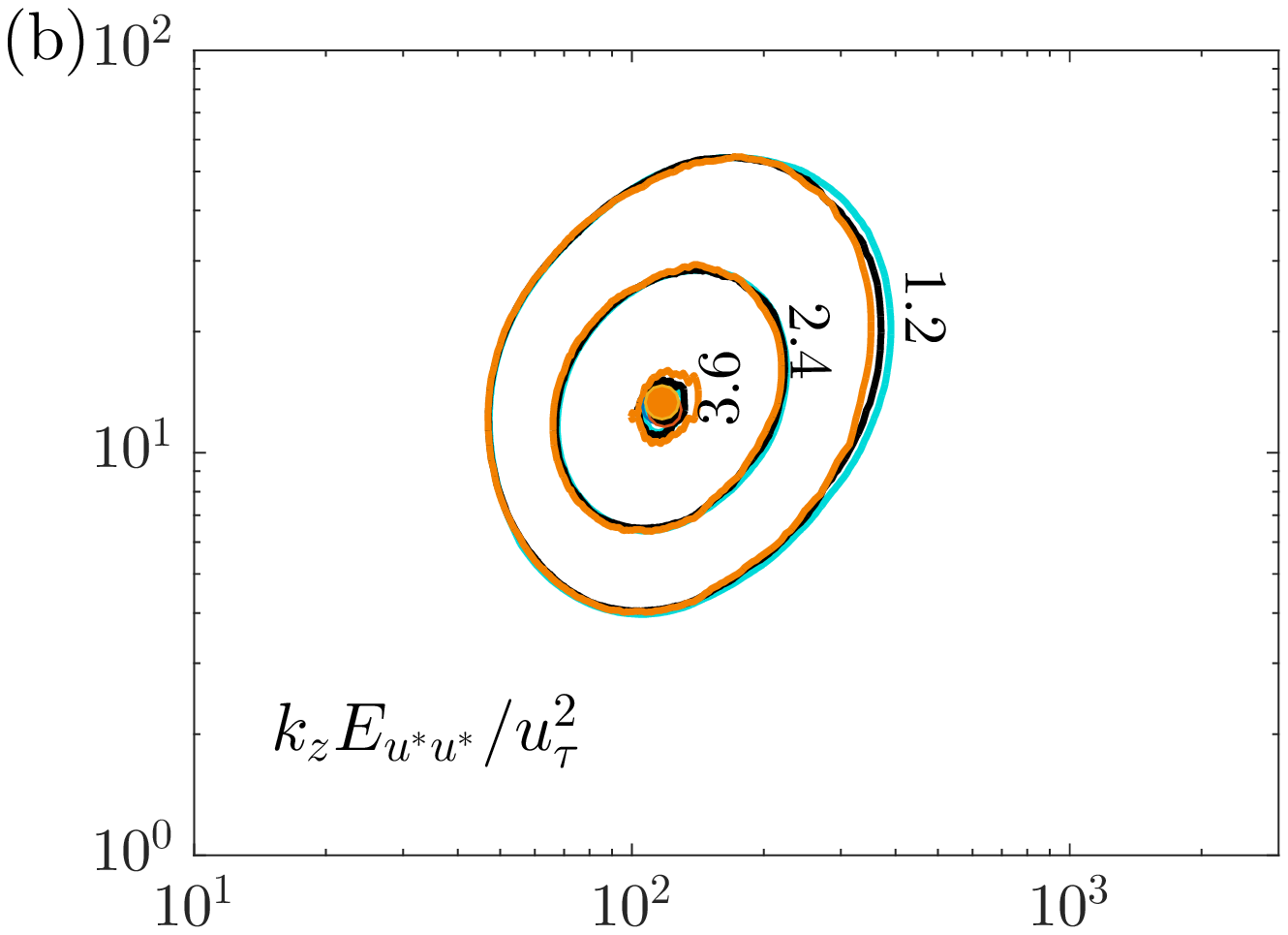}}
\end{minipage}
\vfill
\begin{minipage}{0.49\linewidth}
\centerline{\includegraphics[width=\textwidth]{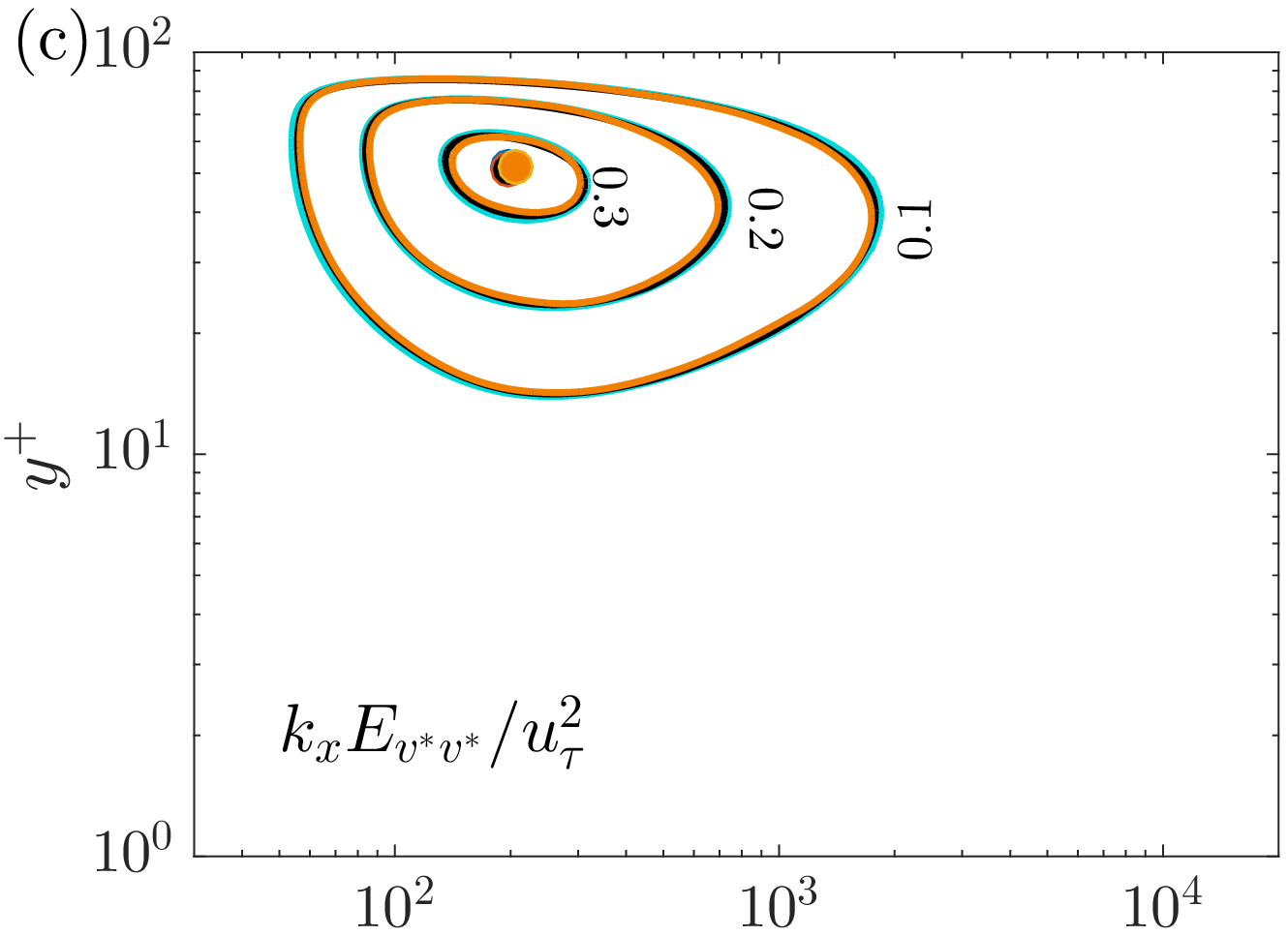}}
\end{minipage}
\hfill
\begin{minipage}{0.49\linewidth}
\centerline{\includegraphics[width=\textwidth]{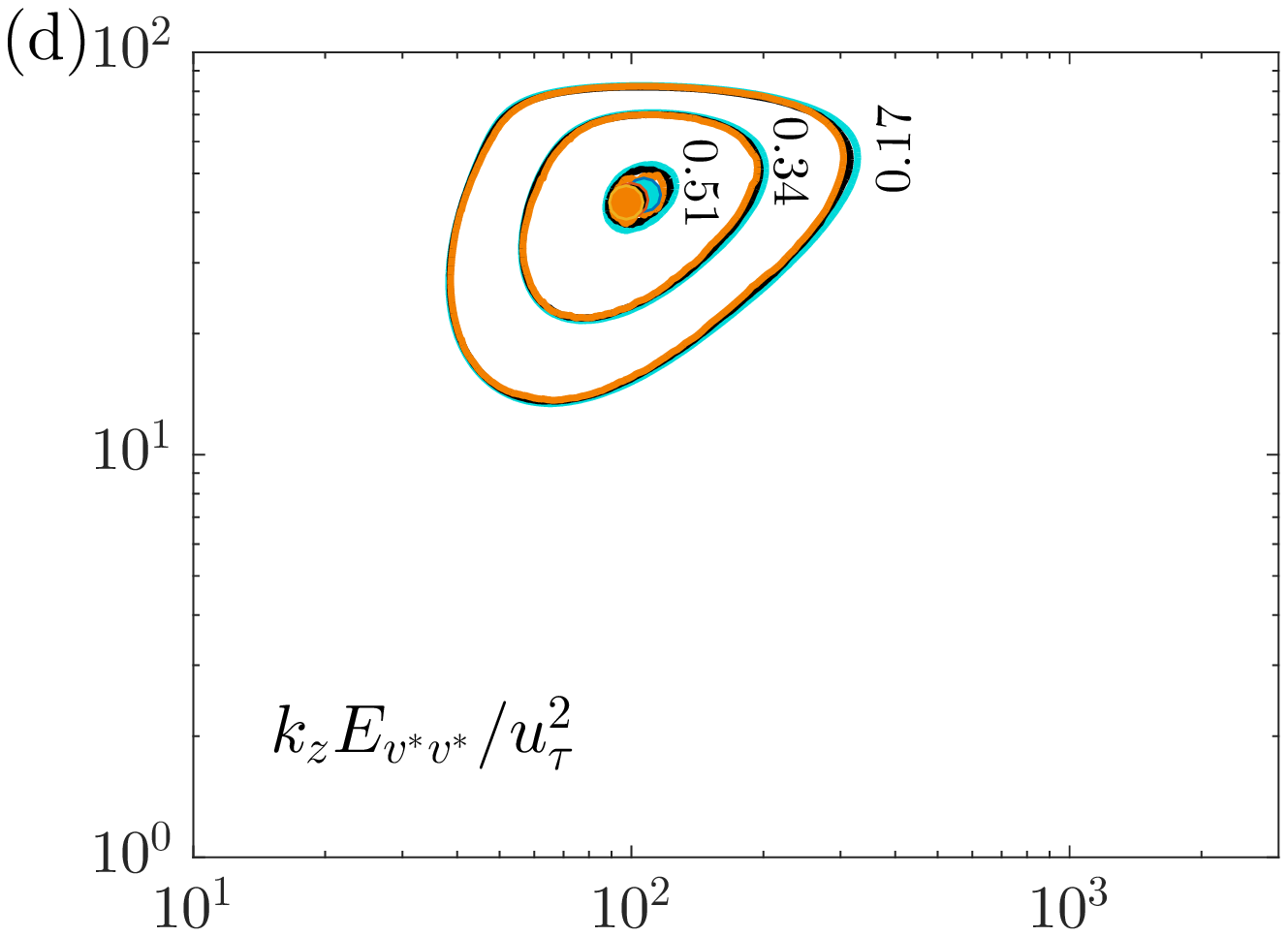}}
\end{minipage}
\begin{minipage}{0.49\linewidth}
\centerline{\includegraphics[width=\textwidth]{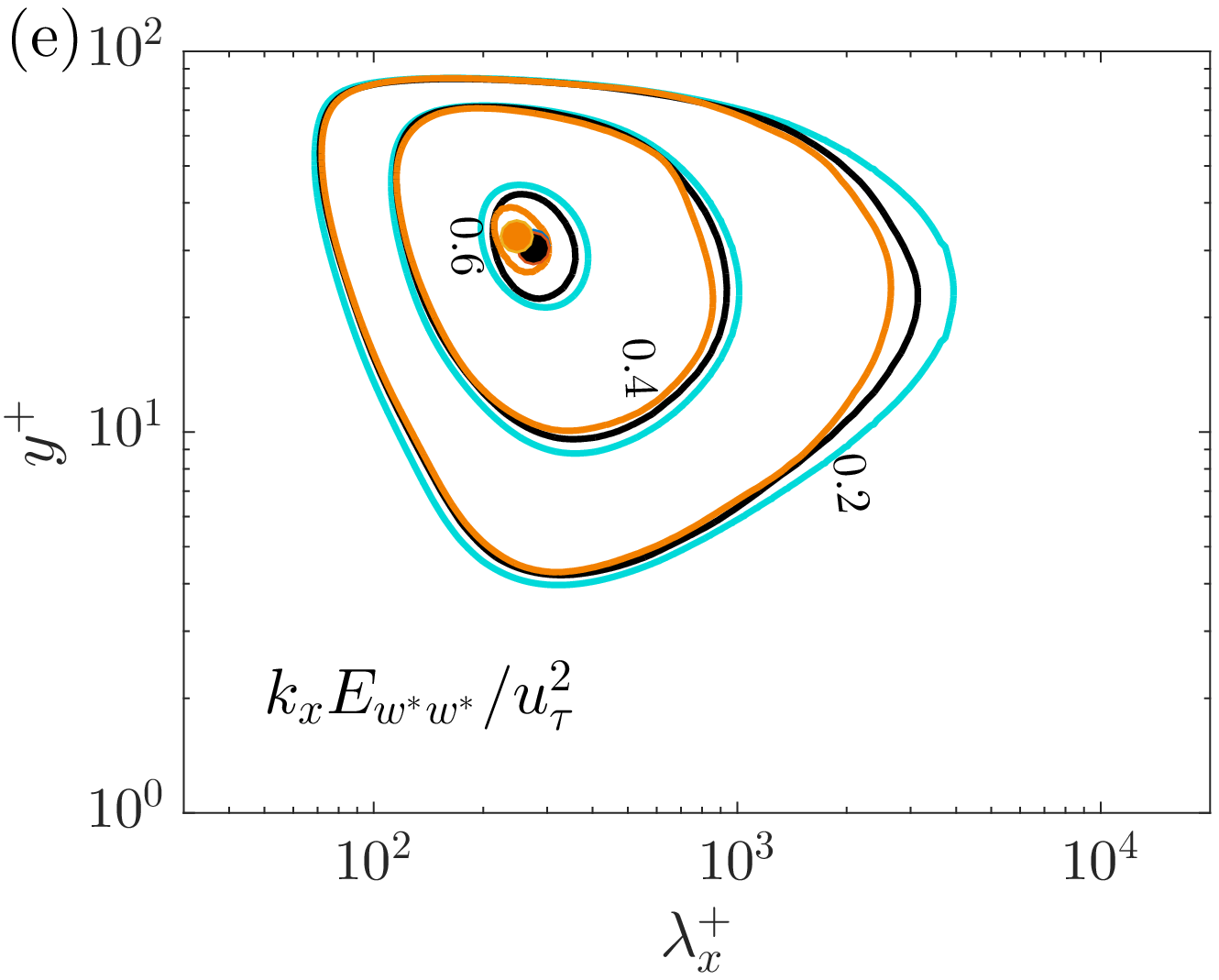}}
\end{minipage}
\hfill
\begin{minipage}{0.49\linewidth}
\centerline{\includegraphics[width=\textwidth]{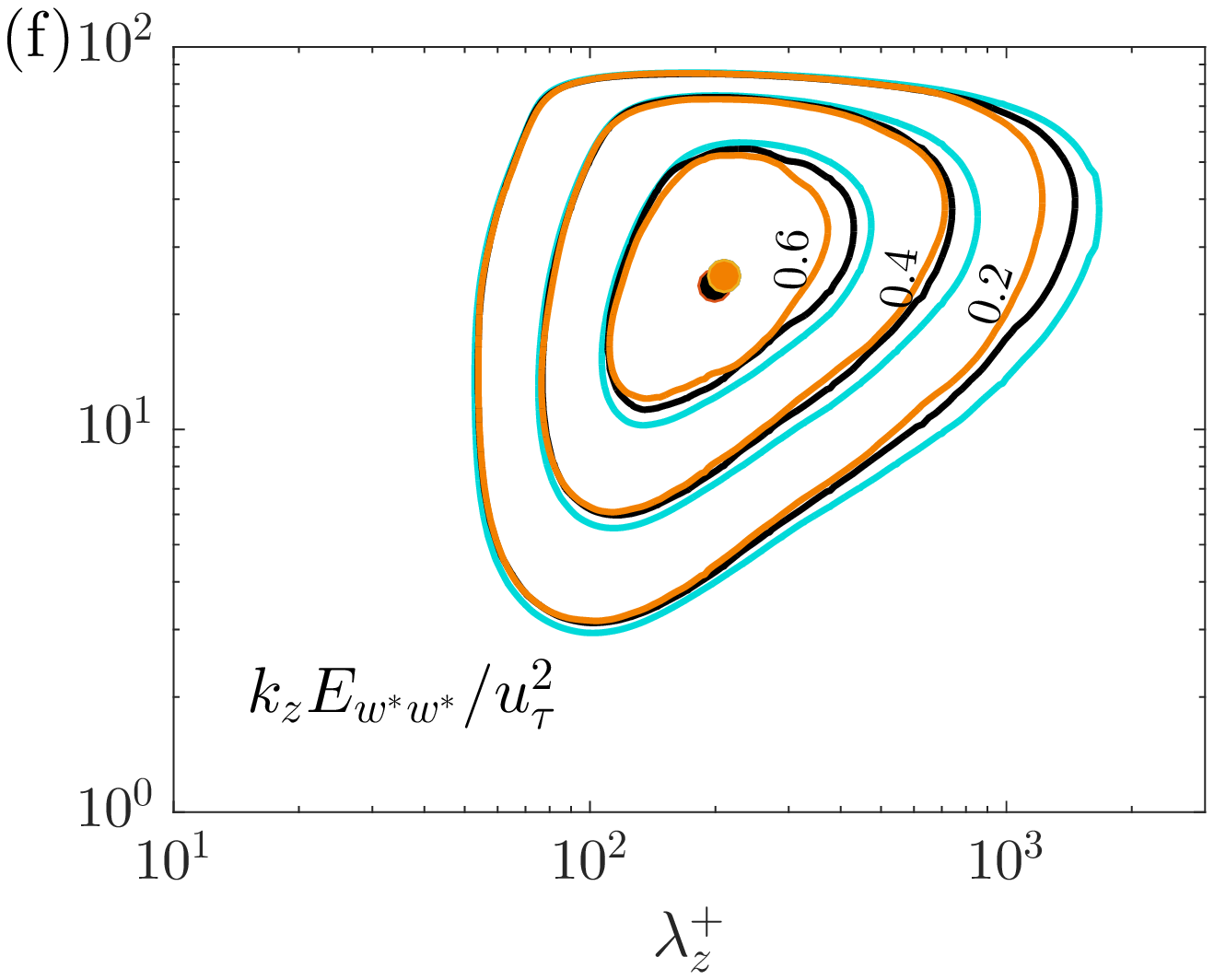}}
\end{minipage}
\caption{Pre-multiplied energy spectra of the three near-wall Reynolds-number-independent velocity components at $Re_\tau=1000$, 2000 and 5200. (a) $k_x E_{u^*u^*}/u_\tau^2$, (b) $k_z E_{u^*u^*}/u_\tau^2$, (c) $k_x E_{v^*v^*}/u_\tau^2$, (d) $k_z E_{v^*v^*}/u_\tau^2$, (e) $k_x E_{w^*w^*}/u_\tau^2$ and (f) $k_z E_{w^*w^*}/u_\tau^2$. The locations of spectral peaks have been marked with symbols. Refer to table~\ref{tab:tab1} for the line colors.}
\label{fig:fig7}
\end{figure}

Moreover, we present evidence of the extracted Reynolds-number-independent near-wall motions in the scale space through pre-multiplied streamwise and spanwise energy spectra of all the three velocity components at the Reynolds numbers $Re_\tau=1000\sim5200$ with $y_O^+ = 100$, as shown in \textcolor{black}{FIG}~\ref{fig:fig7}. It is clearly seen that the spectra of the extracted $u^*$ and $v^*$ collapse excellently at the three Reynolds numbers. The inner peaks of the streamwise and spanwise $u^*$-spectra locate at $y^+=10\sim20$, $\lambda_x^+\sim O(10^3)$ and $\lambda_z^+\sim O(10^2)$, which is well in accordance with experimental observations of near-wall streaks \cite[]{Kline1967,Smith1983,Kim1987}. For the wall-normal components $v^*$, the spectral peaks locate at $y^+\approx50$, $\lambda_x^+=200\sim300$ and $\lambda_z^+\approx100$.
In accordance with the integrated spanwise velocity intensity $\langle w^{*2} \rangle$, it is also found there exists slight Reynolds number dependence in the pre-multiplied spectra of $w^*$, as displayed in \textcolor{black}{FIG}~\ref{fig:fig7} (e, f). 
The discrepancy is principally located at large wavelengths, i.e., $\lambda_x^+>O(10^3)$ and $\lambda_z^+>200\sim300$. 
% This may imply that the influence of a small portion of outer motions still exists in spanwise velocity. 
Here we claim that this discrepancy is only marginal and could be neglected. The spectral peaks of $w^*$ locate at similar wall-normal height and wavelengths with $v^*$, namely, $y^+=30\sim40$, $\lambda_x^+=200\sim300$ and $\lambda_z^+\approx200$. Therefore, the wall-normal positions of ($v^*$, $w^*$) spectral peaks are much higher than that of $u^*$, while their streamwise and spanwise wavelengths are much smaller, which is consistent with the characteristics of the near-wall inner spectral peaks before decomposition (\textcolor{black}{FIG}~\ref{fig:fig1}). It is also noted that the $u^*$ and $w^*$ spectra can well penetrate into $y^+<10$ while the $v^*$ spectra are mainly located at $y^+>10$. This is due to the impermeable condition or the blocking effect of the wall-normal velocity at the wall \cite[]{Perry1982,Yang2018hierarchical}.

\begin{figure}
\centering
\begin{minipage}{0.49\linewidth}
\centerline{\includegraphics[width=\textwidth]{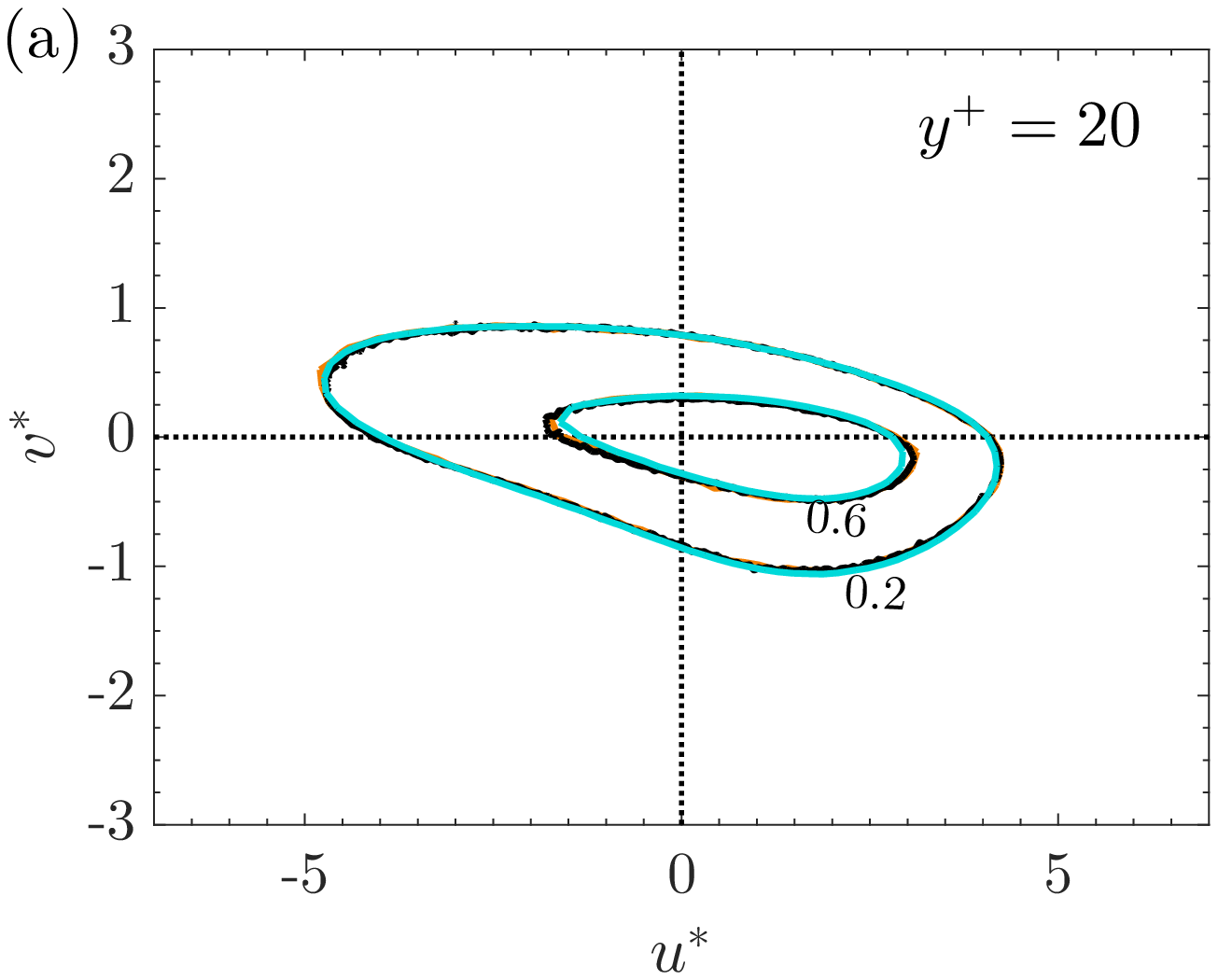}}
\end{minipage}
\hfill
\begin{minipage}{0.49\linewidth}
\centerline{\includegraphics[width=\textwidth]{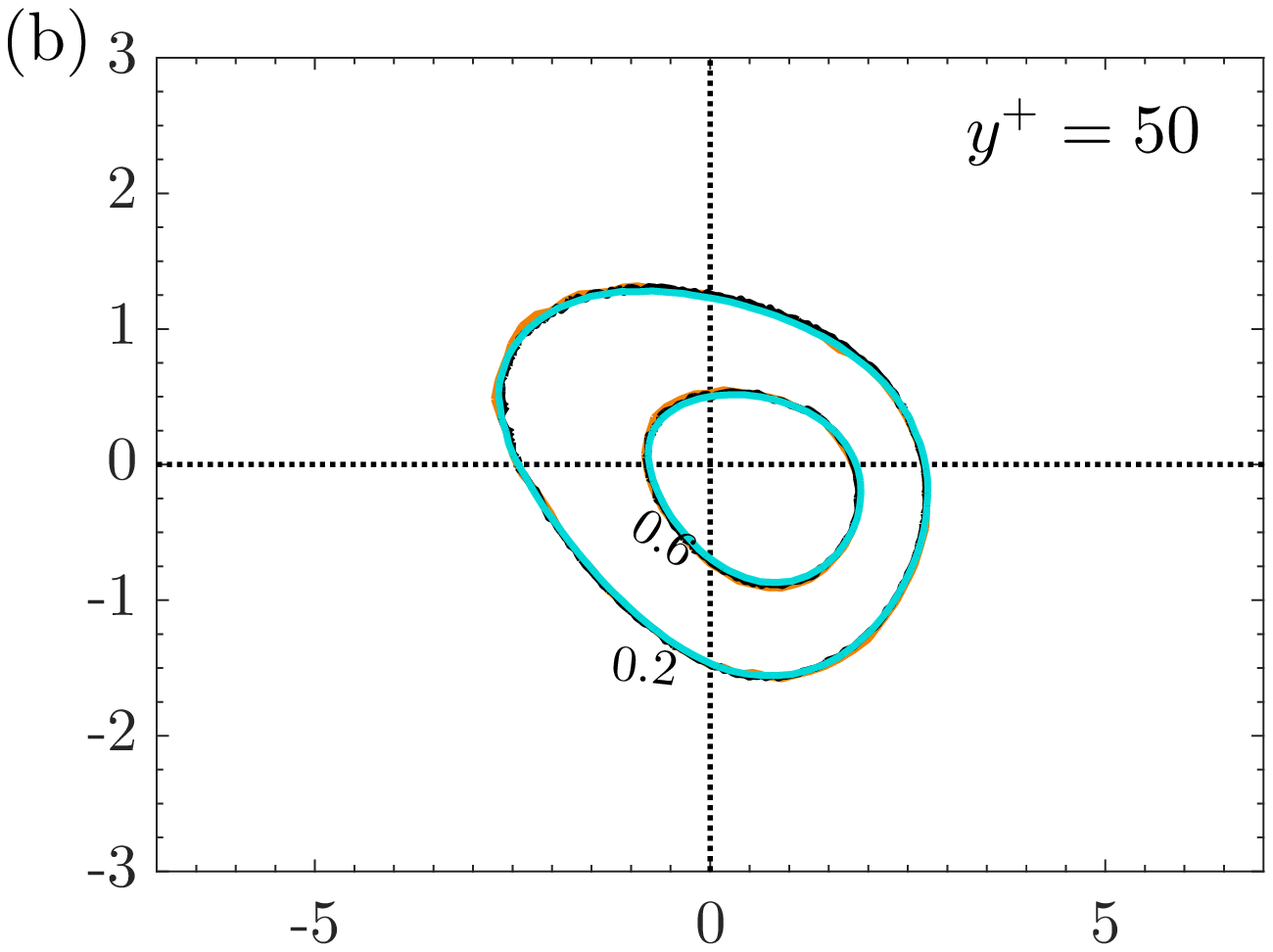}}
\end{minipage}
\vfill
\begin{minipage}{0.49\linewidth}
\centerline{\includegraphics[width=\textwidth]{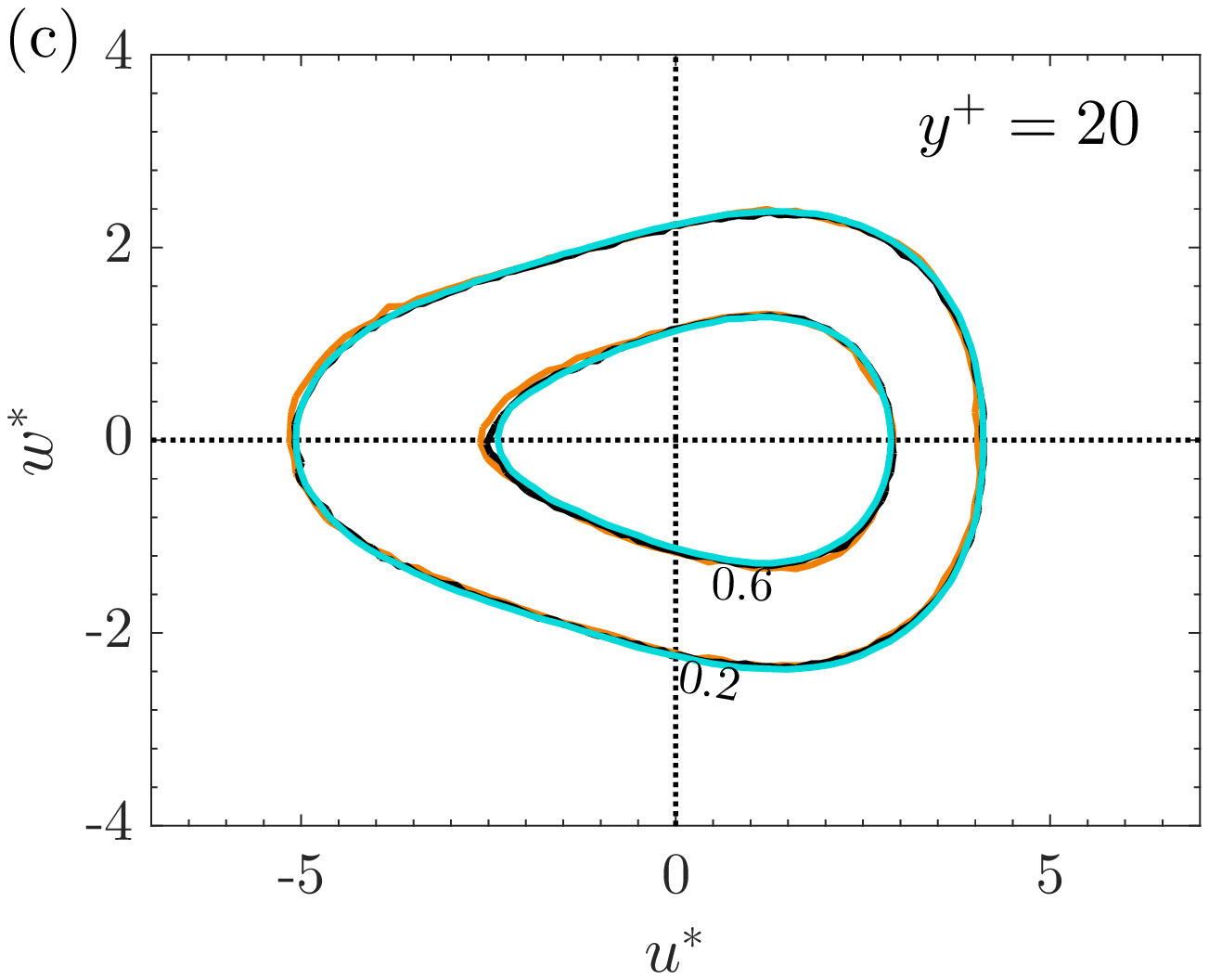}}
\end{minipage}
\hfill
\begin{minipage}{0.49\linewidth}
\centerline{\includegraphics[width=\textwidth]{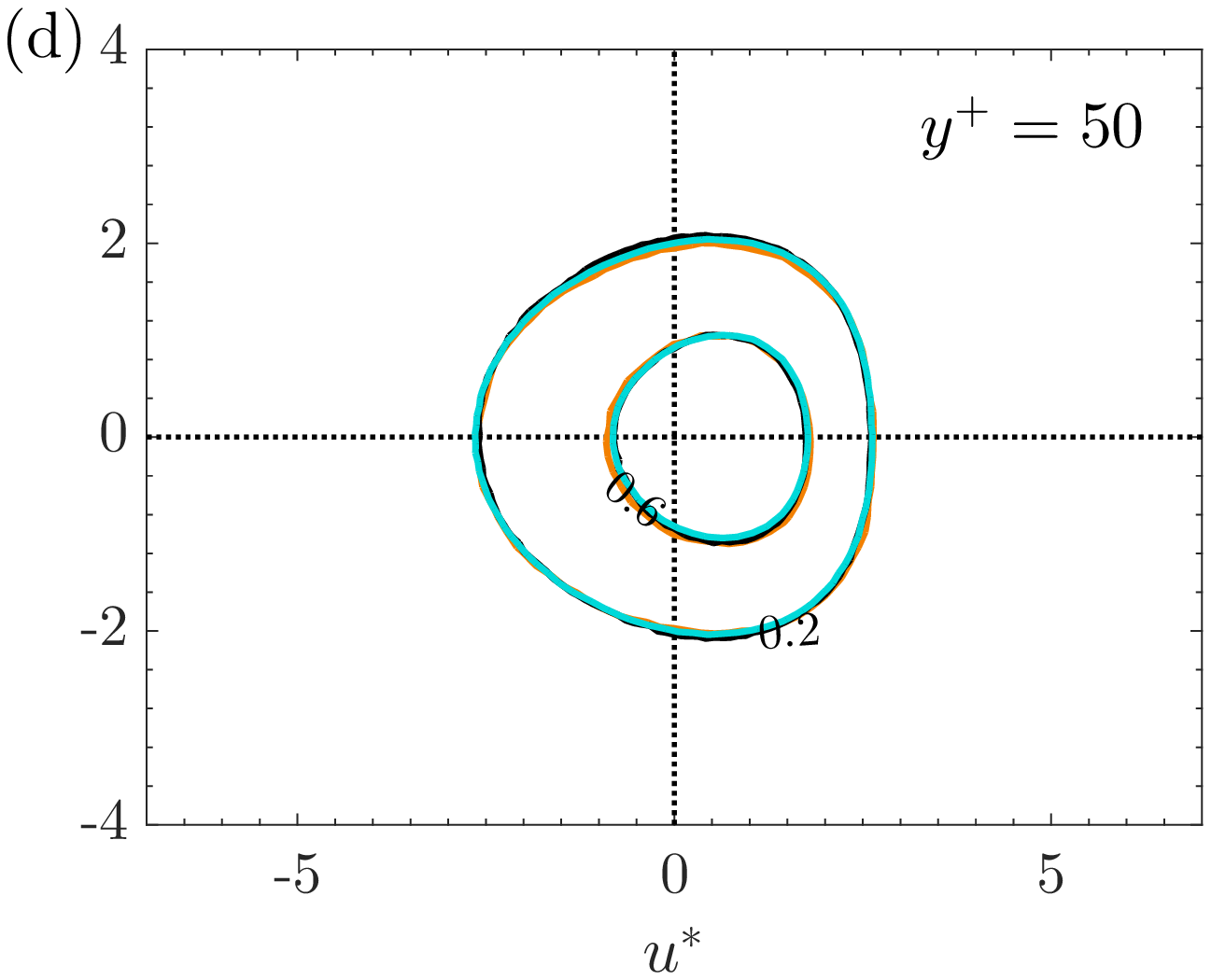}}
\end{minipage}

\caption{Joint p.d.f.s $P(u^*, v^*)$) (a, b) and $P(u^*, w^*)$ (c, d) at $y^+$ = 20 (a, c) and $y^+$ = 50 (b, d) in the fully developed turbulent channel flows at the Reynolds numbers $Re_\tau=1000\sim5200$. The contour levels are normalized by the peak value of the p.d.f. at $Re_\tau=1000$ of the two heights. Refer to table~\ref{tab:tab1} for the line colors.}
\label{fig:fig8}
\end{figure}

% While the $\langle w^{*2} \rangle$ of the universal signals has the worse Reynolds number universality than the $\langle u^{*2} \rangle$ and $\langle v^{*2} \rangle$ due to the spanwise fluctuations sign, consistent with the results of Fig. \ref{fig:fig13}. It is expected that the spanwise universal velocity behaves the deviation. 
% This is because the wall-normal distance of $y_O^+ = 100$ is the lower bound of the logarithmic region, below this position, the velocity fluctuations can feel all the superposition and modulation of the attached eddies located in logarithmic or outer region. 

Furthermore, we present joint p.d.f.s $P(u^*,v^*)$ and $P(u^*, w^*)$ of the extracted near-wall Reynolds-number-independent velocities at the Reynolds number $Re_\tau=1000\sim5200$, which is displayed in \textcolor{black}{FIG}~\ref{fig:fig8}. 
% The joint p.d.f.s of the universal velocity were directly calculated using the data obtained by the method of section 3.1. 
The comparisons at the two wall-normal heights $y^+=20$ and $y^+=50$ exhibit remarkable coincidence of the joint p.d.f.s of the extracted Reynolds-number-independent velocity components at the three Reynolds numbers. 
% The two sets of signals resulted in almost identical at the position of $y^+ = 20$ and $y^+ = 50$, for (a,c) and (b,d) respectively. 
The major axis of $P(u^*, v^*)$ is inclined in the Q2-Q4 direction, indicating higher probabilities of the ejection and sweep motions. By comparing $P(u^*, v^*)$ at the two heights, i.e. \textcolor{black}{FIG}~\ref{fig:fig8} (a) and (b), it can be seen that the inclination of the major axis is shallower at $y^+=20$, suggesting the ejecting or sweeping of the turbulent motions occurs at a smaller angle nearer the wall, similar to \textcolor{black}{FIG}~\ref{fig:fig2} (a, b). 
In addition, the joint p.d.f. $P(u^*, w^*)$ is shown in \textcolor{black}{FIG}~\ref{fig:fig8} (c, d), which is symmetric about $w^* = 0$ axis, also similar to \textcolor{black}{FIG}~\ref{fig:fig2} (c, d). 
% Note that we have normalized the contour level using the peak value of the p.d.f for $Re_\tau = 1000$, more results of PDFs presented below are the same contour levels. 
% Anyway, these current results can shed light on the issue of Reynolds-number independence of universal signals between $Re_\tau = 1000$ to $Re_\tau = 5200$. It is readily acknowledged that the results presented herein, while encouraging, pertain to this Reynolds number range, and can, therefore, not be viewed as offering a universal representation across a wide range of Reynolds numbers. The DNS is limited to the higher Reynolds number, however, the study aim to identify whether the low-Reynolds number can show the same universality is also significantly important. 

\begin{figure}
\centering
\begin{minipage}{0.49\linewidth}
\centerline{\includegraphics[width=\textwidth]{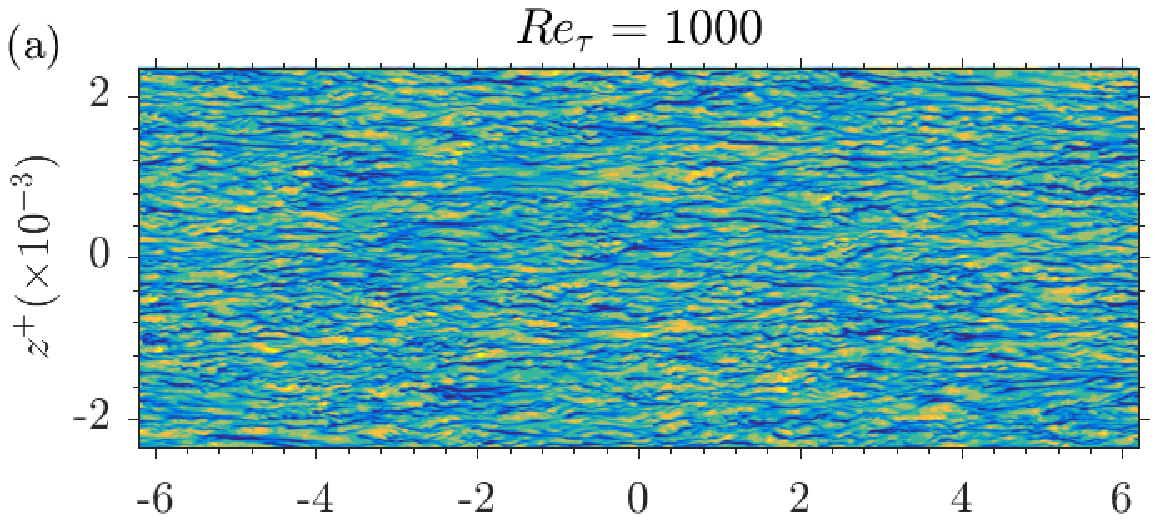}}
\end{minipage}
\hfill
\begin{minipage}{0.49\linewidth}
\centerline{\includegraphics[width=\textwidth]{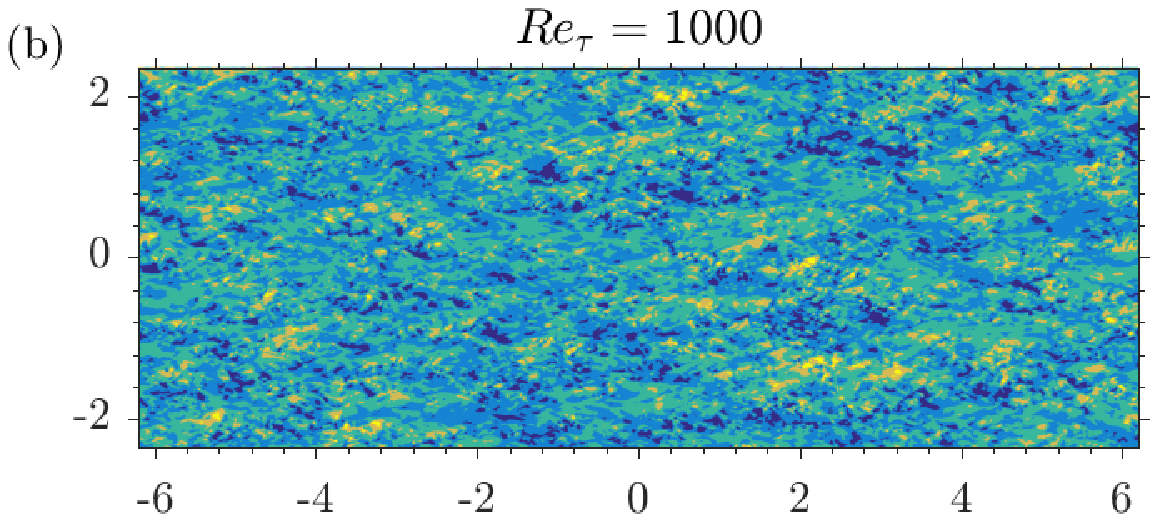}}
\end{minipage}
\vfill
\begin{minipage}{0.49\linewidth}
\centerline{\includegraphics[width=\textwidth]{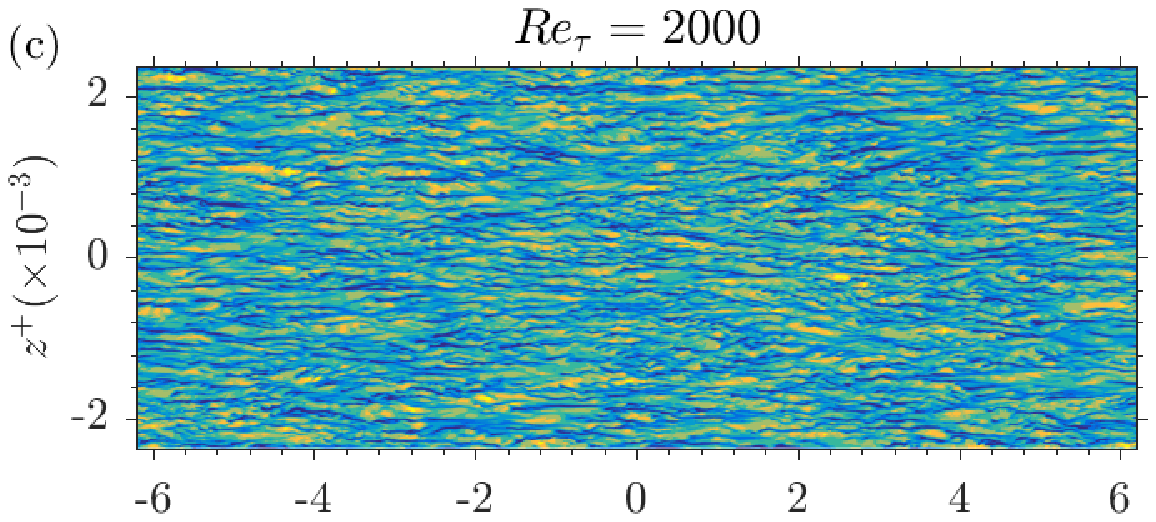}}
\end{minipage}
\hfill
\begin{minipage}{0.49\linewidth}
\centerline{\includegraphics[width=\textwidth]{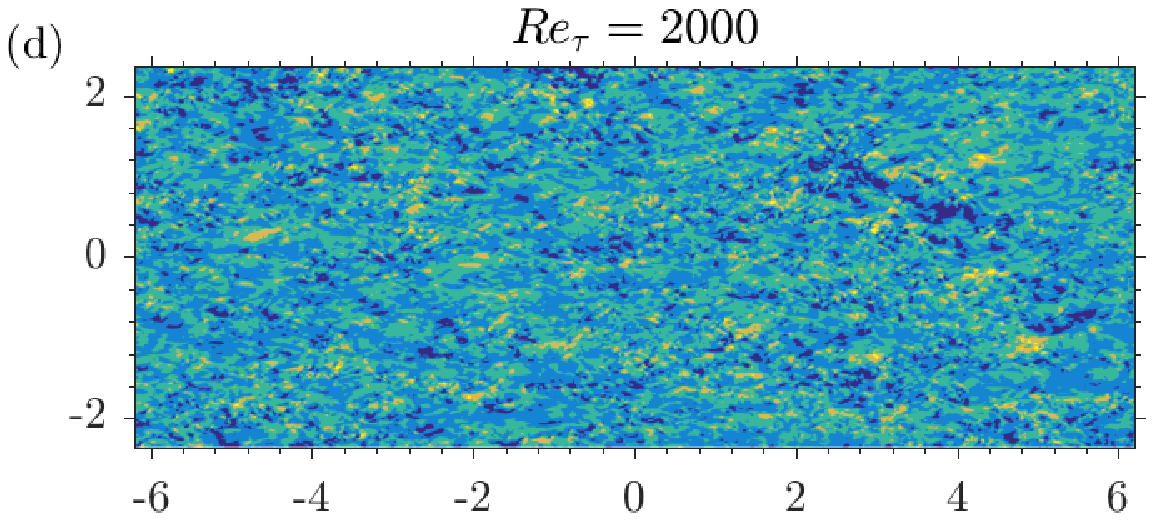}}
\end{minipage}
\vfill
\begin{minipage}{0.49\linewidth}
\centerline{\includegraphics[width=\textwidth]{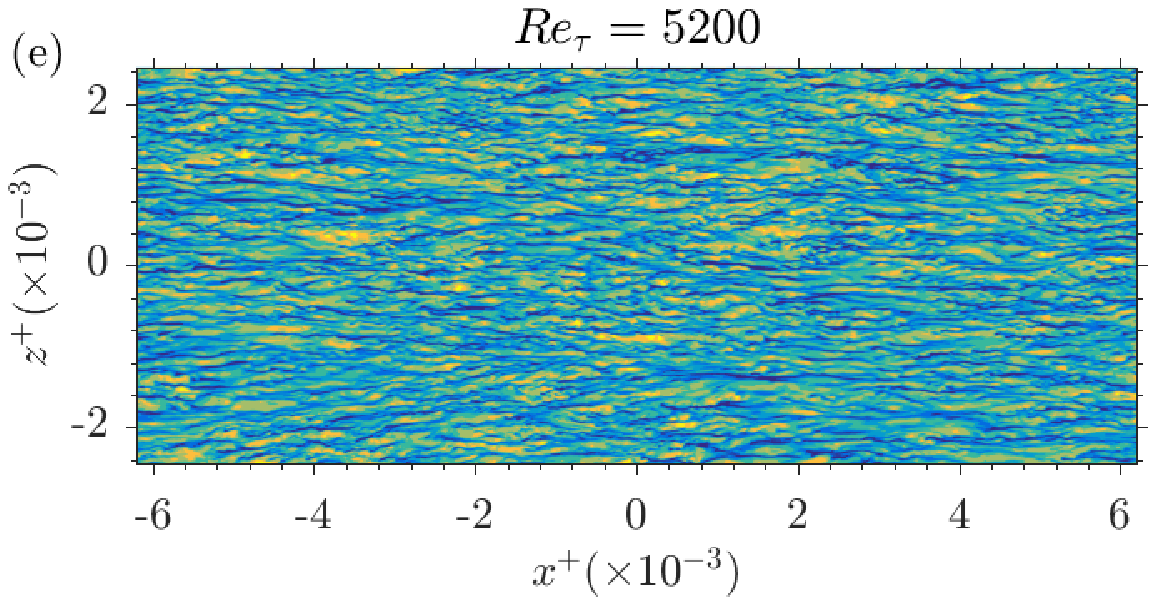}}
\end{minipage}
\hfill
\begin{minipage}{0.49\linewidth}
\centerline{\includegraphics[width=\textwidth]{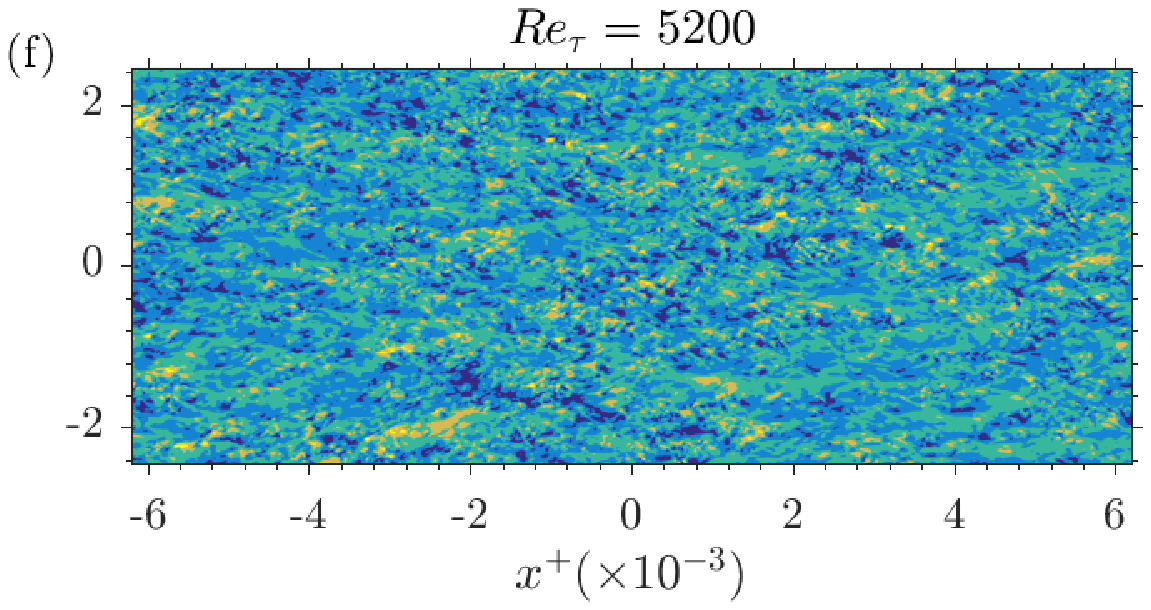}}
\end{minipage}
\vfill
\begin{minipage}{0.49\linewidth}
\centerline{\includegraphics[width=\textwidth]{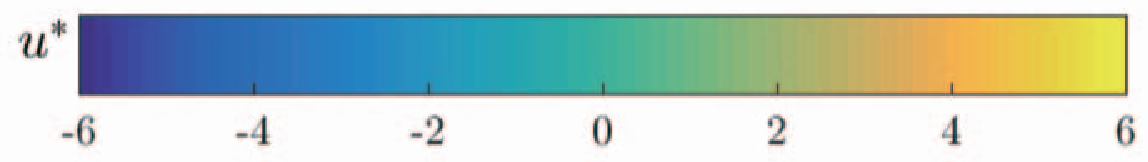}}
\end{minipage}
\hfill
\begin{minipage}{0.49\linewidth}
\centerline{\includegraphics[width=\textwidth]{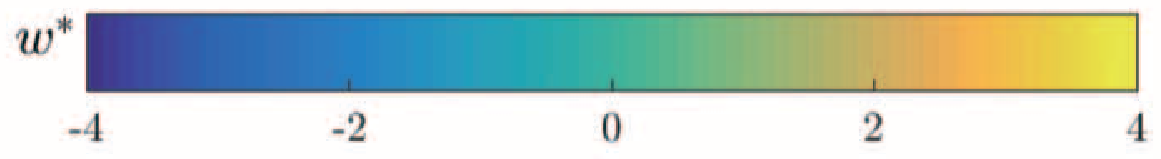}}
\end{minipage}
\caption{Plane snapshots of the near-wall Reynolds-number-independent streamwise velocity fields at $y^+ = 15$ (a,c,e) and Reynolds-number-independent spanwise velocity fields at $y^+ = 25$ (b,d,f) for $Re_\tau=1000,2000,5200$.}
\label{fig:fig9}
\end{figure}

% \subsubsection{Coherent structures}
So far, Reynolds-number-invariant statistics of the decomposed small-scale near-wall motions have been thoroughly demonstrated, in the Reynolds number range of $Re_\tau=1000\sim5200$. Now we turn to demonstrate the corresponding instantaneous flow snapshots and reveal the dominant coherent structures composing the Reynolds-number-independent near-wall flow. \textcolor{black}{FIG}~\ref{fig:fig9} displays $x-z$ plane snapshots of the Reynolds-number-independent streamwise velocity at $y^+=15$, where the full streamwise velocity fluctuation intensity is approximately the maximum. It is seen that, as Reynolds number increases (i.e., \textcolor{black}{FIG}~\ref{fig:fig9} (a,c,e)), the general streaky features of the flows are quite similar, indicating that not only the statistics, but also the instantaneous fields of $u^*$ exhibit good Reynolds-number-independent behavior. Furthermore, instantaneous $w^*$ fields are shown on \textcolor{black}{FIG}~\ref{fig:fig9} (b,d,f) at the $y^+=25$, which shows that there is no evident residual large-scale footprints.
% Instan $v^*$ structures are also very similar at the three Reynolds numbers, which are not shown for saving the space. 

\begin{figure}
\centering
\begin{minipage}{0.49\linewidth}
\centerline{\includegraphics[width=\textwidth]{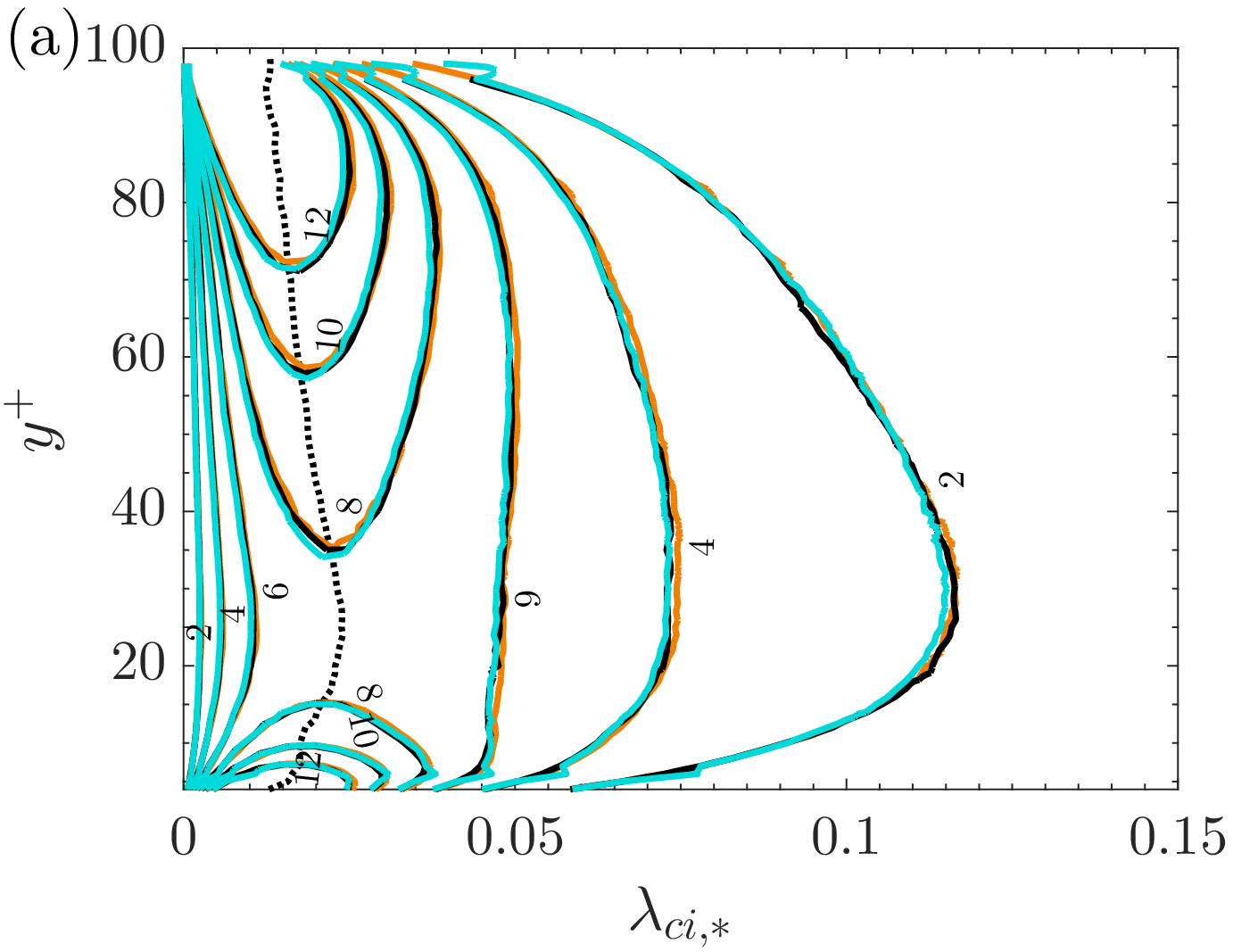}}
\end{minipage}
\hfill
\begin{minipage}{0.49\linewidth}
\centerline{\includegraphics[width=\textwidth]{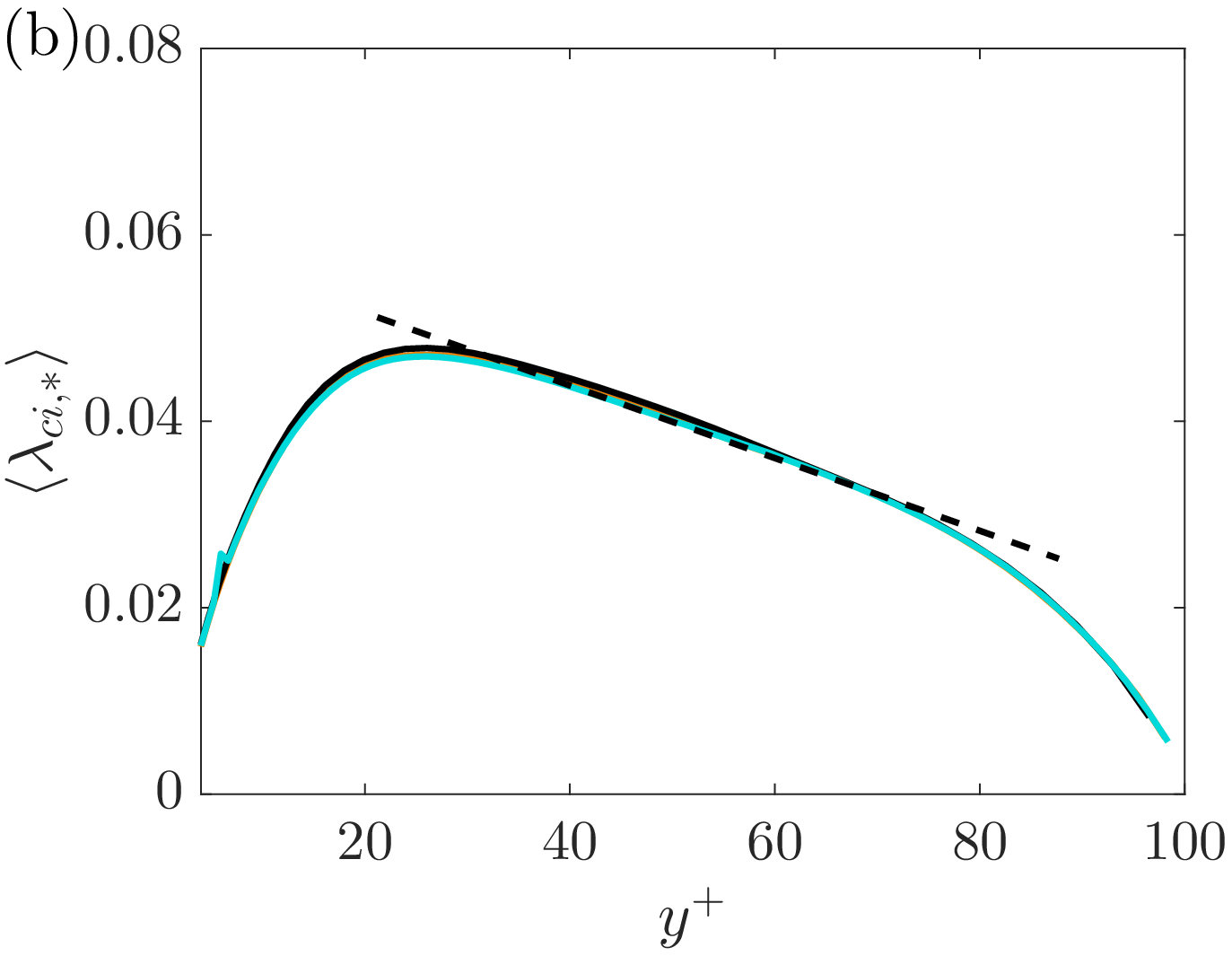}}
\end{minipage}
\caption{P.d.f.s of $\lambda_{ci,*}$ (a) and mean swirling strength $\langle \lambda_{ci,*} \rangle$ (b) as functions of $y^+$ at $Re_\tau = 1000$, 2000 and 5200. The dashed line indicates a linear variation of $\langle \lambda_{ci,*} \rangle$ with $y^+$. Refer to table~\ref{tab:tab1} for the line colors.}
\label{fig:fig10}
\end{figure}

In \textcolor{black}{FIG}~\ref{fig:fig10}, we show Reynolds-number invariance of vortical statistics of the Reynolds-number-independent flow fields. The $\lambda_{ci}$ criterion \cite[]{Zhou1999} is employed for the vortex identification, which is defined as the imaginary part of the complex eigenvalue of velocity gradient tensor and usually referred to the local swirling strength. The p.d.f.s of $\lambda_{ci,*}$ at the three Reynolds numbers, i.e., $Re_\tau=1000, 2000$ and 5200, are plotted in \textcolor{black}{FIG}~\ref{fig:fig10} (a), which demonstrates excellent agreement among the three distributions. Meanwhile, the wall-normal variation of the most probable value of $\lambda_{ci,*}$ is represented by the black dotted line, the maximum of which is located around $y^+ = 25$. 
% Both \cite{nagaosa2003statistical} and \cite{Wu2006} found that Q (a criterion based on the second invariant of the velocity gradient tensor) and $\lambda_{ci}$ normalized with its root-mean-square (rms) at a given wall-normal location yielded nearly-identical probability density functions of, insensitive to $Re$.
% As for the universal signal here, without dividing by the root mean square, the distribution of the swirl strength $\lambda_{ci,*}$ of the universal signal is also independent of the Reynolds number and vertical height.
Furthermore, \textcolor{black}{FIG}~\ref{fig:fig10} (b) exhibits the mean swirling strength profiles. It is seen that, the mean swirling strength increases first and then decreases with $y^+$, and the maximum also appears at $y^+ = 25$. Excellent Reynolds-number invariance is observed. An approximately linear variation of the mean swirling strength $\langle \lambda_{ci,*} \rangle$ with $y^+$ in the range of $y^+ \approx 30\sim75$ is also denoted in the \textcolor{black}{FIG}~\ref{fig:fig1}. Since the swirling strength is defined in terms of velocity gradient, it is of higher order than velocity itself, and the comparison here further strengthens the reliability of the current extraction scheme for the Reynolds-number-independent velocity field. 
% At $y^+ = 30-75$, the mean swirling strength almost linearly decreases with the increase of $y^+$, then the mean swirling strength decays rapidly from $y^+ = 75$ to 100. 

\begin{figure}
\centering
\begin{minipage}{0.49\linewidth}
\centerline{\includegraphics[width=\textwidth]{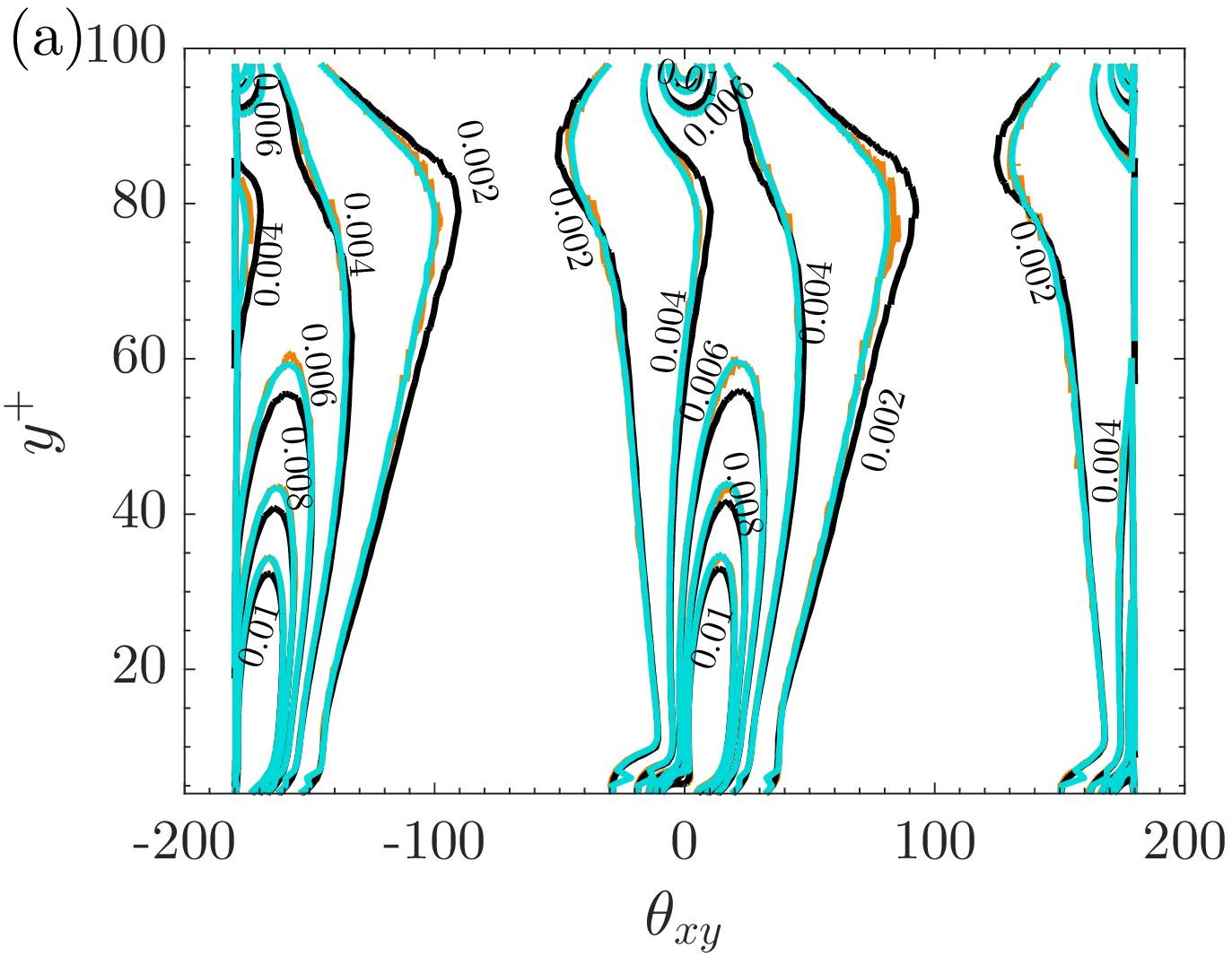}}
\end{minipage}
\hfill
\begin{minipage}{0.49\linewidth}
\centerline{\includegraphics[width=\textwidth]{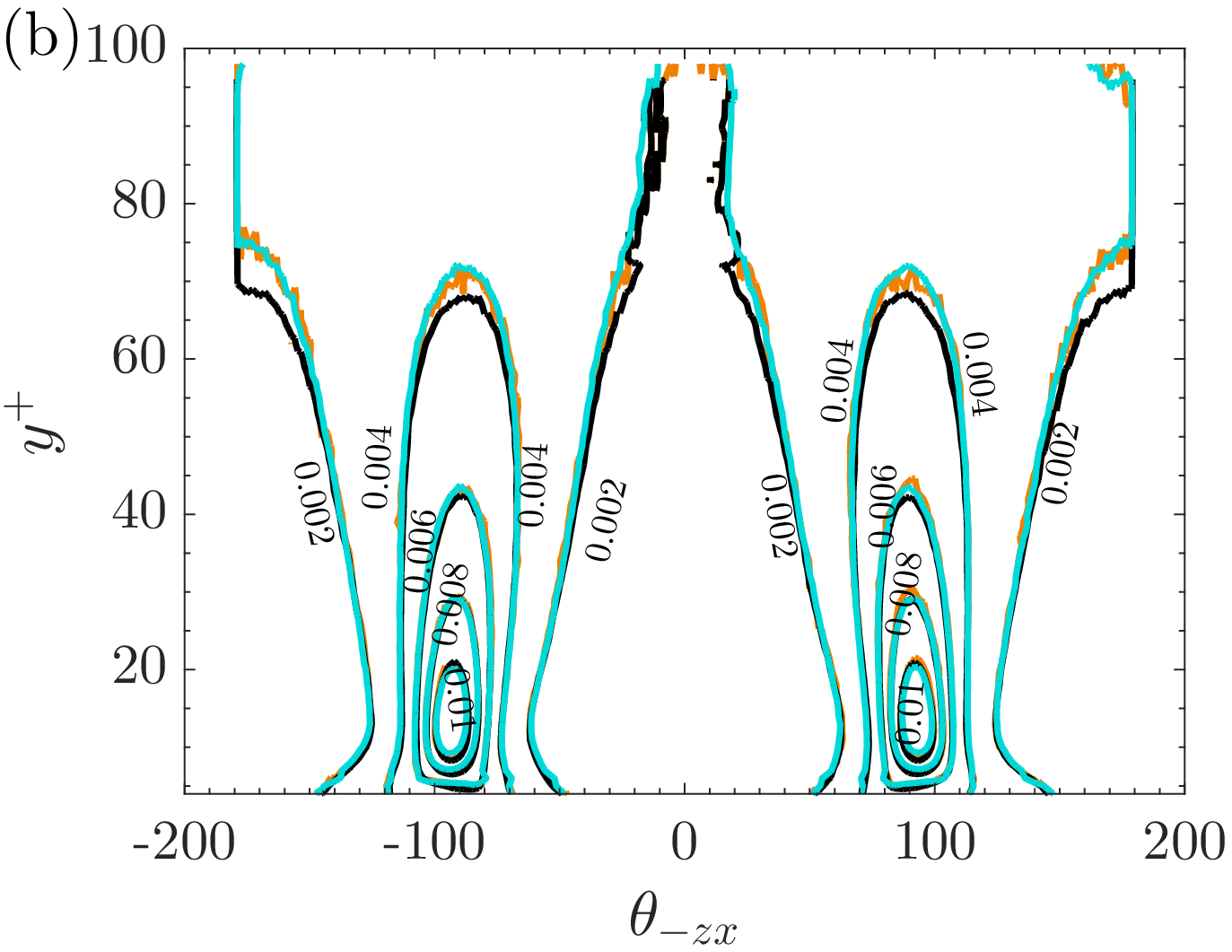}}
\end{minipage}
\caption{P.d.f.s of $\theta_{xy,*}$ (a) and $\theta_{-zx,*}$ (b) as functions of $y^+$ at $Re_\tau = 1000$, 2000 and 5200. Refer to table~\ref{tab:tab1} for the line colors.}
\label{fig:fig11}
\end{figure}

Besides swirling strength $\lambda_{ci}$, vortex orientation is another important aspect for the characterization of vortical structures.
Zhou \emph{et al.} \cite{Zhou1999} suggested that local swirling flow will be stretched
or compressed along the direction of the real eigenvector $\mathbf{\Lambda}_r$ of the velocity gradient tensor. Gao \emph{et al.} \cite{Gao2011} employed $\mathbf{\Lambda}_r$ to identify the local vortex orientations. Recently, Wang \emph{et al.} \cite{Wang2019experimental} analyzed the vortex geometries and topologies in turbulent boundary layers measured by tomographic particle image velocimetry using the same method. 
Following the above studies, in this work, we also identify the vortex orientations through the real eigenvector $\Lambda_r$ of the velocity gradient tensor. For the details of the method, one could see Gao \emph{et al.} \cite{Gao2011} and Wang \emph{et al.} \cite{Wang2019experimental}. 
% Next, we compare the vortex orientations in the universal near-wall flows. 
\textcolor{black}{FIG}~\ref{fig:fig11} illustrates the p.d.f.s of the vortex orientations at $Re_\tau=1000 \sim 5200$, where the $\Lambda_r$ vector is projected onto the $x-y$ plane and $x-z$ plane separately. In the $x-y$ plane, the angle between the projected vector and the $x$ axis is denoted by $\theta_{xy}$. In
the $x-z$ plane, the angle between the projected $\Lambda_r$ vector and the negative $z$ axis is indicated by $\theta_{-zx}$. \textcolor{black}{FIG}~\ref{fig:fig11} clearly shows that the p.d.f.s collapse excellently, i.e., demonstrating Reynolds number invariance of vortex orientations.
Wang \emph{et al.} \cite{Wang2019experimental} reported that the near-wall vortex orientations from the full velocity fields are also nearly independent of Reynolds number in the range of $Re_\tau=1238\sim3081$. They attributed it to that the Reynolds number only has evident influence on large-scale flow structures, while the small-scale vortical structures are likely independent of the Reynolds number. 

% This proves that the removal of the large-scale superposition signal makes the streamwise vortex inclination angle become smaller, which symbolize the streamwise inclined structure change back to the streamwise direction.

In summary, here we have successfully demonstrated that truly Reynolds-number-independent near-wall turbulent motions that are independent of outer influences have been successfully extracted via the inner-outer interaction hypothesis, the extraction scheme (\ref{eqn:equ9}-\ref{eqn:equ10}) and the outer reference height $y_O^+=100$, in the Reynolds number range of $Re_\tau=1000\sim5200$. Plenty of evidence, i.e., integrated statistics, spectra, joint p.d.f. as well as instantaneous flow fields, has been provided. 
% In the study of \cite{Hwang2013near}, the near-wall motions wider than $\lambda_z^+\approx100$ were removed via the spanwise minimum flow unit. It was found that the filtered near-wall streamwise velocity fluctuations at $y^+ \le 40$ are well collapsed with the viscous scaling at the Reynolds numbers up to $Re_\tau=660$. \cite{Yin2017near} extended this approach to higher Reynolds numbers, i.e., $Re_\tau=1000\sim4000$, and obtained similar findings. We shortly comment here, that both \cite{Hwang2013near} and \cite{Yin2017near} isolated healthy and Reynolds-number-invariant near-wall turbulent fluctuations at $y^+ < \sim50$ and $\lambda_z^+ \le 100$, whereas the present methodology could finally extract universal near-wall turbulence in a higher wall-normal range ($y^+ \le 100$) and without spanwise wavelength restriction.

% This can be explained, first all the data points satisfying $\lambda_{ci} > \lambda_{ci,thre}$ (the threshold be chosen as the most probable value of swirl strength) are counted to calculate the p.d.f.s of the vortex orientation angles. While, we find the swirl strength range of the footprint of outer large-scale fluctuations on the near-wall turbulence fluctuations is too narrow to affect the vortex structure inclination statistics. 

\section{Low-Reynolds-number effect}

% Based on the assumption of wall-attached eddies \cite[]{Perry1982} and reducing the outer input reference height of predictive inner-outer (PIO) model to $y_O^+ = 100$, then, we use sufficient evidence to prove the universal property of the universal signals at Reynolds number 1000 to 5200. Although there is no logarithmic region in the case of low Reynolds number below the $Re_\tau = 667$ \cite[]{Hutchins2007}, we still think that the PIO model is applicable, to observe the universal signal results of $Re_\tau$ from 180 to 600. 
Some studies have reported the existence of the low-Reynolds-number effect \textcolor{black}{($180 < Re_\tau < 1000$)}, that near-wall turbulence statistics can not be well scaled by the viscous units at low Reynolds numbers \cite[]{Wei1989Reynolds,Antonia1992Low,Antonia1994Low}. It is unclear whether this anomalous scaling is due to the effect of outer footprints, which should not be very strong at low Reynolds number in our view. In this part, we will inspect whether the small-scale near-wall velocity fields extracted by (\ref{eqn:equ9}-\ref{eqn:equ10}) could also be Reynolds-number-independent in the fully developed low-Reynolds-number turbulent channel flows, e.g., $Re_\tau<1000$.  

\begin{figure}
\centering
\begin{minipage}{0.325\linewidth}
\centerline{\includegraphics[width=\textwidth]{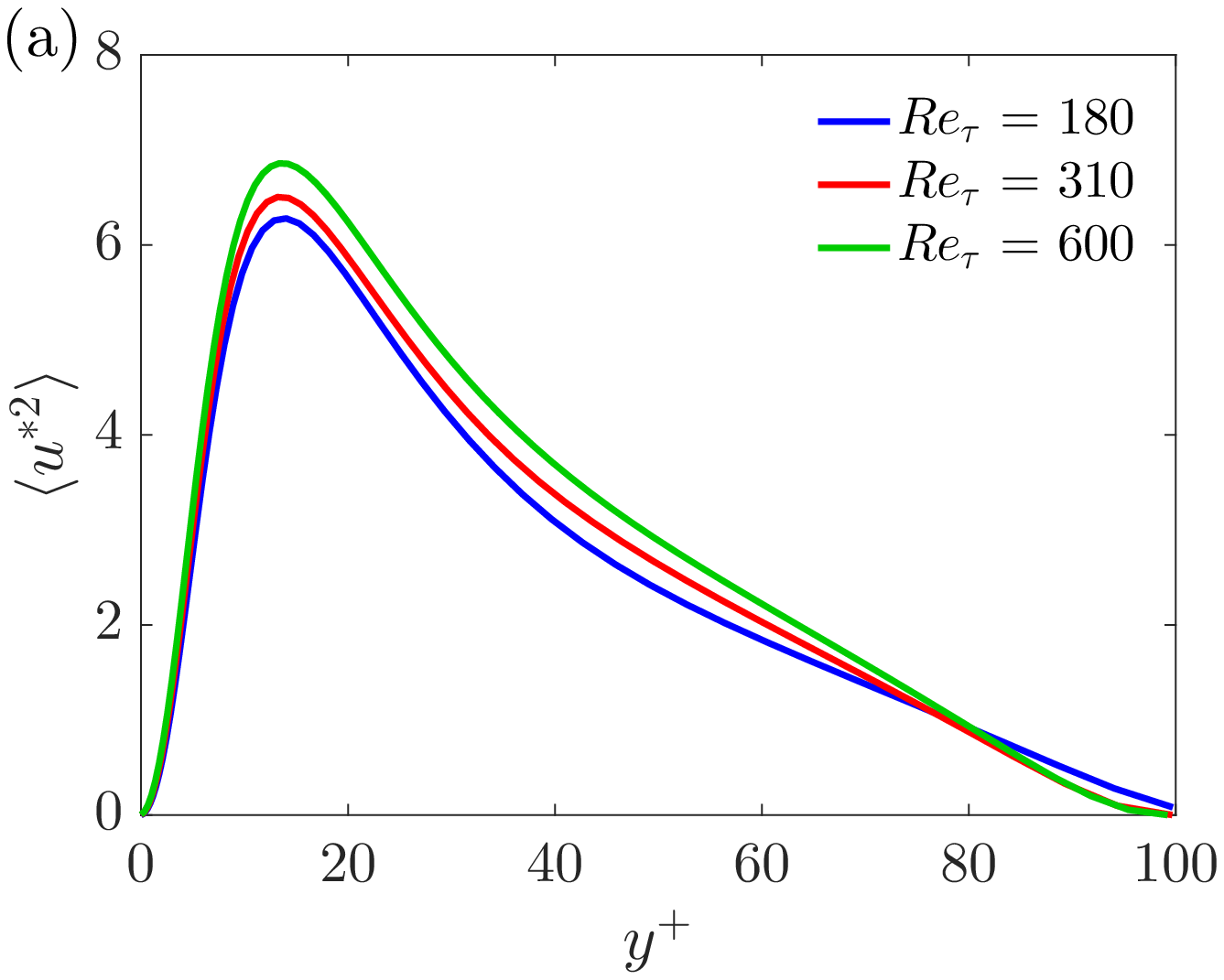}}
\end{minipage}
\hfill
\begin{minipage}{0.325\linewidth}
\centerline{\includegraphics[width=\textwidth]{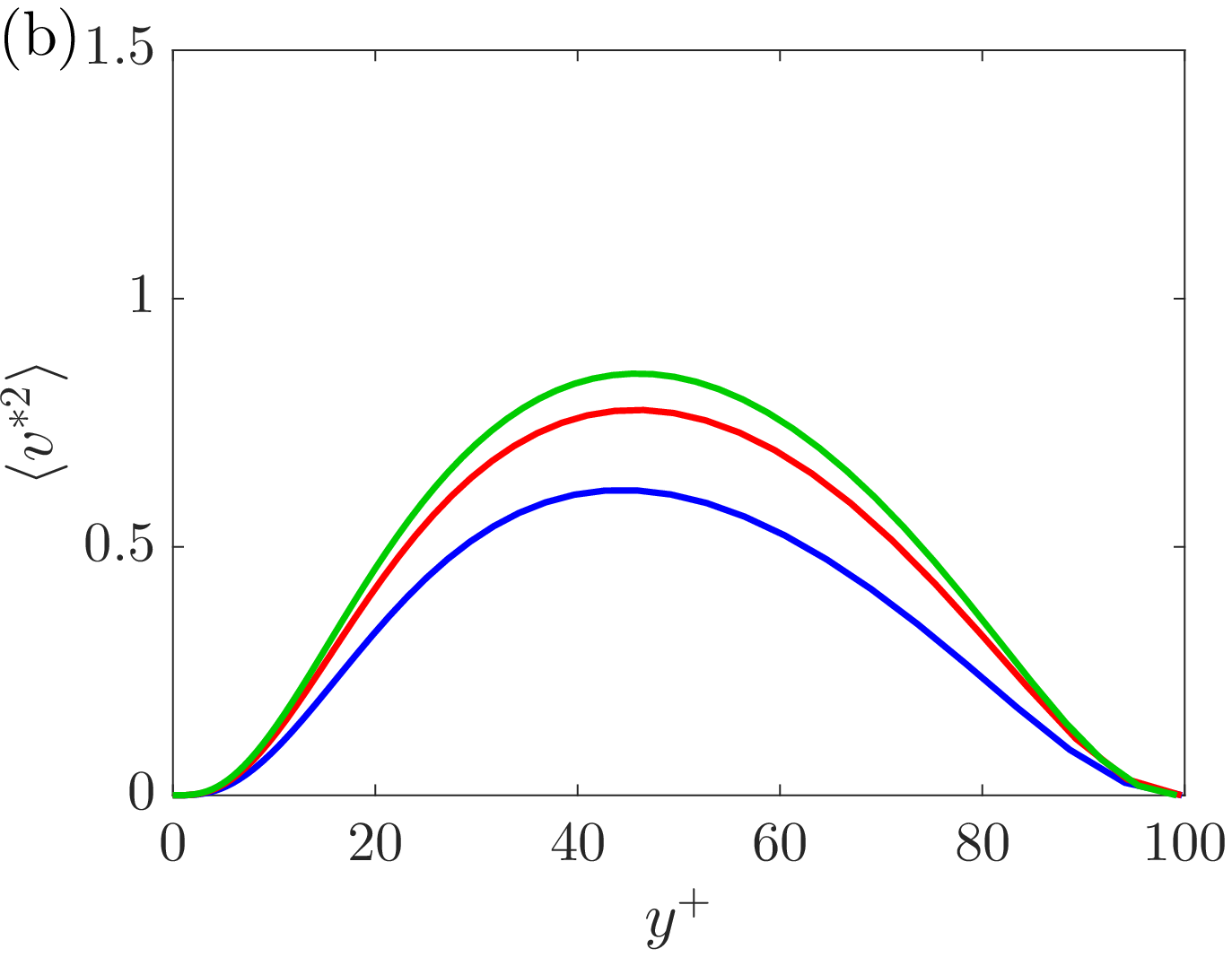}}
\end{minipage}
\hfill
\begin{minipage}{0.325\linewidth}
\centerline{\includegraphics[width=\textwidth]{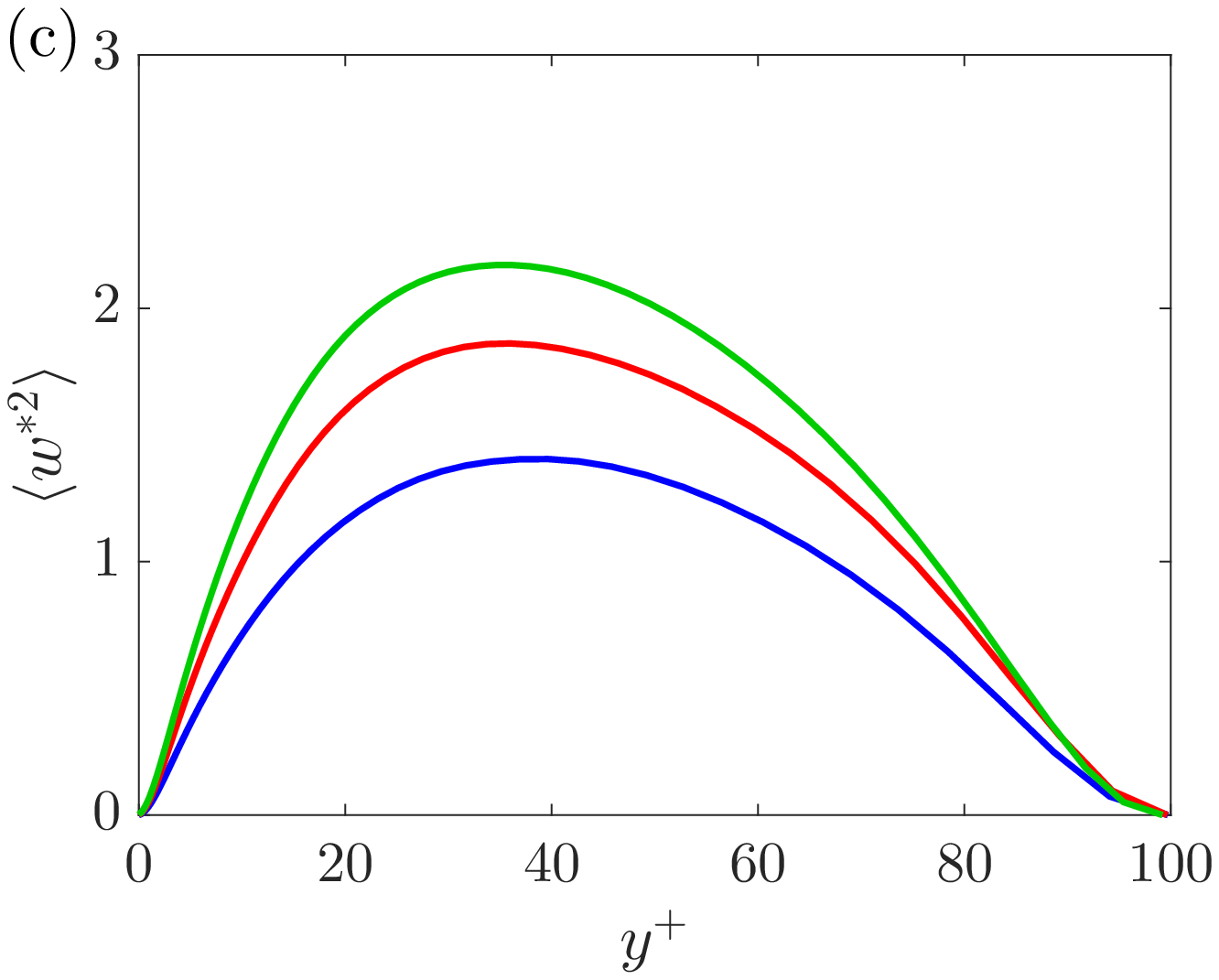}}
\end{minipage}
\caption{The intensities of the extracted $u^*$ (a), $v^*$ (b) and $w^*$ (c) at $Re_\tau=180, 310$ and 600. Refer to table~\ref{tab:tab1} for the line colors.}
\label{fig:fig12}
\end{figure}

The intensities of the extracted $u^*$, $v^*$ and $w^*$ in the context of the inner-outer interaction model are shown in \textcolor{black}{FIG}~\ref{fig:fig12}.
% Each subfigure contains three lines, $Re_\tau = 180$ in blue, $Re_\tau = 310$ in red and $Re_\tau = 600$ in green. 
It is seen that the streamwise turbulence intensity $\langle u^{*2}\rangle$ slightly increases with $Re_\tau$ and the intensity peaks locate at $y^+ \approx 15$ at all the three Reynolds numbers, as displayed in \textcolor{black}{FIG}~\ref{fig:fig12} (a). 
% Compared to the $Re_\tau$ of 600, the $Re_\tau$ of 180 and 310 have similar $\langle u^{*2}\rangle$ values. 
The Reynolds number dependence is also evident from the wall-normal and spanwise turbulence intensities, i.e., $\langle v^{*2}\rangle$ and $\langle w^{*2}\rangle$, as shown in \textcolor{black}{FIG}~\ref{fig:fig12} (b) and (c). The wall-normal peak locations of $\langle v^{*2}\rangle$ and $\langle w^{*2}\rangle$ are $y^+\approx40\sim50$ and $y^+\approx30\sim40$, respectively. Therefore, the extracted $\langle u^{*2}\rangle$, $\langle v^{*2}\rangle$ and $\langle w^{*2}\rangle$ are not Reynolds-number-independent, instead they show definite Reynolds number dependence. Then, we will denote it as low Reynolds number effect hereafter, which should not be confounded with that in the literature \cite[]{Antonia1992Low,Antonia1994Low}, since the large scales were present and small scales were not separated from the total fluctuations. {In addition, we also tried smaller $y_o^+$ at these low Reynolds numbers, and the results are presented in Appendix \ref{sec:appenC}. It is found that the low-Reynolds-number effect still exists.}
% \textcolor{black}{The reasion that give rise to the low-Reynolds number effect maybe because the universal signal is still in the development stage and has not reached stability in low Reynolds number.}
 
\begin{figure}
\centering
\begin{minipage}{0.49\linewidth}
\centerline{\includegraphics[width=\textwidth]{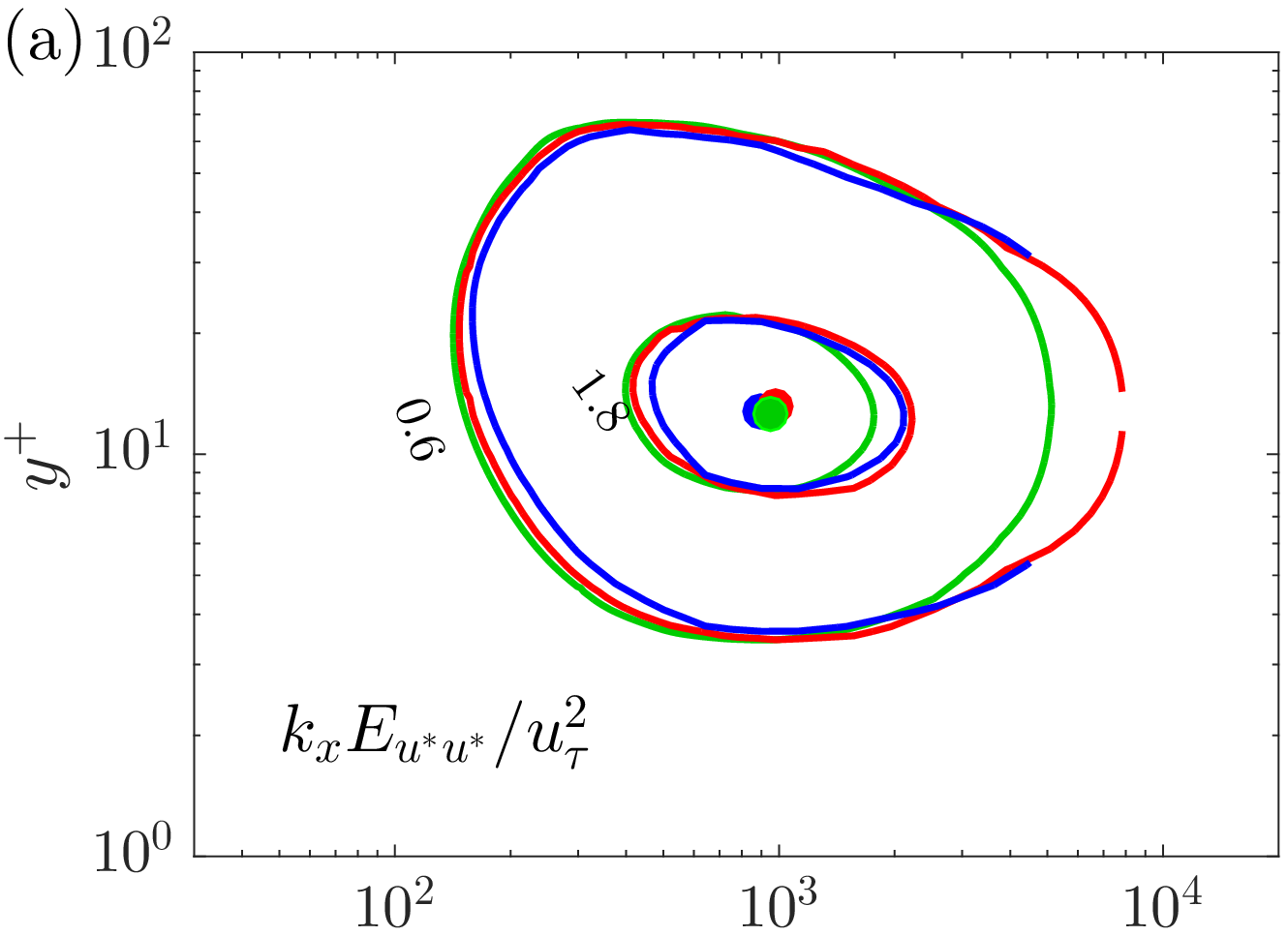}}
\end{minipage}
\hfill
\begin{minipage}{0.49\linewidth}
\centerline{\includegraphics[width=\textwidth]{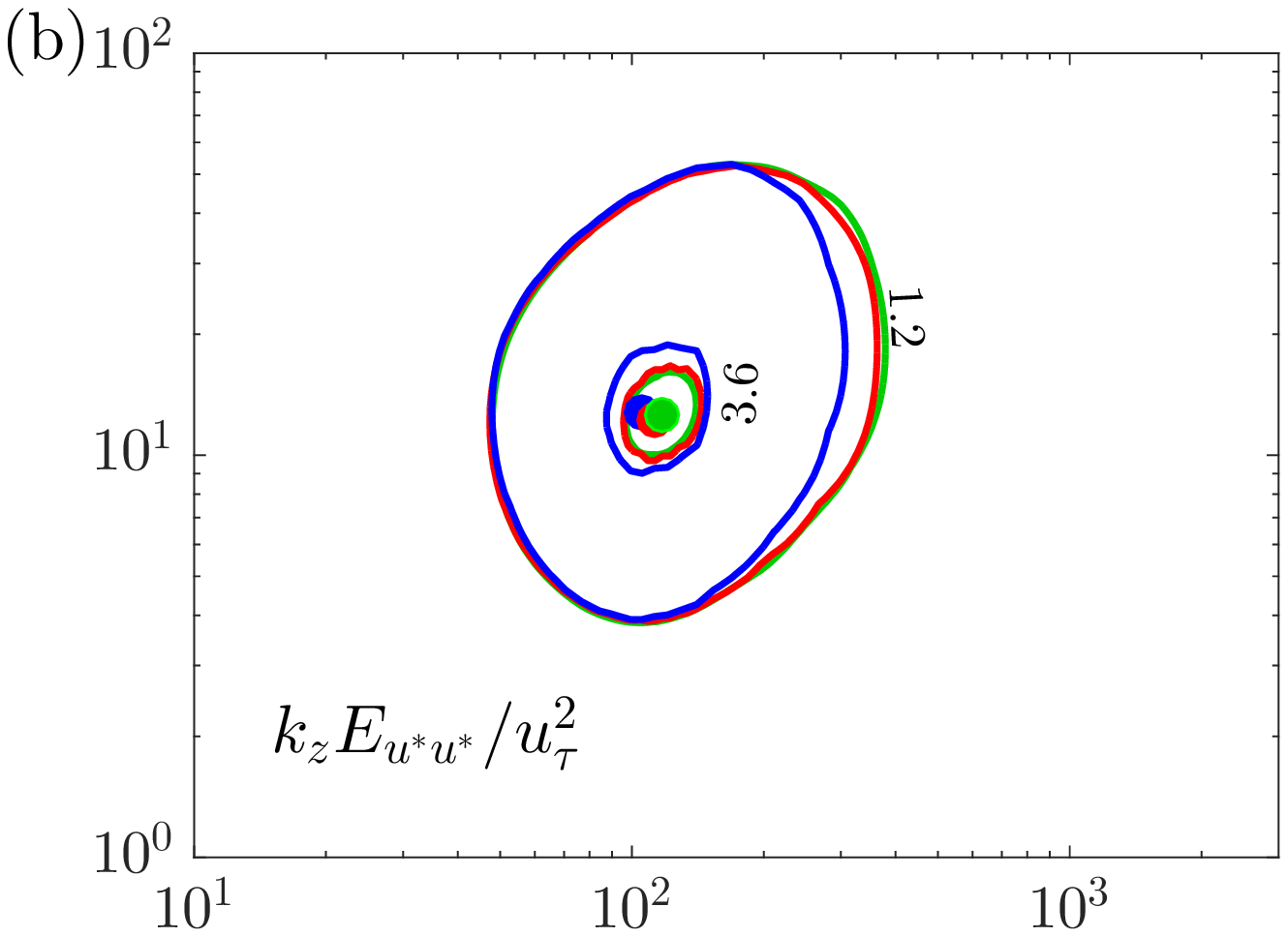}}
\end{minipage}
\vfill
\begin{minipage}{0.49\linewidth}
\centerline{\includegraphics[width=\textwidth]{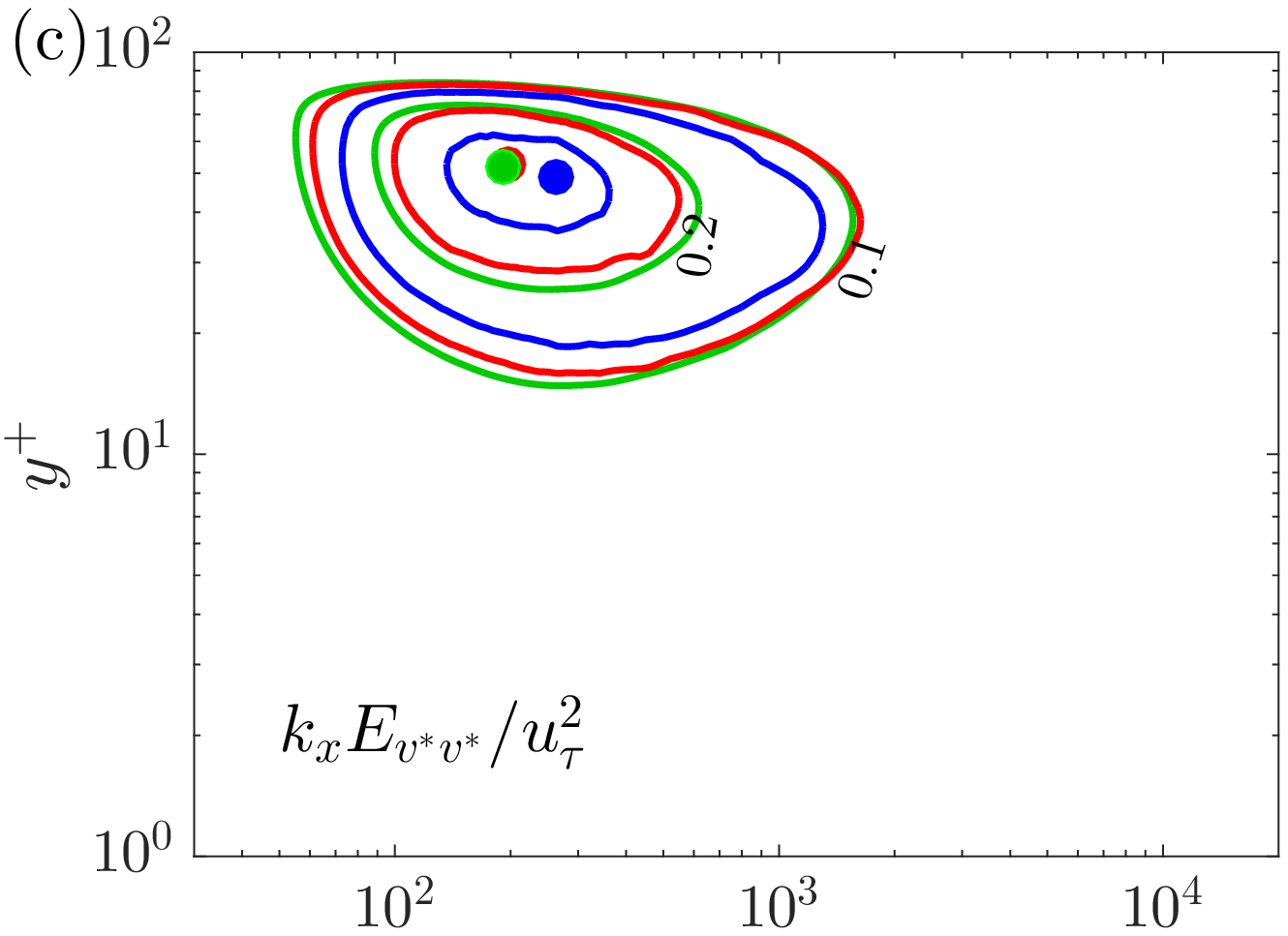}}
\end{minipage}
\hfill
\begin{minipage}{0.49\linewidth}
\centerline{\includegraphics[width=\textwidth]{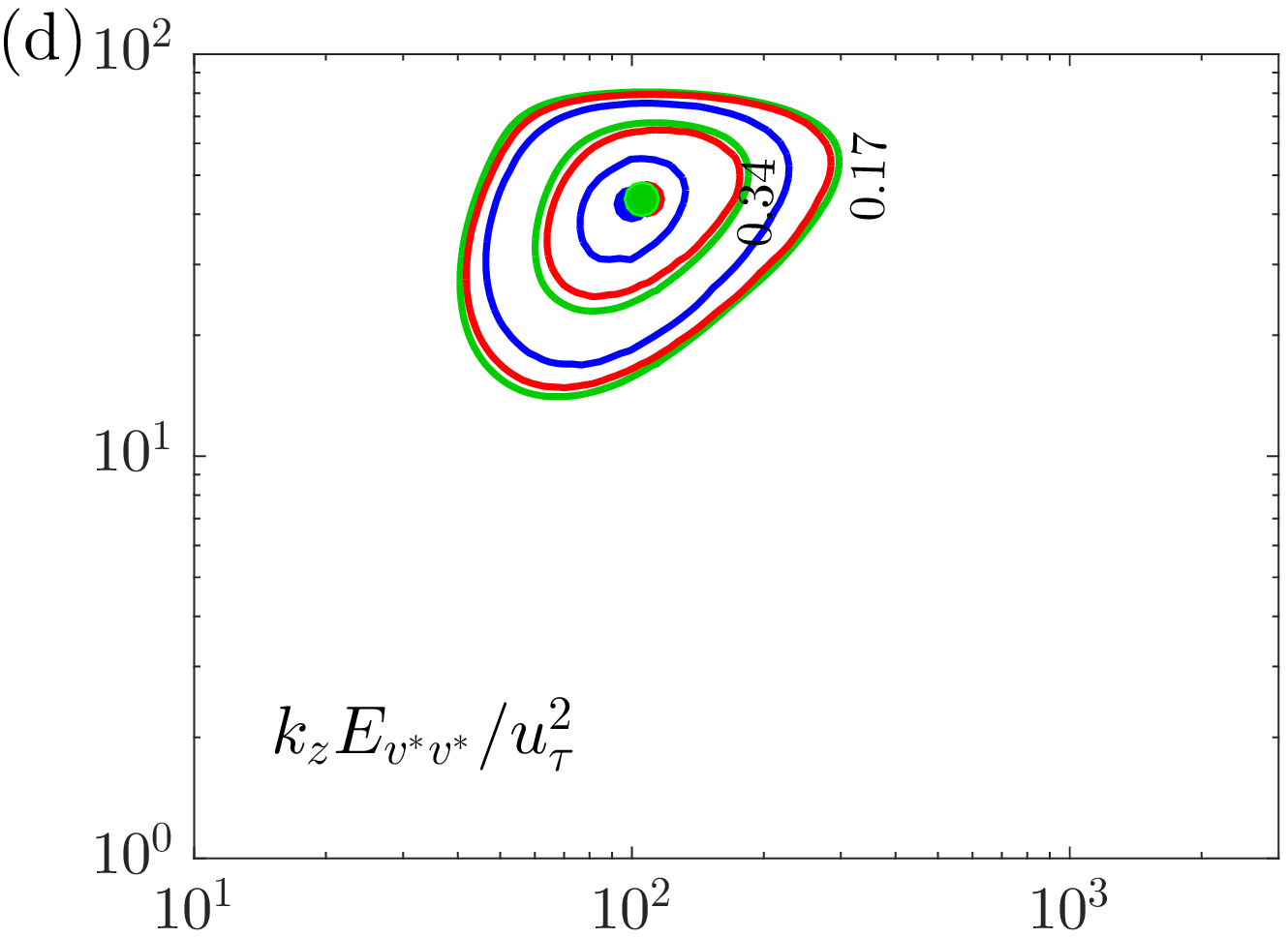}}
\end{minipage}
\begin{minipage}{0.49\linewidth}
\centerline{\includegraphics[width=\textwidth]{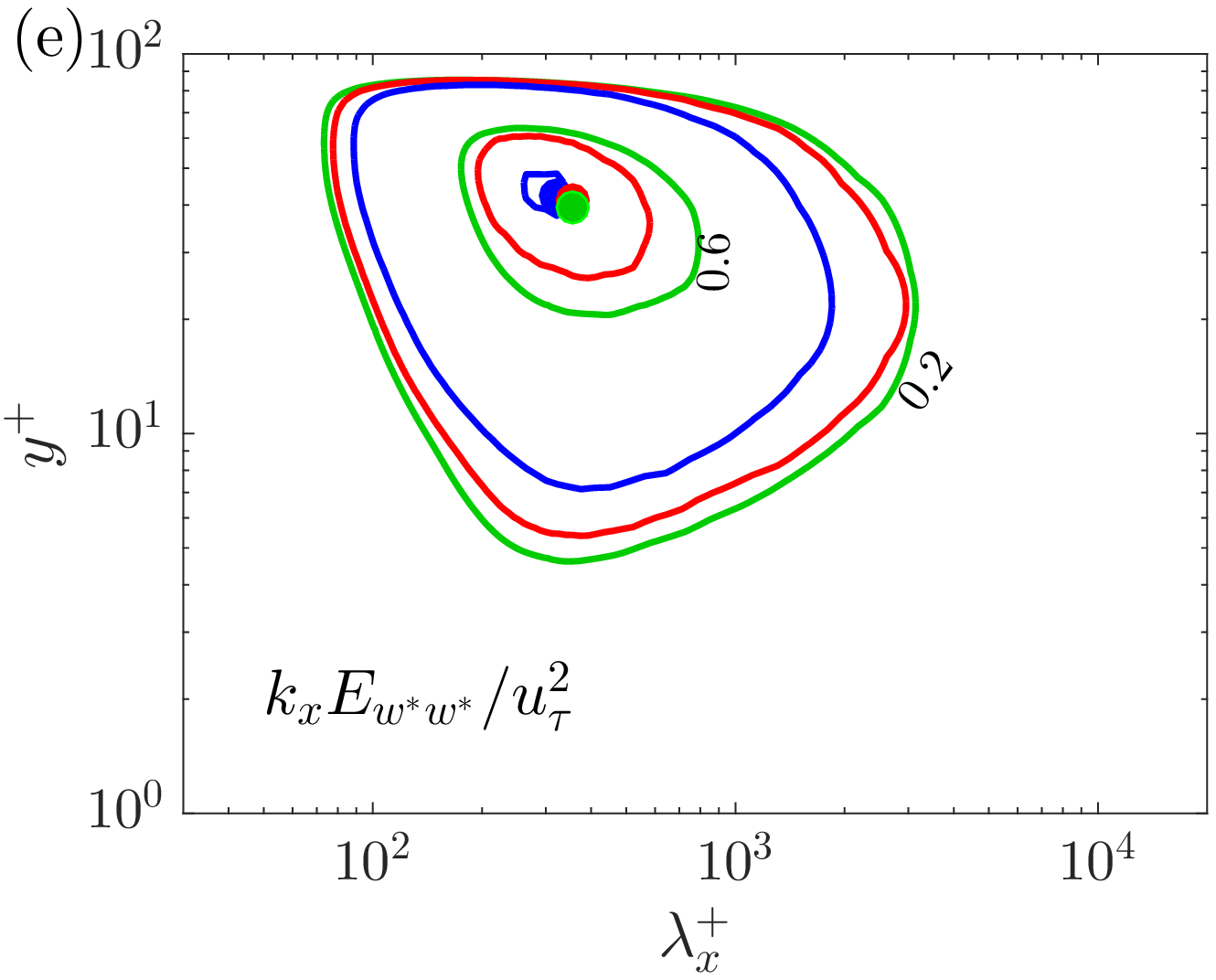}}
\end{minipage}
\hfill
\begin{minipage}{0.49\linewidth}
\centerline{\includegraphics[width=\textwidth]{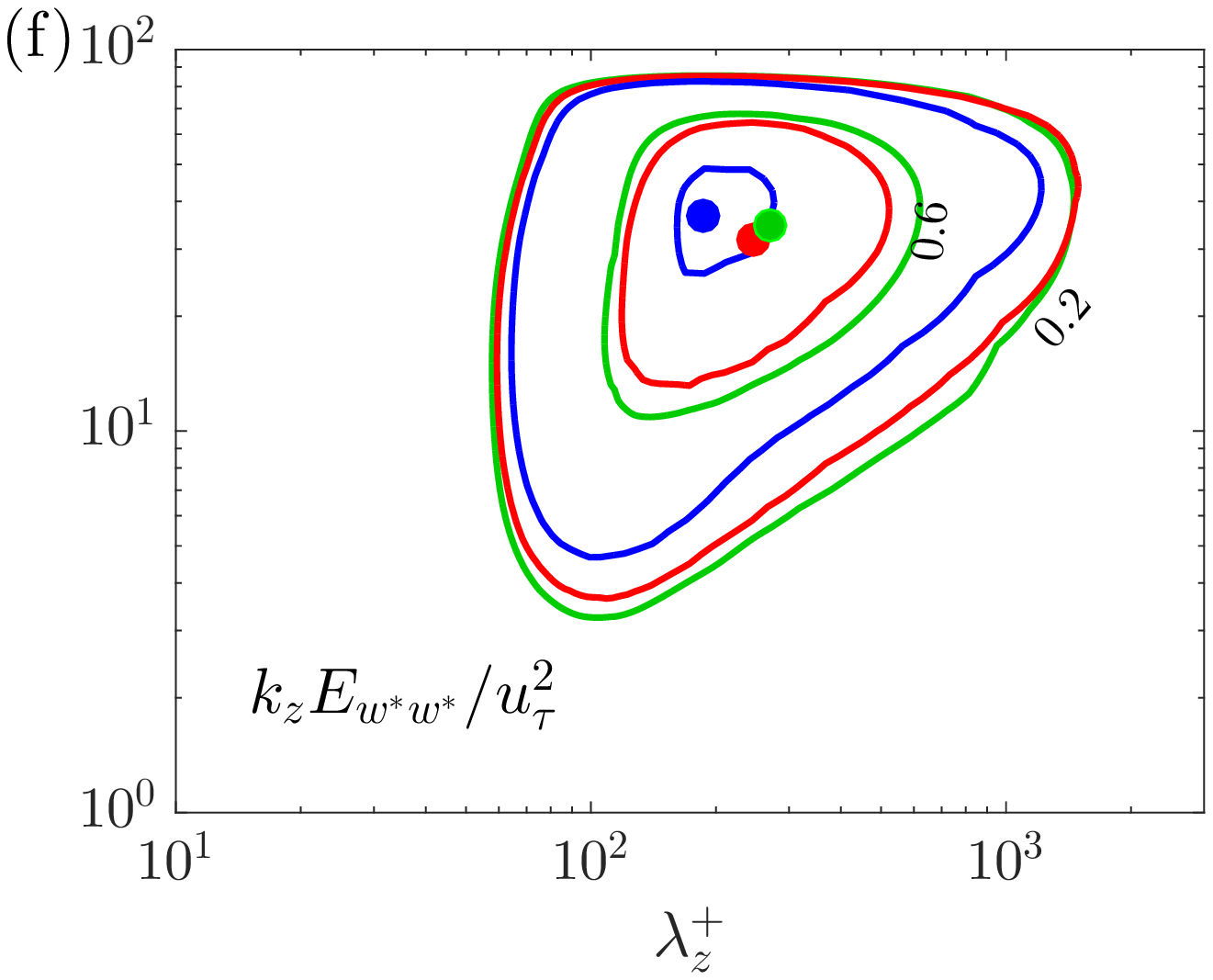}}
\end{minipage}
\caption{Pre-multiplied energy spectra of the three near-wall small-scale velocity components at $Re_\tau=$180, 310 and 600. (a) $k_x E_{u^*u^*}/u_\tau^2$, (b) $k_z E_{u^*u^*}/u_\tau^2$, (c) $k_x E_{v^*v^*}/u_\tau^2$, (d) $k_z E_{v^*v^*}/u_\tau^2$, (e) $k_x E_{w^*w^*}/u_\tau^2$ and (f) $k_z E_{w^*w^*}/u_\tau^2$. \textcolor{black}{The locations of spectral peaks have been marked with symbols.} Refer to table~\ref{tab:tab1} for the line colors.}
\label{fig:fig13}
\end{figure}

Furthermore, similar to \textcolor{black}{FIG}~\ref{fig:fig13}, we show the one-dimensional streamwise and spanwise pre-multiplied energy spectra of the three small-scale velocity components at the low Reynolds numbers in \textcolor{black}{FIG}~\ref{fig:fig13}, with the viscous-scaled wavelength ($\lambda_x^+$ or $\lambda_z^+$) and the wall-normal height $(y^+)$. For the contour level, we try to keep consistent with \textcolor{black}{FIG}~\ref{fig:fig13}, but for clarity, only two levels between zero and the peak values are shown. 
As displayed in \textcolor{black}{FIG}~\ref{fig:fig13} (a,b), the near-wall spectral peaks of the streamwise and spanwise pre-multiplied spectra locate at $y^+=10\sim20$ with $\lambda_x^+\sim O(10^3)$ and $\lambda_z^+\sim O(10^2)$, which are similar with those in \textcolor{black}{FIG}~\ref{fig:fig1} and \textcolor{black}{FIG}~\ref{fig:fig13}. 
However, it may not be claimed with confidence that the spectral contours are perfectly scaled by viscous units. In fact, \textcolor{black}{FIG}~\ref{fig:fig13} (a) shows that the streamwise wavelength of the inner peak is slightly larger at lower Reynolds numbers, which is consistent with our previous simulations at much lower Reynolds numbers \cite[]{Hu2018energy}. Meanwhile, the spanwise wavelength of the inner peak is larger at higher Reynolds number, see \textcolor{black}{FIG}~\ref{fig:fig13} (b). This may indicate that as Reynolds number increases, the near-wall small-scale $u^*$ structures tend to be shorter and wider, in the low Reynolds number regime. 
% The main difference is concentrated in the large-scale part, i.e.$\lambda_x^+ > 2000$ and $\lambda_z^+ > 150$. 
Moreover, in \textcolor{black}{FIG}~\ref{fig:fig13} (c-f), the streamwise and spanwise pre-multiplied energy spectra of $v^*$ and $w^*$ are presented, in which the wall-normal locations of the inner spectral peaks are also consistent with the undecomposed ones, as in \textcolor{black}{FIG}~\ref{fig:fig1} (c-f), say $y^+ = 30 \sim 70$. Compared to the $u^*$-spectra, the $v^*$- and $w^*$-spectra exhibit much stronger Reynolds number dependence. Particularly, as Reynolds number increases, the spectral energy increases accordingly, resulting in stronger integrated $\langle v^{*2}\rangle$ and $\langle w^{*2}\rangle$, as shown in \textcolor{black}{FIG}~\ref{fig:fig12} (b) and (c). Since the spanwise and wall-normal velocity components are closely related to vortical structures, it may imply that the strength of near-wall small-scale quasi-streamwise vortices increases with Reynolds number in the low $Re_\tau$ regime ($Re_\tau<1000$), and can reach a fully-developed asymptotic status once $Re_\tau \ge 1000$. It should be mentioned that Antonia \& Kim \cite{Antonia1994Low} also attributed the low-Reynolds-number effect to an increase in strength of the quasi-streamwise vortices in the buffer layer, other than the average diameter or average location.

\begin{figure}
\centering
\begin{minipage}{0.49\linewidth}
\centerline{\includegraphics[width=\textwidth]{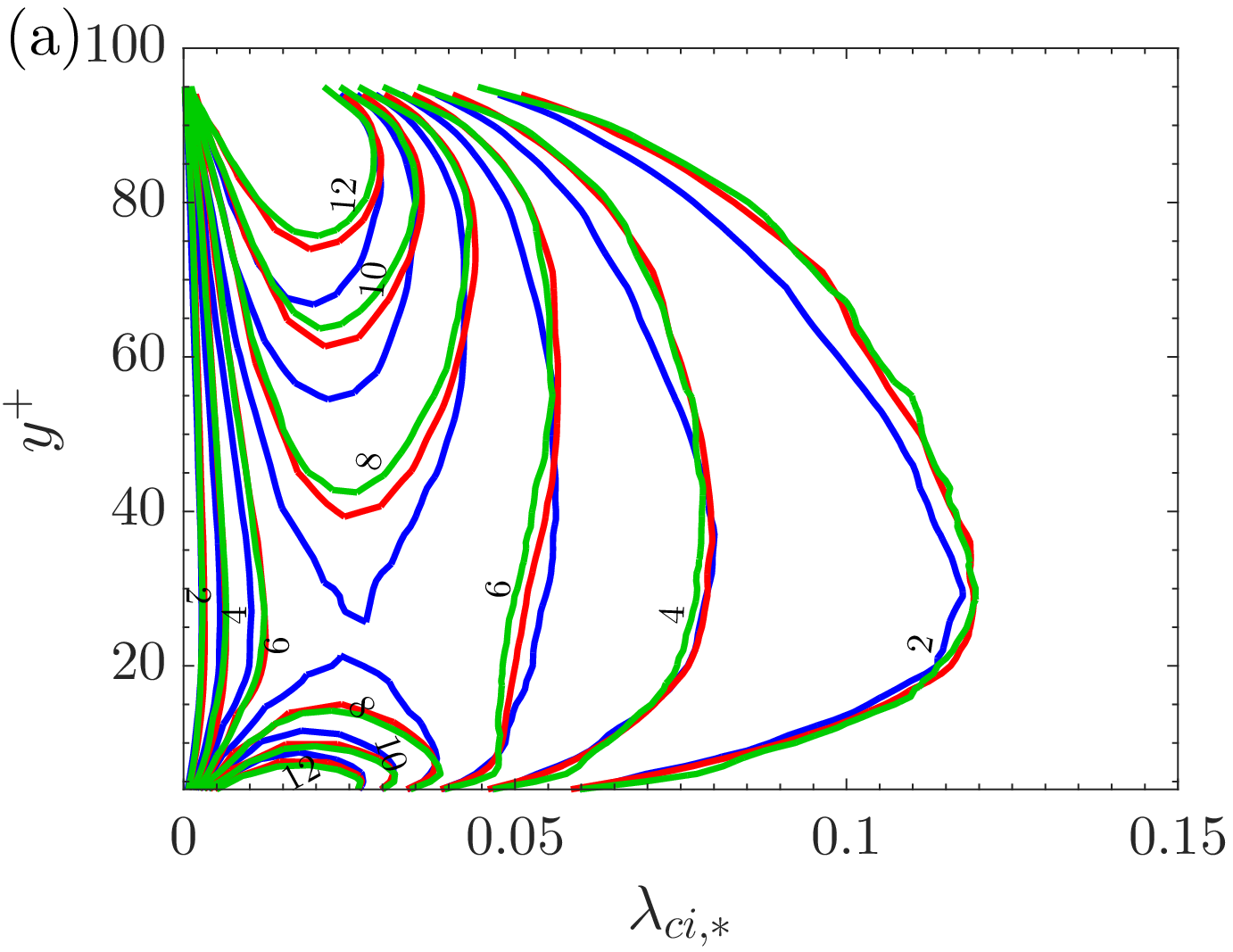}}
\end{minipage}
\hfill
\begin{minipage}{0.49\linewidth}
\centerline{\includegraphics[width=\textwidth]{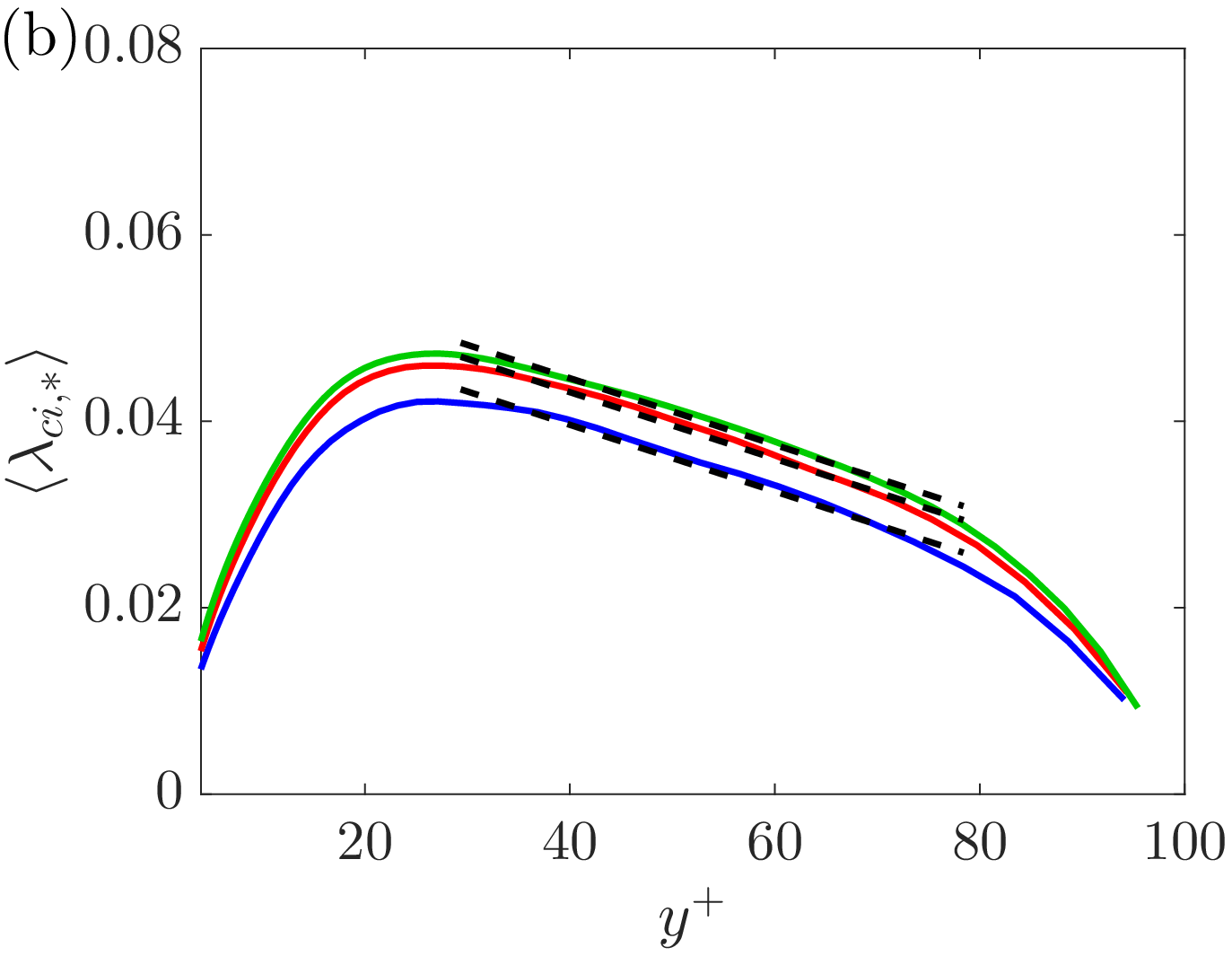}}
\end{minipage}
\caption{P.d.f.s of $\lambda_{ci,*}$ (a) and mean swirling strength $\langle \lambda_{ci,*} \rangle$ (b) as functions of $y^+$ at $Re_\tau$ = 180, 310 and 600. Refer to table~\ref{tab:tab1} for the line colors.}
\label{fig:fig14}
\end{figure}

Finally, we illustrate the low-Reynolds-number effect on the swirling strength statistics. The p.d.f. distributions of $\lambda_{ci,*}$ at $Re_\tau=180$, 310 and 600 in the near-wall region are  compared in \textcolor{black}{FIG}~\ref{fig:fig14} (a). The shapes of the p.d.f.s at the different Reynolds numbers are generally similar.
However, it can be clearly seen that, as Reynolds number increases, the probability of large swirling strength gradually becomes higher, indicating stronger vortical strength at higher $Re_\tau$. 
% Although the p.d.f.s results are different, a good consistency in the variation of the contour map with the $y^+$ can be observed among this Reynolds number. These results provide important clues for the Reynolds number effect on the vortex swirling strength.
Besides, \textcolor{black}{FIG}~\ref{fig:fig14} (b) demonstrates the mean swirling strength profiles at the three low Reynolds numbers. The swirling strength increases first then decreases with $y^+$, and the maximum value appears at $y^+ \approx 25$, which is similar to the higher Reynolds number cases, as in \textcolor{black}{FIG}~\ref{fig:fig10} (b). However, the absolute value of $\lambda_{ci,*}$ is found to increase with $Re_\tau$ and be consistent with the p.d.f. result.
Moreover, the linear relationship also exists and almost has an identical slope from $Re_\tau=$180 to 5200 in the range of $y^+ = 30\sim75$. In addition, the mean inclination and the corresponding p.d.f. of the near-wall small-scale vortex structures are found to be basically independent of Reynolds number, which are not shown here for saving space.

In summary, we have applied the decomposition scheme (\ref{eqn:equ9}-\ref{eqn:equ10}) to the low-Reynolds-number turbulent channels at $Re_\tau=180\sim600$, and the extracted near-wall small-scale velocity fields ($u^*$, $v^*$ and $w^*$) are discovered to be Reynolds number dependent. The main mechanism may be the strengthening of the near-wall quasi-streamwise vortical structures. And it will induce the augments of the velocity fluctuations, as well as the widening and shortening of the near-wall streaks.

% \subseber ction{Flow structures}

\section{Characteristics of outer footprints}

In this part, we present the characteristics of the near-wall footprints of outer turbulent motions, i.e., $(u_L, v_L, w_L)$, also known as the superposition effect \cite[]{Hutchins2007,Marusic2010,Mathis2011}, as well as their interactions with the near-wall small-scale motions, which is quantified by the amplitude modulation of large scales to small scales.

\begin{figure}
\centering
\begin{minipage}{0.49\linewidth}
\centerline{\includegraphics[width=\textwidth]{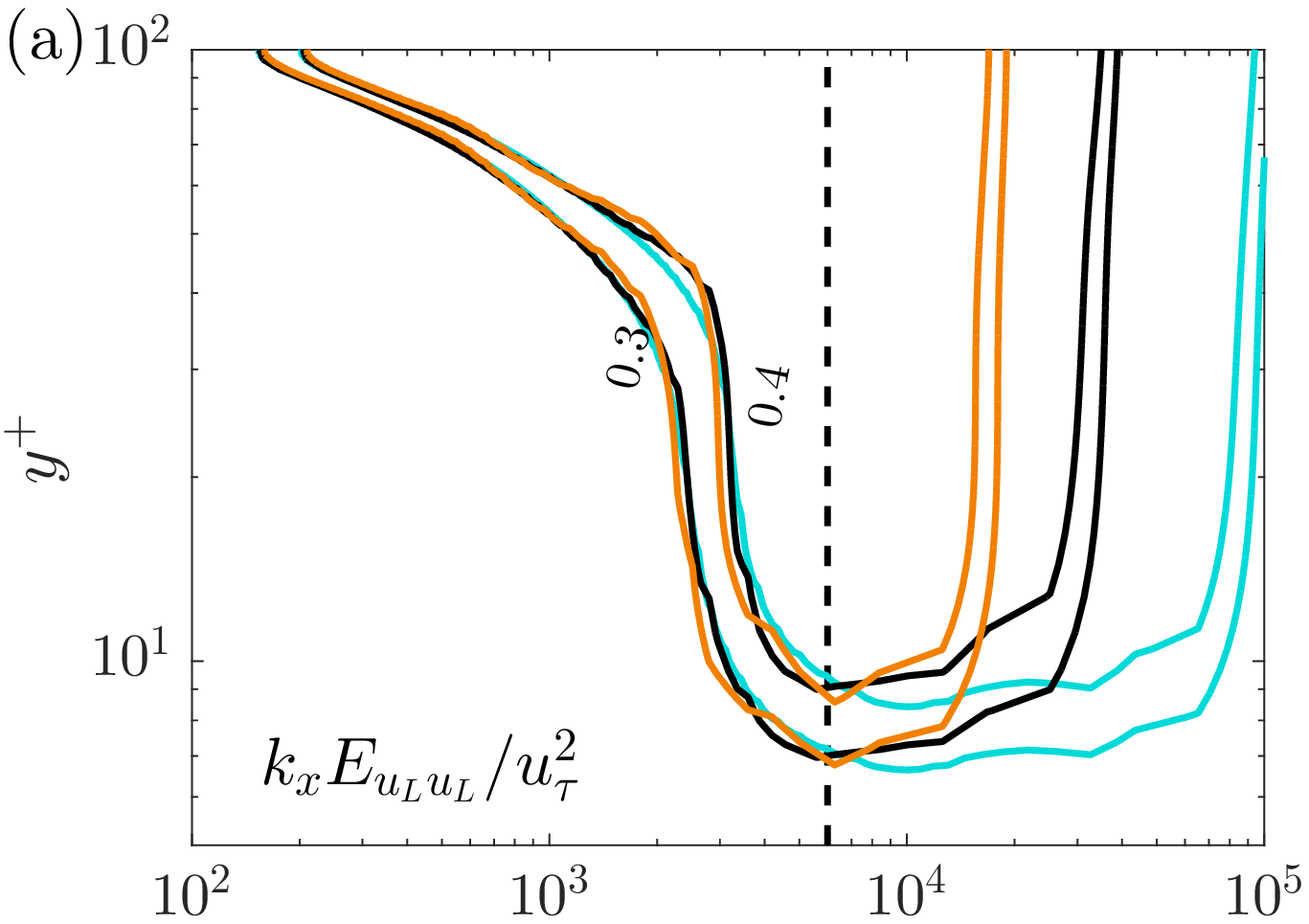}}
\end{minipage}
\hfill
\begin{minipage}{0.49\linewidth}
\centerline{\includegraphics[width=\textwidth]{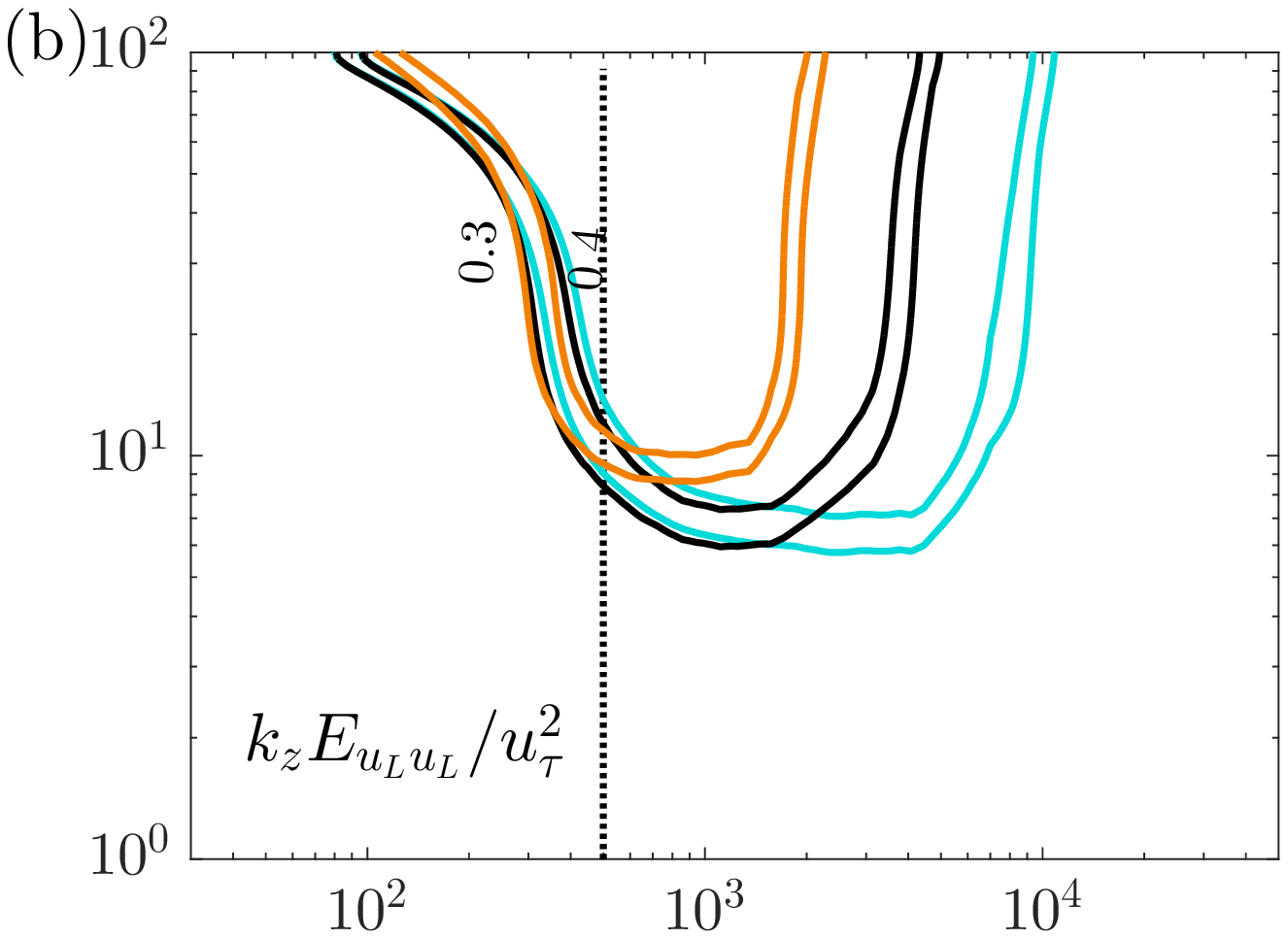}}
\end{minipage}
\vfill
\begin{minipage}{0.49\linewidth}
\centerline{\includegraphics[width=\textwidth]{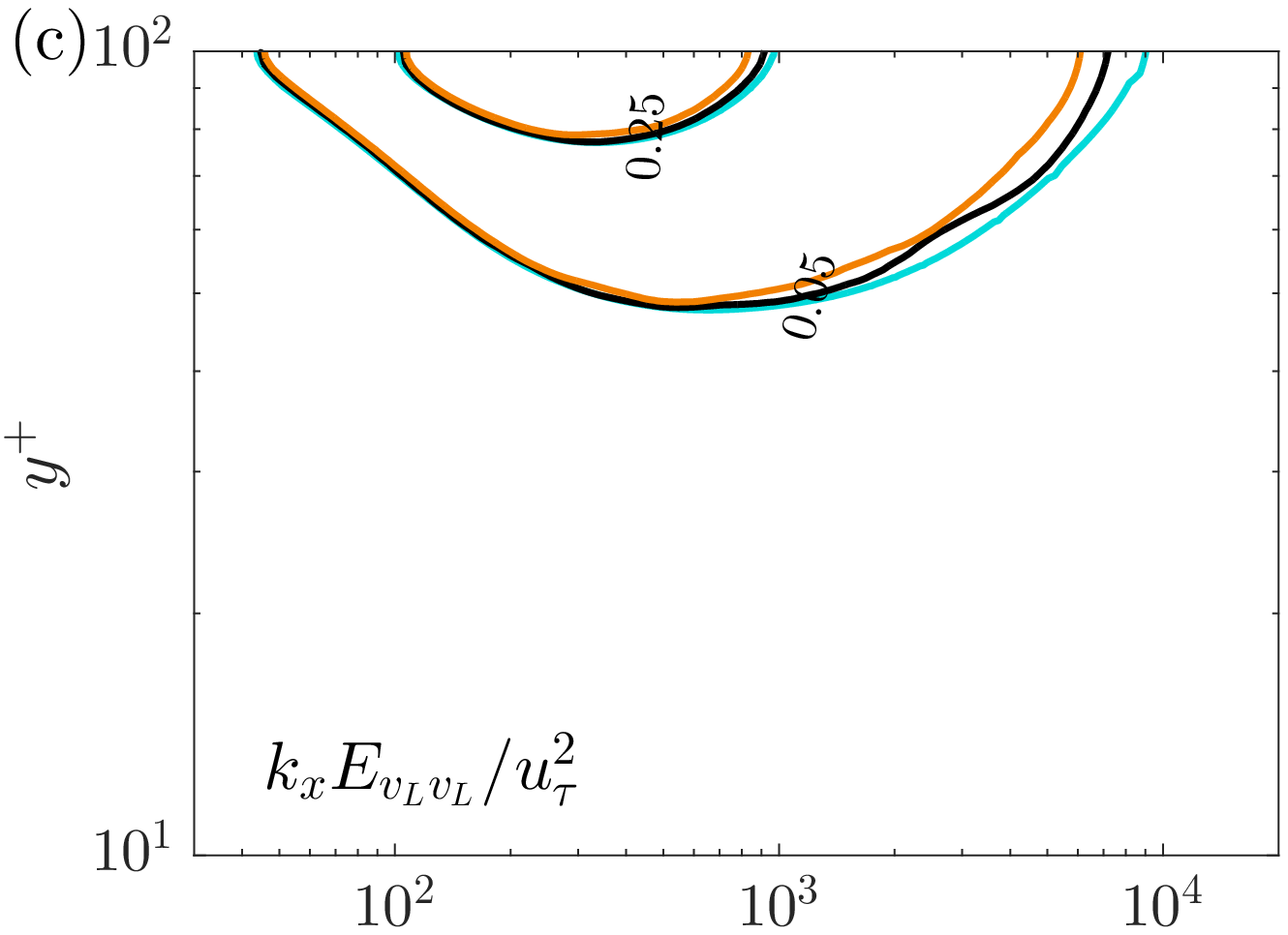}}
\end{minipage}
\hfill
\begin{minipage}{0.49\linewidth}
\centerline{\includegraphics[width=\textwidth]{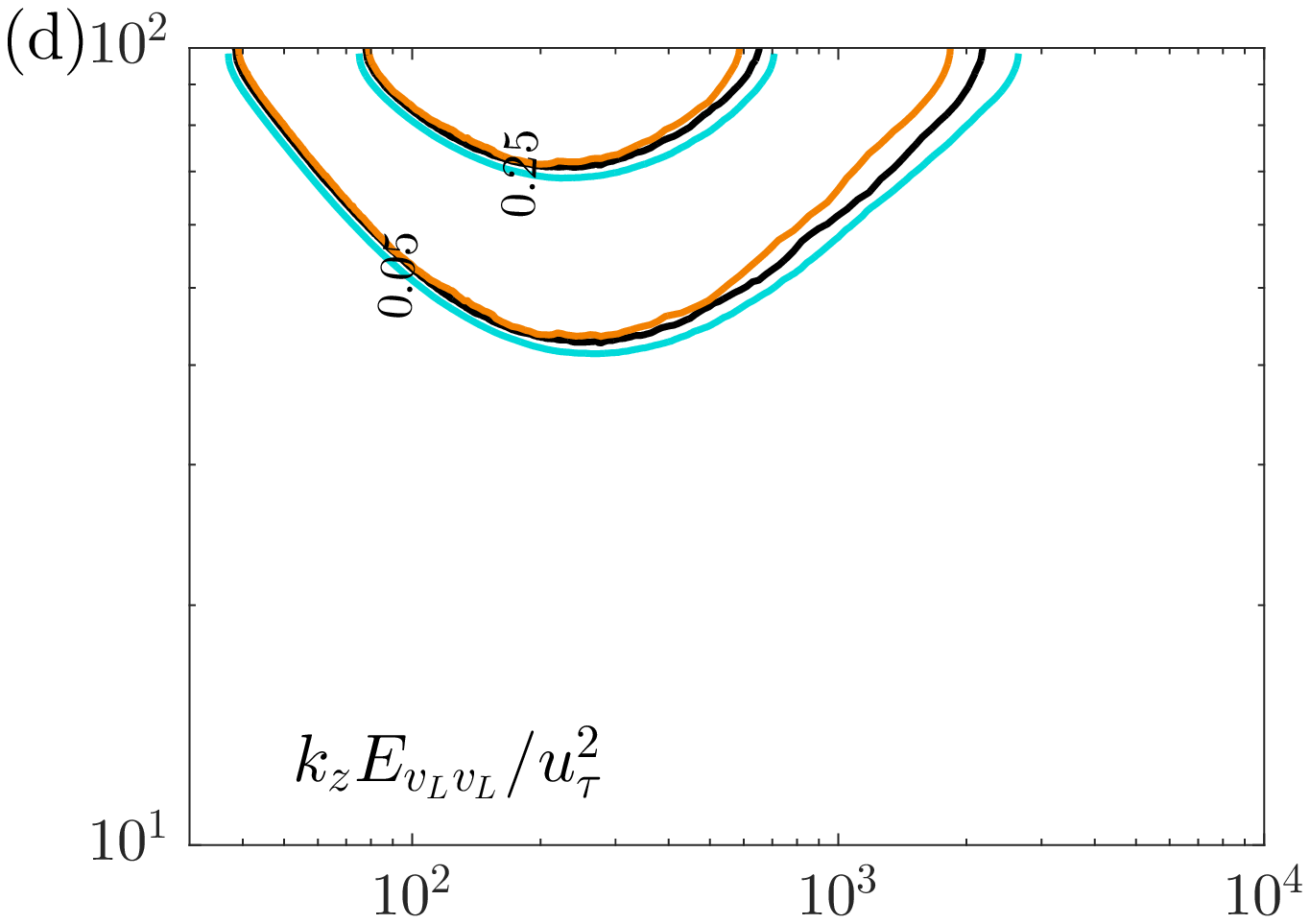}}
\end{minipage}
\vfill
\begin{minipage}{0.49\linewidth}
\centerline{\includegraphics[width=\textwidth]{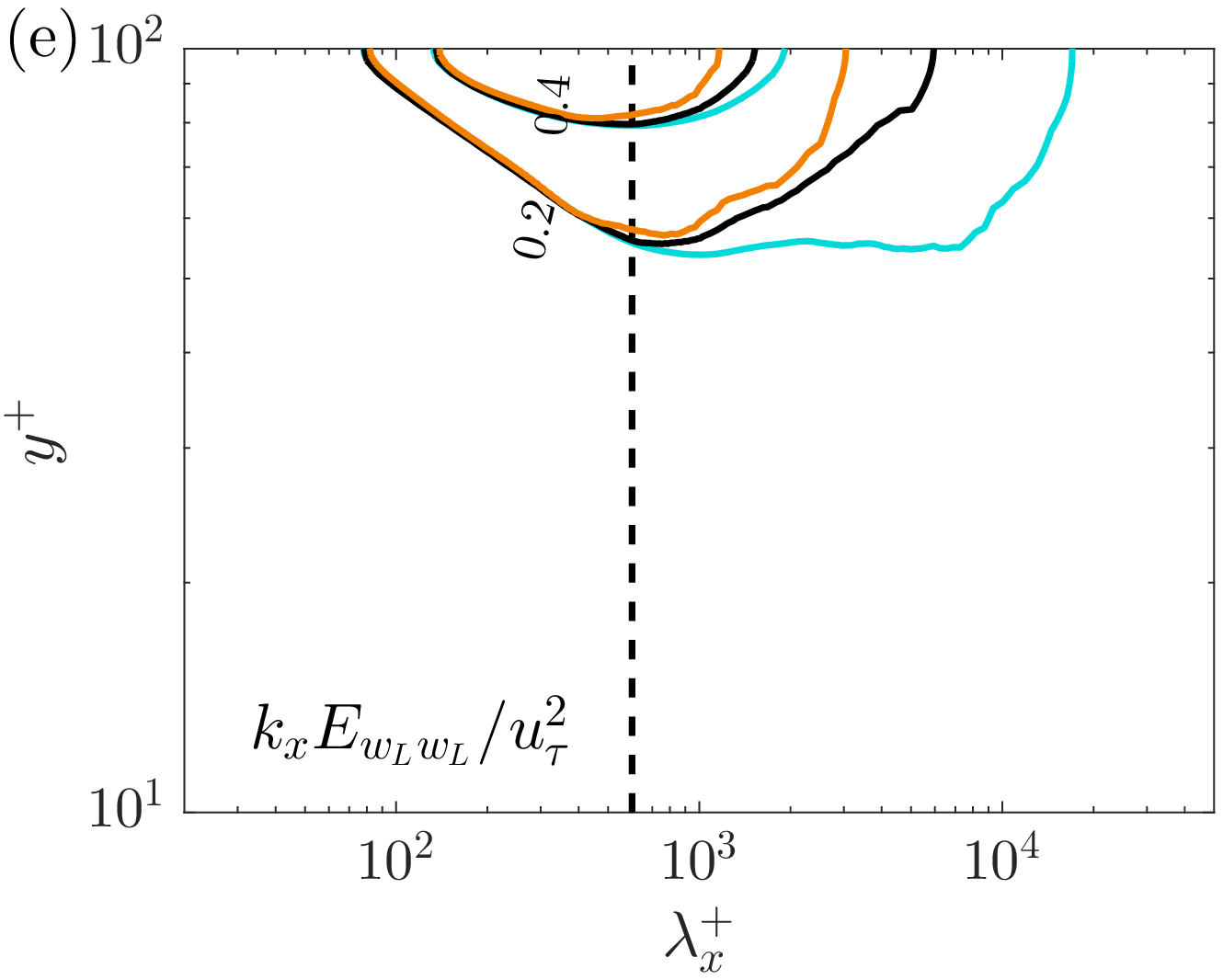}}
\end{minipage}
\hfill
\begin{minipage}{0.49\linewidth}
\centerline{\includegraphics[width=\textwidth]{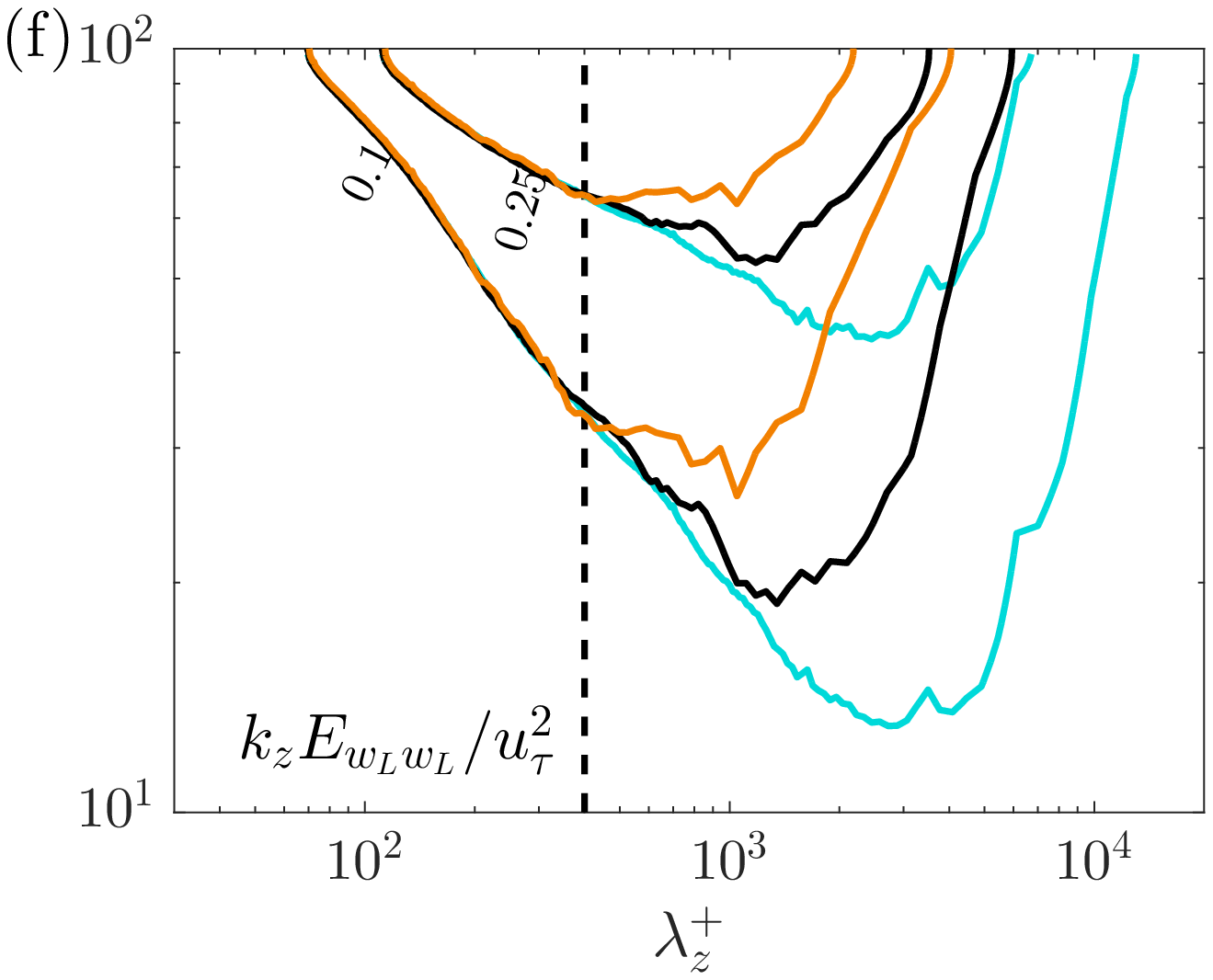}}
\end{minipage}
\caption{Streamwise (a,c,e) and spanwise (b,d,f) pre-multiplied energy spectra of the outer footprint velocity components $(u_L, v_L, w_L)$ with the viscous-scaled wavelength $\lambda_x^+$ and $\lambda_z^+$. Refer to table~\ref{tab:tab1} for the line colors.}
\label{fig:fig15}
\end{figure}

We scrutinize the spectral energy distributions of the outer footprints in the near-wall region. The streamwise and spanwise pre-multiplied energy spectra of all the three velocity components with the viscous scaling, are given in \textcolor{black}{FIG}~\ref{fig:fig15}. In \textcolor{black}{FIG}~\ref{fig:fig15} (a) and (b), it is shown that the TKE spectral distribution of $u_L$ seems to obey the viscous scaling very well in the range of $\lambda_x^+ < 6000$ and $\lambda_z^+<500$ approximately at $Re_\tau\ge1000$. 
% With the outer scaling, we can see that the $u_L$ spectra collapse with each other very well for $\lambda_x/\delta > 6$ at $Re_\tau\ge1000$, see figure~\ref{fig:fig15} (b). 
% Generally, scaling of the 'footprint' large-scale energy spectra agree notably well full streamwise velocity energy spectra. 
Furthermore, the pre-multiplied energy spectra of the wall-normal velocity footprints $v_L$ are displayed in \textcolor{black}{FIG} \ref{fig:fig15} (c,d). 
% It is seen that, clear Reynolds number dependence of the $v_L$ spectra exists with the outer scaling, see figure~\ref{fig:fig15} (d). 
It is seen that excellent collapse is found with the viscous scaling with $\lambda_x^+ < 1000$ and $\lambda_z^+<300$ at $Re_\tau\le1000$.
% , as displayed in figure~\ref{fig:fig15} (c). 
The pre-multiplied spectra of the spanwise velocity footprints $w_L$ are shown in \textcolor{black}{FIG}~\ref{fig:fig15} (e,f). In the wavelength region of $\lambda_x^+ \le 600$ and $\lambda_z^+<400$ at $Re_\tau\le1000$, it can be clearly seen that the spectra are well collapsed with the viscous-scaled wavelength. 
% However, from figure~\ref{fig:fig15} (f), we may not confirm a collapse of the $w_L$ spectra with the outer-scaled wavelength, as good as the streamwise one. 
% The excellent collapse of $u_L$, $v_L$ and $w_L$ spectra with the viscous-scaled wavelength in the small-scale regimes, may imply possible relations of the small-scale parts of the outer footprints (as well as the corresponding outer motions) with the viscous-dominated near-wall turbulence. 
% Somewhat unexpectedly, however, there is obvious deviation for Reynolds number 5200 and unreasonable for this trend, shown in figure \ref{fig:fig15} (f). The deviation may be because the amount of 5200 data is too few to get fully converge. 

\begin{figure}
\centering
\begin{minipage}{0.49\linewidth}
\centerline{\includegraphics[width=\textwidth]{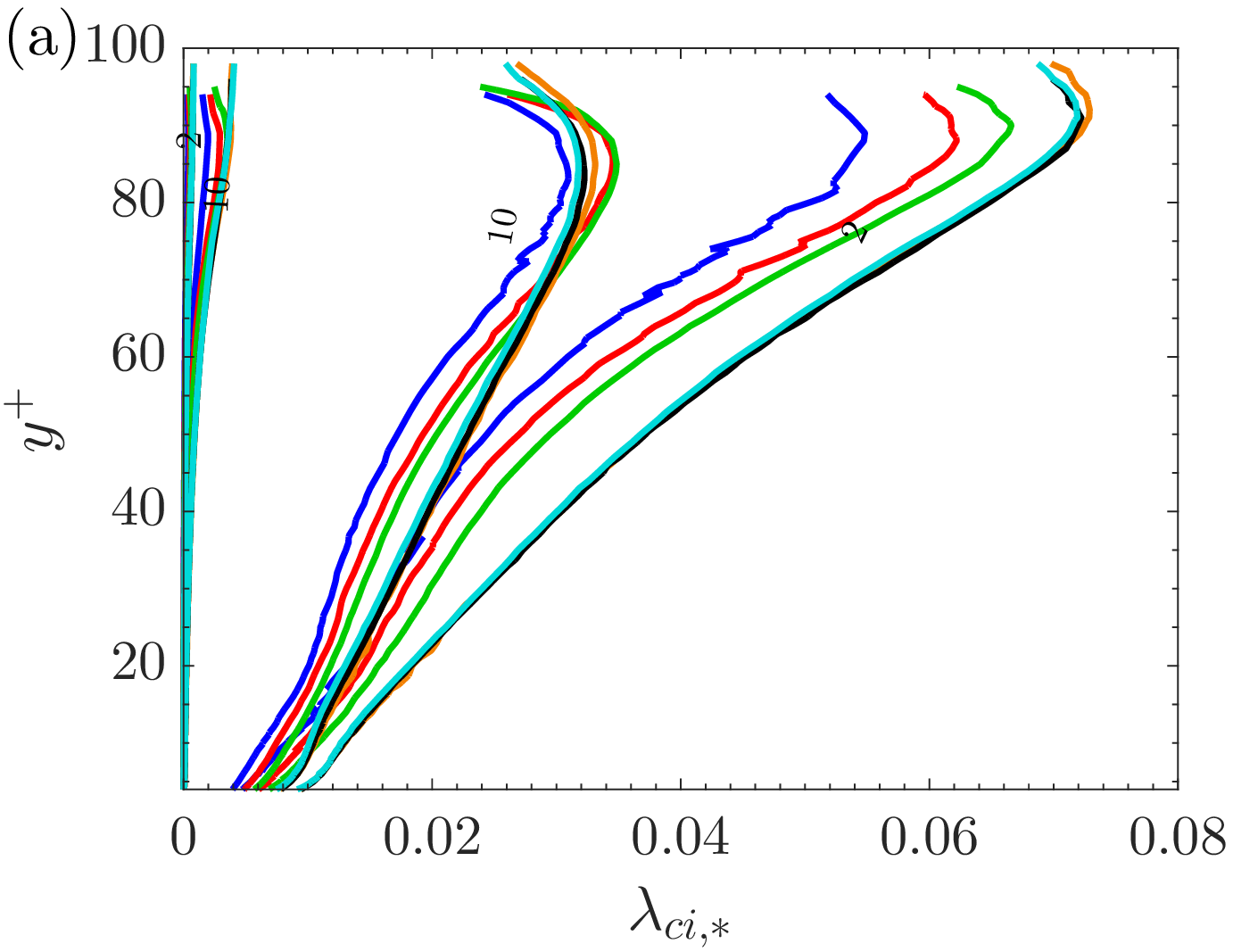}}
\end{minipage}
\hfill
\begin{minipage}{0.49\linewidth}
\centerline{\includegraphics[width=\textwidth]{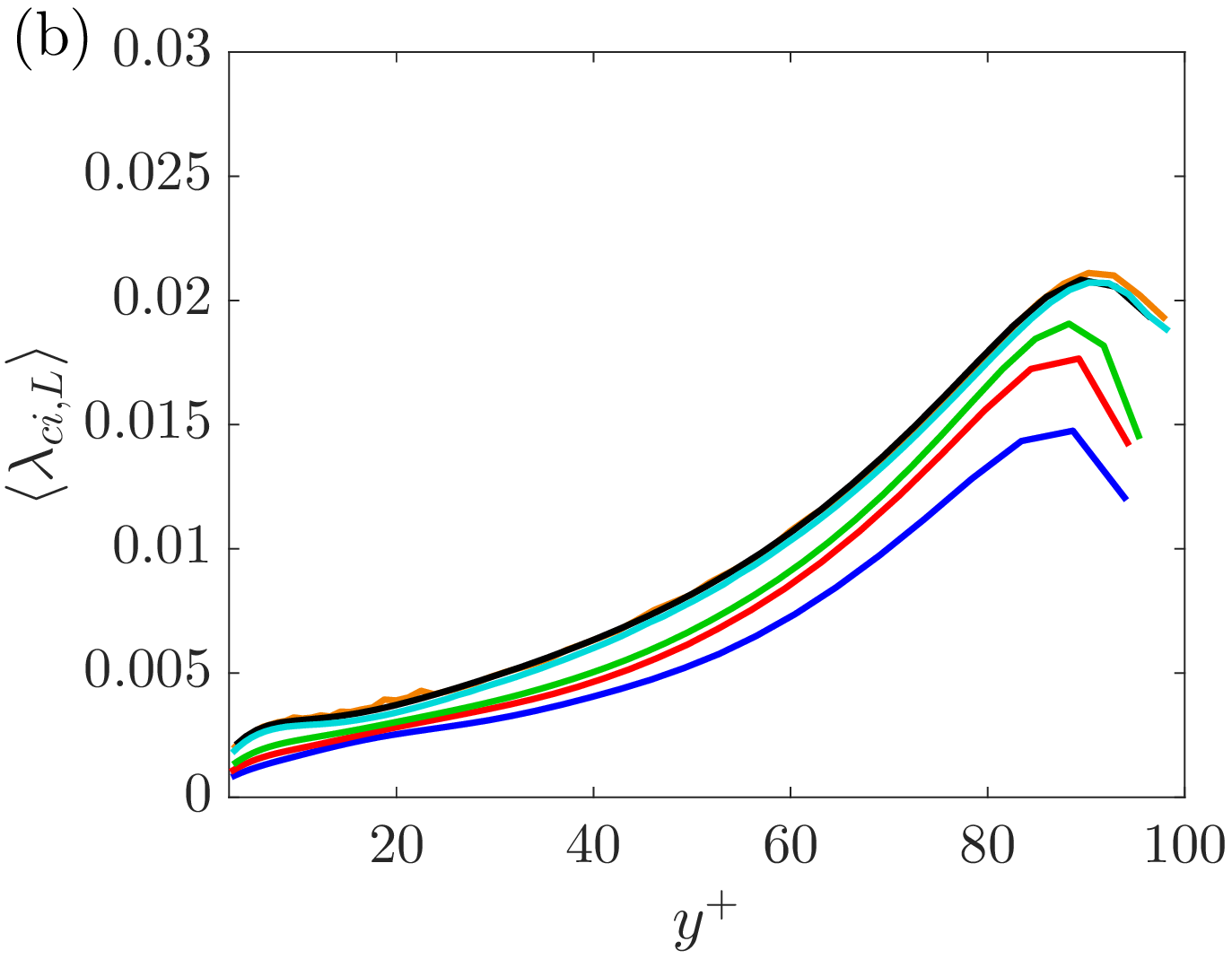}}
\end{minipage}
\caption{P.d.f.s (a) and mean profiles (b) of $\lambda_{ci,L}$ as functions of $y^+$ at the Reynolds numbers from 180 to 5200. Refer to table~\ref{tab:tab1} for the line colors.}
\label{fig:fig16}
\end{figure}

At last, we demonstrate the Reynolds-number effect on the swirling strength statistics of outer footprint fields.
The p.d.f. distributions of $\lambda_{ci,L}$ at the Reynolds numbers from 180 to 5200 in the near-wall region are compared in \textcolor{black}{FIG}~\ref{fig:fig16} (a). The contours of the p.d.f.s exhibit excellent collapse at the Reynolds numbers of 1000, 2000 and 5200. However, the swirling strengths at lower Reynolds numbers show definite $Re_\tau$ dependence, and decreases as Reynolds number decreases. 
\textcolor{black}{FIG}~\ref{fig:fig16} (b) displays the mean swirling strength profiles at different Reynolds numbers. It can be seen that, the mean swirling
strength $\langle \lambda_{ci,L} \rangle$ is generally larger at higher $y^+$, while only decays at about $y^+>90$ possibly due to the  decrease of the gradient of $u_{i,L}$ near $y^+=100$. At the three higher Reynolds numbers, i.e., $Re_\tau=1000$, 2000 and 5200, the $\langle \lambda_{ci,L} \rangle$ profiles collapse very well which is consistent with \textcolor{black}{FIG}~\ref{fig:fig16} (a), but not the case at lower Reynolds numbers which is increasing with the Reynolds number. 

\section{Concluding remarks}\label{sec:con}

In this work, we present a scaling based decomposition methodology of three-dimensional turbulence velocities into small-scale and large-scale components in the near-wall region at $y^+<100$. The method is principally based on the refined PIO model of Baars \emph{et al.} \cite{Baars2016}. However, a significant difference is that we use $y_O^+=100$ instead of $y^+_O=3.9\sqrt{Re_\tau}$ as the reference height for evaluating outer footprints. Reynolds-number-invariant small-scale turbulent motions are then extracted at $1000 \le Re_\tau \le 5200$ with plenty of evidences, including integrated intensities, spectra and joint p.d.f.s of velocity fluctuations, as well as vortex swirling quantities. 
% Both the viscous-scaled intensities and length scales of the outer footprints increase with Reynolds number.
Finally, it is discovered that a small-scale part of the outer footprint can also be well scaled by the viscous units, as well as the vortical statistics.

% We know $y^+ = 100$ usually as the lower bound of the logarithmic region \citep{garcia2013off, mizuno2013wall, tuerke2013simulations}. Moreover, \cite{Hutchins2007} suggested the logarithmic region will only exist when
% \begin{equation}
%   0.15\delta > 100 \frac{\nu}{u_\tau}
%     \label{equ:equ8}
% \end{equation}
% giving a lower Reynolds number limit for logarithmic behaviour of the mean
% velocity profile. This gives
% \begin{equation}
%   Re_\tau = \frac{\delta u_\tau}{\nu} > 667
%     \label{equ:equ9}
% \end{equation}
% In the log-region, eddy sizes scale with their distance from the wall and fluid motions in the inertial range are self-similar, as is also the case for structures in the log-region consistent with the attached-eddy hypothesis scenario proposed by \cite{Townsend1976}, which suggests that the near-wall region will fell wall-parallel motions due to all attached eddies that reside above that point (including superstructures).

\begin{figure}
% \centering
% \begin{minipage}{0.7\linewidth}
% \centerline{\includegraphics[width=\textwidth]{LCS_100.eps}}
% \end{minipage}
% \vfill
\begin{minipage}{\linewidth}
\centerline{\includegraphics[width=\textwidth]{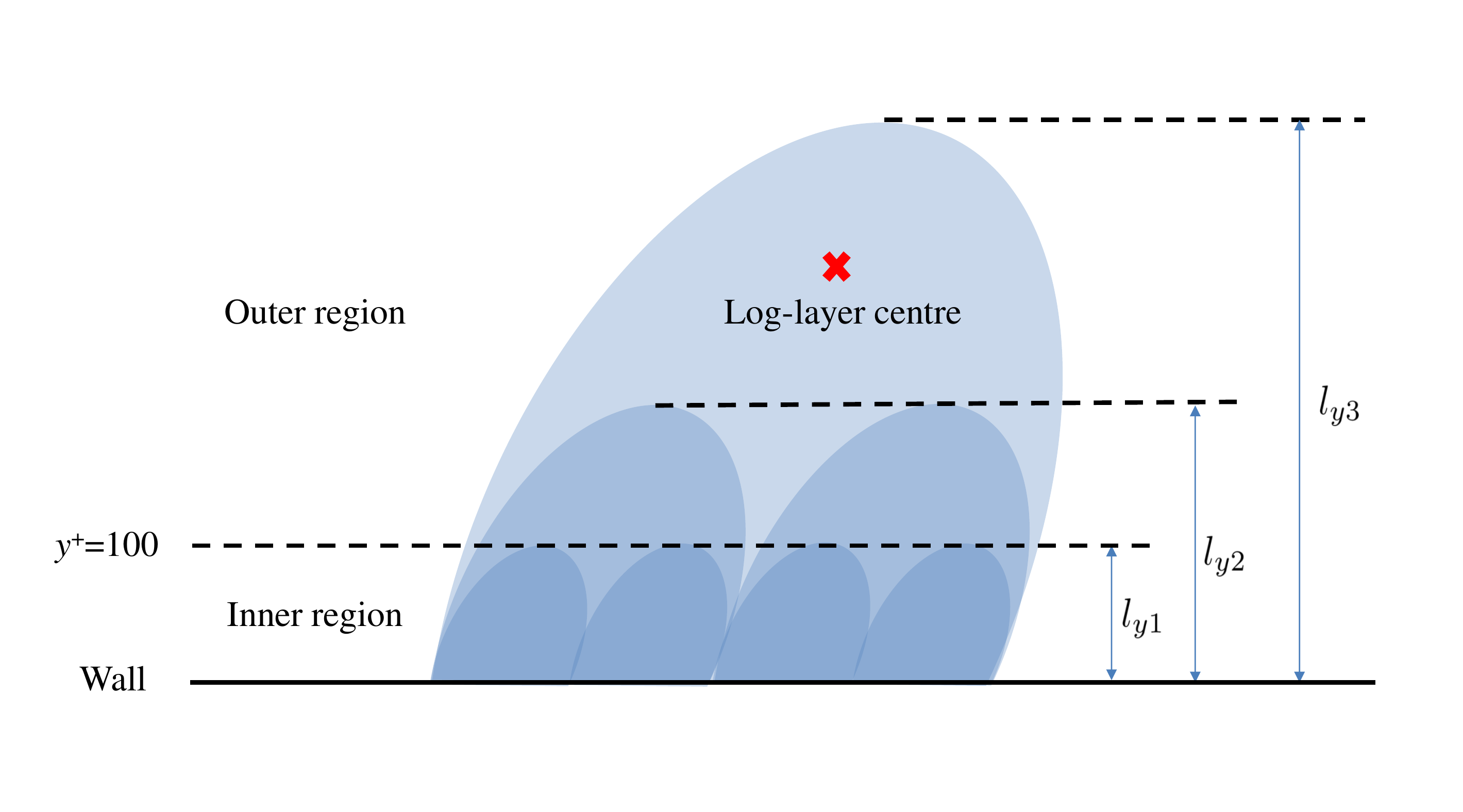}}
\end{minipage}
\caption{Sketch of a hierarchical structure of attached eddies with three levels.}
\label{fig:AE}
\end{figure}

The reason for the improved scaling collapse can be simply attributed to that the eddies with size of $100 < l_y^+ < 3.9\sqrt{Re_\tau}$ should also be responsible and incorporated for near-wall footprints, and $y^+\approx100$ can be regarded as one 'critical' dividing height of inner and outer regions in the context of inner-outer interactions. A recent experimental investigation in open channel flows also supports it \cite[]{Duan2020characteristics}. 
This can be further elucidated by the properties of attached eddies and detached eddies \cite[]{Perry1995,Baars2020a,Hu2020AEH} that both have coherence with the wall. The detached eddies in the context of Hu \emph{et al.} \cite{Hu2020AEH} are longer than the attached eddies and peaked at the centre of the logarithmic layer approximately. This is why previous studies commonly used $y^+_O = 3.9\sqrt{Re_\tau}$, where the outer spectral peak resides \cite[]{Mathis2011,Baars2016}. However, the attached eddies are more populated near the wall and lead to the logarithmic decay of $\langle u^{+2} \rangle$. As illustrated in \textcolor{black}{FIG}~\ref{fig:AE}, if we use $y^+_O\sim\sqrt{Re_\tau}$, only the contribution of the largest attached eddies is included. In order to take into account smaller eddies, it is required to let $y^+_O=l_{y1}^+$, i.e., the size of the smallest attached eddies. According to the present finding, it is suggested that $l_{y1}^+=y^+_O\approx100$, which is consistent with previous conjecture that the smallest attached eddies should be on the order of 100 viscous units in height \cite[]{Perry1982}.

At lower Reynolds numbers of $Re_\tau < 1000$, we find that the extracted small-scale velocity and swirling-strength statistics can not be scaled by the viscous units. The intensities of these turbulence quantities increase with Reynolds number, showing a developing trend of near-wall small-scale turbulence. After examining the vortical structures, it is revealed that the overall swirling strength of the small-scale motions is enhanced at larger $Re_\tau$, which is consistent with the vortex strengthening mechanism \cite[]{Wei1989Reynolds,Antonia1992Low,Antonia1994Low}. 

The Reynolds-number independence of the extracted small-scale motions is not obtained at low Reynolds numbers, thus we may add a restriction for applications of the PIO model as $Re_\tau \ge 1000$. On the other hand, the low-Reynolds-number effect could probably not be a surprise, which can also be observed in \textcolor{black}{FIG}~\ref{fig:fig1} before decomposition, and explained by insufficient separation of inner and outer scales in this range. 
Another related issue is the quasi-steady-quasi-homogeneous (QSQH) theory proposed by Chernyshenko and coworkers \cite[]{chernyshenko2012quasi,zhang2016quasisteady}, in which it is assumed that the small-scale motions vary much faster than the large-scale motions, therefore should be universal if scaled by local large-scale wall shear stress, instead of the mean one in the viscous scaling. \textcolor{black}{Moreover, some studies \cite[]{mizuno2011mean,lozano2019} also showed that velocity fluctuations can be better-scaled accounting by the effect of the mean shear.} In fact, the QSQH theory admits two universalities, i.e., one is Reynolds-number invariance as in the PIO model, and the other is the independence of the small scales scaled by the large-scale wall shear stress. The former one was validated by Chernyshenko and coworkers using spectral cut-off filters \cite[]{Chernyshenko2017extrapolating,Chernyshenko2019large}.The latter has been recently checked by Agostini \& Leschziner \cite{Agostini2019departure} at a single Reynolds number of $Re_\tau \approx 1000$. Each one of the two universalities may not rely on the other. Whether the low-Reynolds-number data satisfies the second universality of the QSQH theory or not could be checked in future. 

% We would like to note that the low-Reynolds-number effect is also consistent with our previous simulations at lower Reynolds numbers \cite[]{Hu2018energy}. 
% Although a wide range of viscous scaling is not established at low Reynolds numbers, it is still speculated that turbulent outer motions may emerge at a quite small Reynolds number as $Re_\tau \approx 100$, since $y^+>100$ is suggested as the outer region here. 

% Finally, we address the characteristics of the outer footprints and inter-scale interactions. Both the viscous-scaled intensities and length scales of the outer footprints increase with Reynolds number. Interestingly, it is also discovered that the small-scale part of the outer footprints can be well scaled by the viscous units, as well as the vortical statistics, indicating possible relations with the near-wall small-scale motions. Another finding is that in average, the statistics of $(u^*, v^*, w^*)$ are almost identical to those of $(u_S^+, v_S^+, w_S^+)$, implying that the average effect of amplitude modulation could be negligible. However, we do not deny that within positive/negative fluctuating regions of the outer footprints, the small-scale intensities would be amplified/depressed, which will be analyzed in the future. 

\section*{Acknowledgement}
Financial supports by grants from the National Natural Science Foundation of China (Nos. 92052202, 11972175, 11490553) are gratefully acknowledged. The authors are grateful for the helpful discussions with X.I.A. Yang, C.-X. Xu, W.-X. Huang, G. Yin and C.-Y. Wang, as well as M. Lee, R. Moser, S. Hoyas, J. Jim\'enez, Z. Wu and C. Meneveau for making the channel DNS data publicly available. R.H. would also like to acknowledge the inspiring communications with W. J. Baars and S. Chernyshenko. 

\section*{Availability of data}
The data that support the findings of this study are available from the corresponding author upon reasonable request.

\appendix

\section{Validation of the present DNS data} \label{sec:appenA}
\begin{figure}
\begin{minipage}{0.49\linewidth}
\centerline{\includegraphics[width=\textwidth]{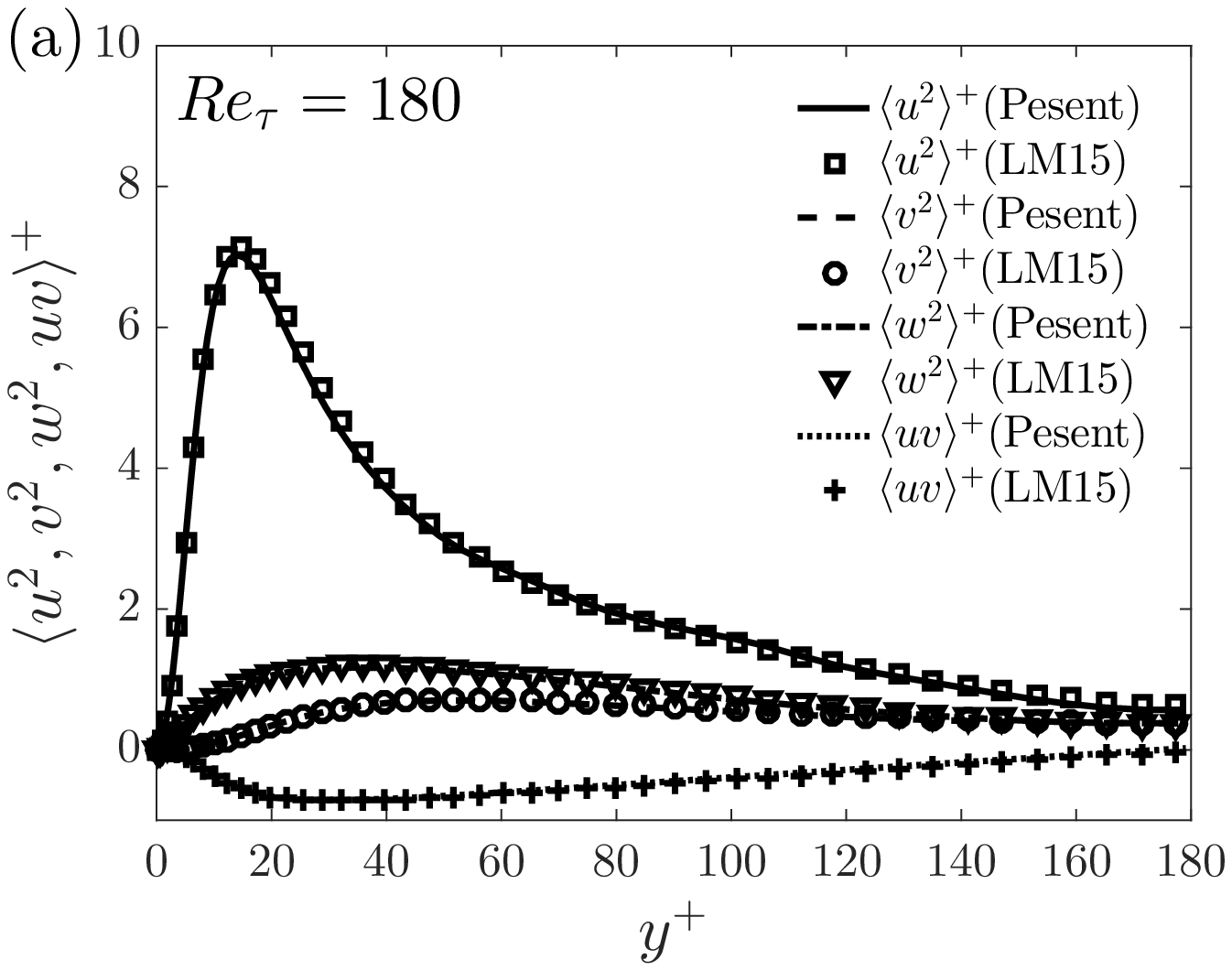}}
\end{minipage}
%%\hfill
\begin{minipage}{0.49\linewidth}
\centerline{\includegraphics[width=\textwidth]{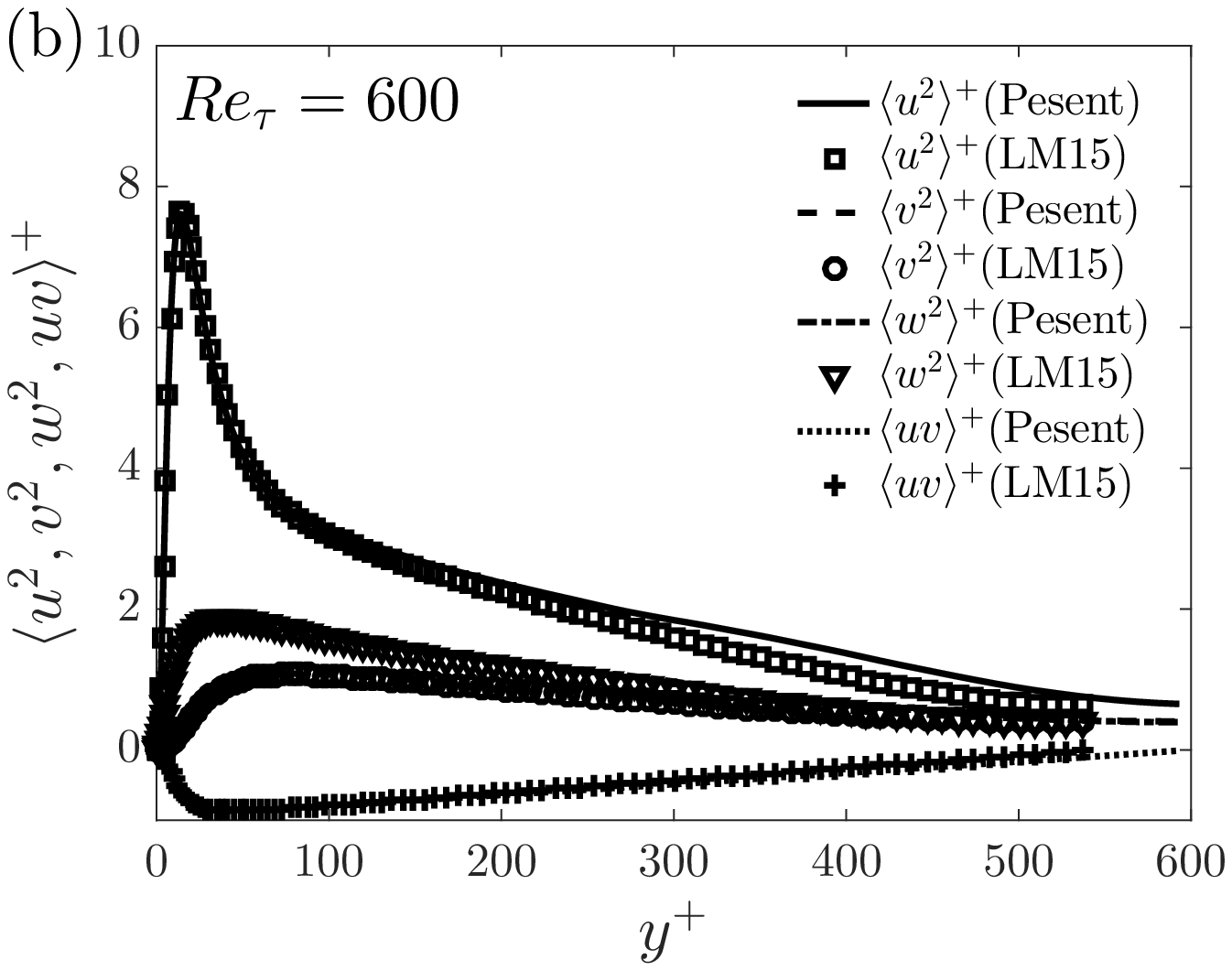}}
\end{minipage}
\caption{Comparison of the turbulence intensities ($\langle u^2, v^2, w^2 \rangle^+$) and Reynolds shear stress $\langle uv \rangle^+$ between the present DNS and \cite{Lee2015} (indicated by LM15) at $Re_\tau$ = 180 (a) and $Re_\tau$ = 600 (b). The LM15 data in (b) is at $Re_\tau$ = 550.}
\label{fig:fig17}
\end{figure}

The low-Reynolds-number data are generated by our own DNS and here compared with the DNS data of \cite{Lee2015} at similar Reynolds numbers to validate the present data quality. The fluctuating intensities of all three velocity components and the Reynolds shear stress are shown in \textcolor{black}{FIG}~\ref{fig:fig17}, where the present results at $Re_\tau$ = 180 and 600 are compared with the data of \cite{Lee2015} at $Re_\tau$ = 180 and 550, respectively. \textcolor{black}{FIG} \ref{fig:fig17} confirms the excellent agreements between the two simulations, verifying the adequacy of the present DNS data. It is also noted that the slightly larger $\langle u^{+2} \rangle$ of the present DNS in (b) is probably due to the higher Reynolds number.

\section{Characteristics of $H_{Lu}$, $H_{Lv}$ and $H_{Lw}$ with different $y^+_O$} \label{sec:appenB}
\begin{figure}
\centering
\begin{minipage}{0.49\linewidth}
\centerline{\includegraphics[width=\textwidth]{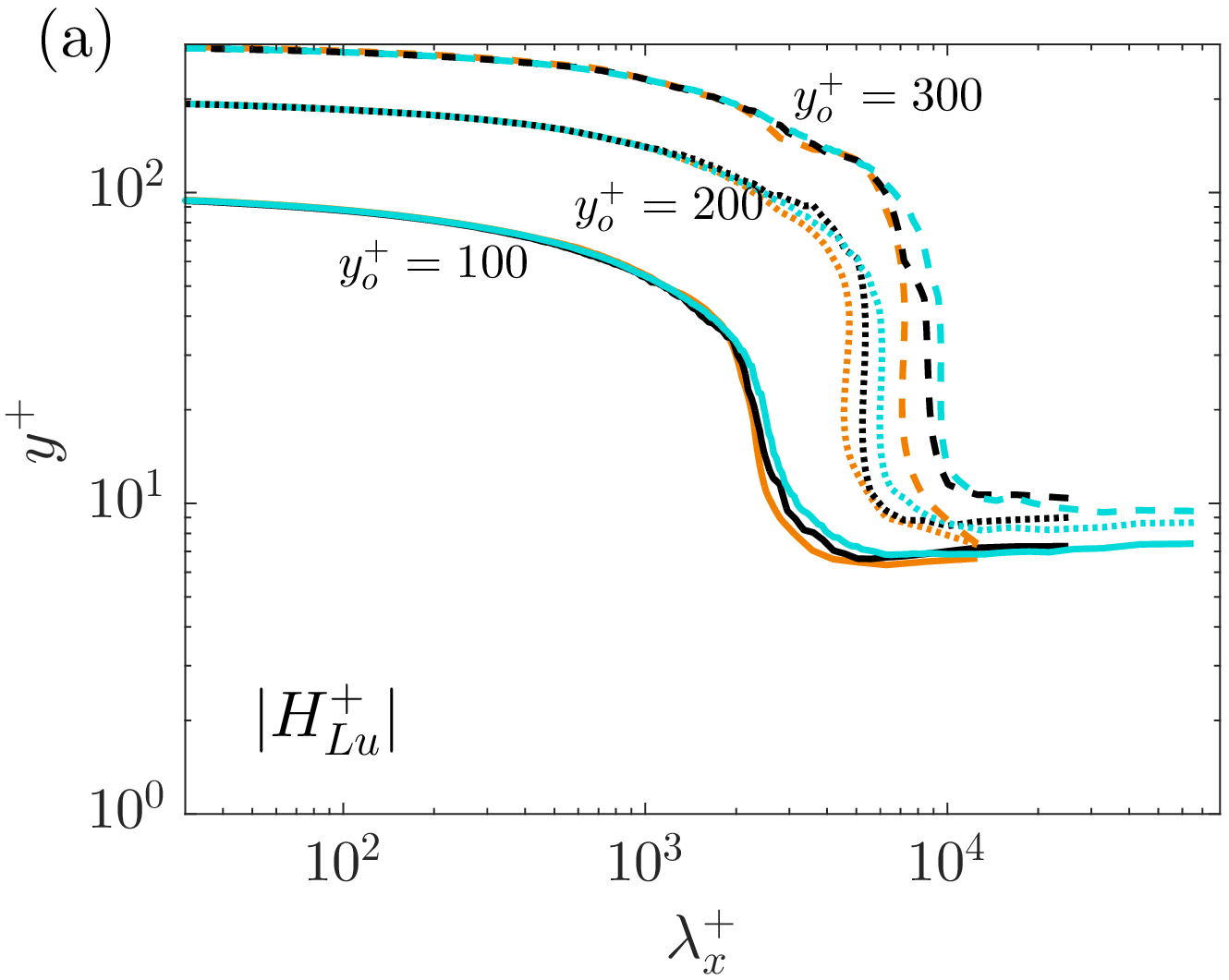}}
\end{minipage}
\hfill
\begin{minipage}{0.49\linewidth}
\centerline{\includegraphics[width=\textwidth]{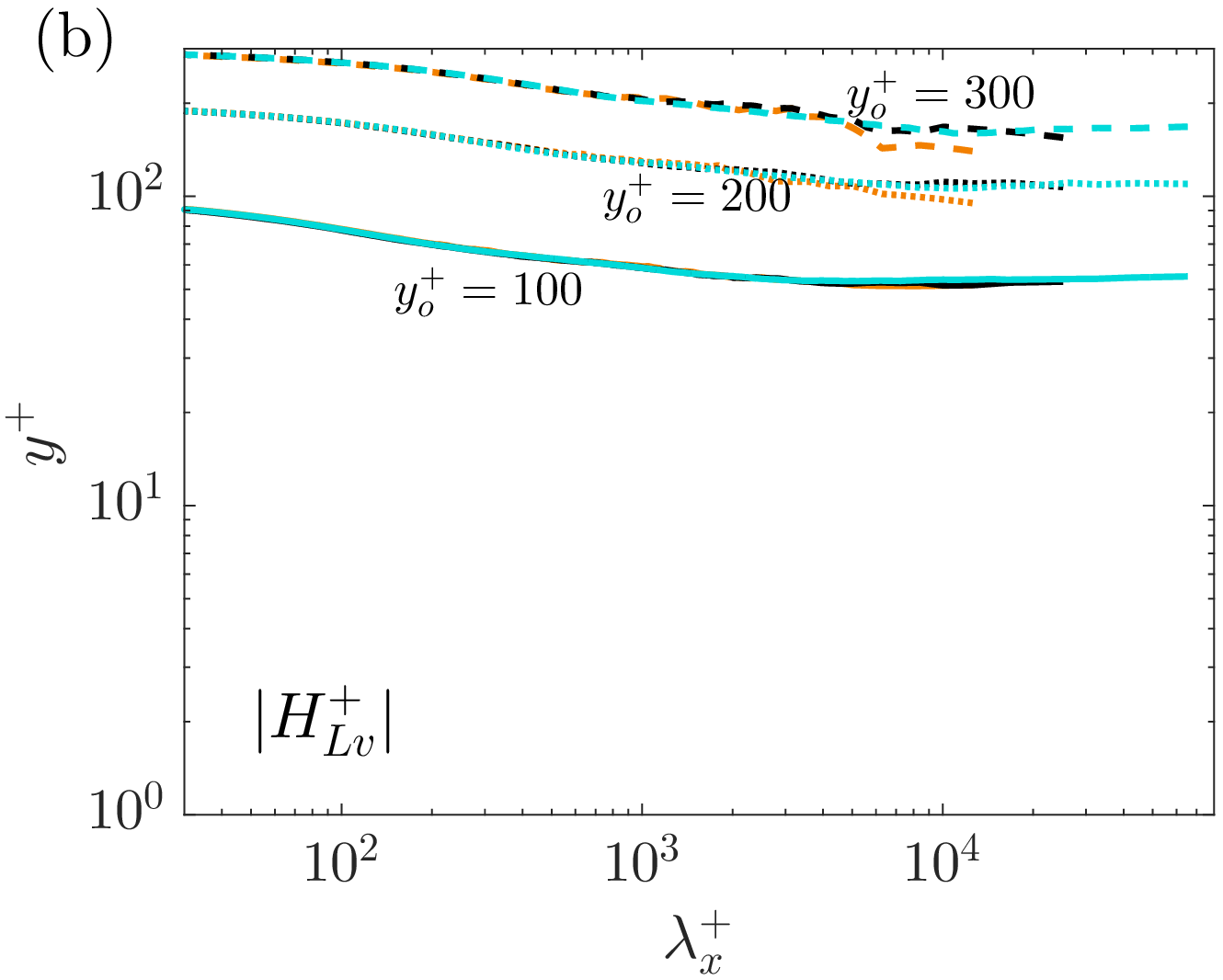}}
\end{minipage}
\vfill
\begin{minipage}{0.49\linewidth}
\centerline{\includegraphics[width=\textwidth]{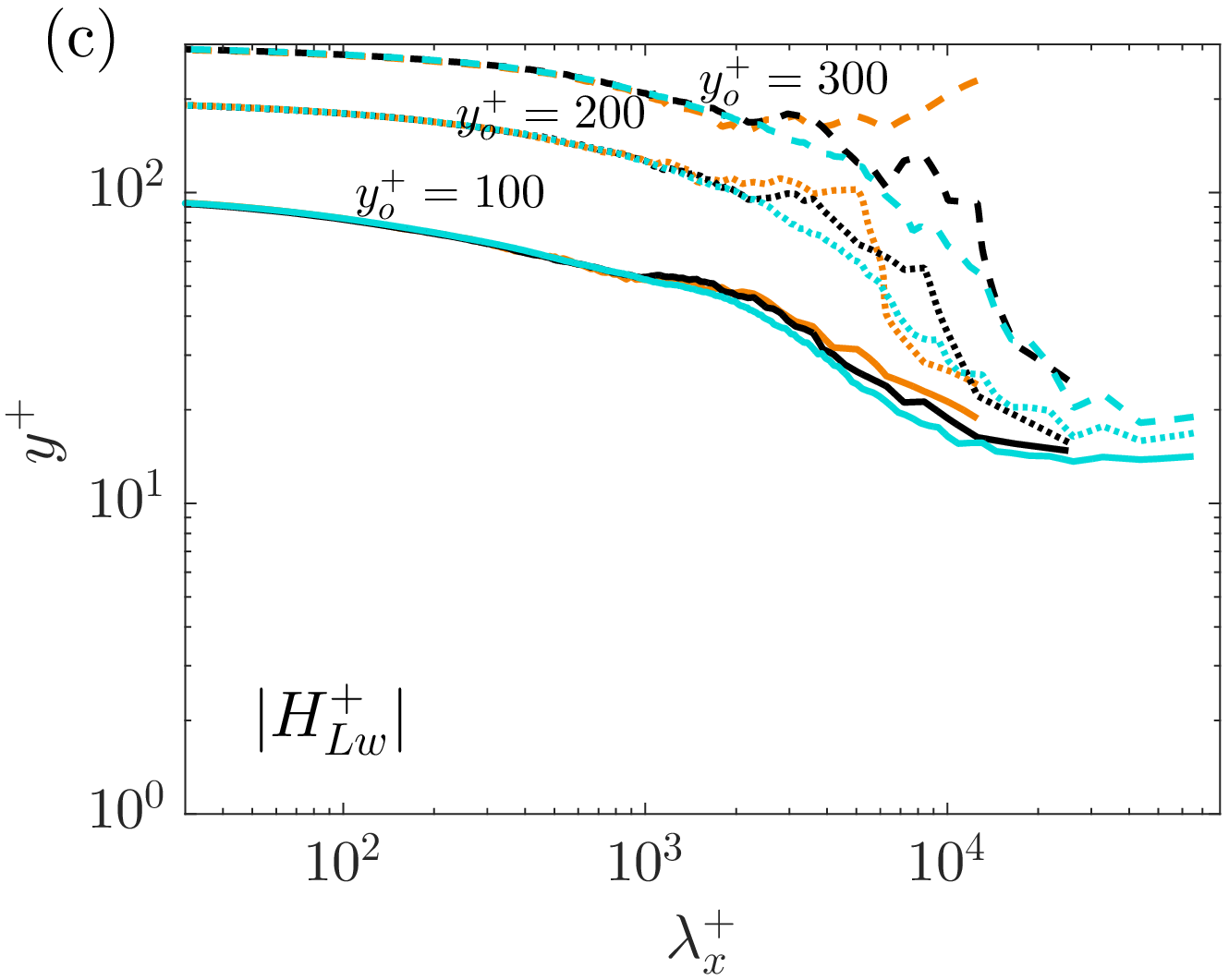}}
\end{minipage}
\caption{Contour lines for the magnitudes of the scale-dependent complex-valued kernel functions, (a) $|H_{Lu}|$, (b) $|H_{Lv}|$ and (c) $|H_{Lw}|$ with a constant value of 0.6, and solid lines for $y_O^+ = 100$, dotted lines for $y_O^+ = 200$, dashed lines for $y_O^+ = 300$. Refer to table~\ref{tab:tab1} for the line colors.}
\label{fig:fig18}
\end{figure}

Here the contour plots of $|H_{Lu}|$, $|H_{Lv}|$ and $|H_{Lw}|$ with $y^+_O=100$, 200 and 300 at $Re_\tau = 1000 \sim 5200$ are compared in \textcolor{black}{FIG}~\ref{fig:fig18}. It can clearly seen that the contour lines of different Reynolds numbers are not well collapsed at large scales with $y^+_O=200$ and 300, while much improved collapse can be found with $y^+_O=100$.

\section{Effect of the outer reference height $y^+_O$ at low Reynolds numbers} \label{sec:appenC}

\begin{figure}
\centering
\begin{minipage}{0.49\linewidth}
\centerline{\includegraphics[width=\textwidth]{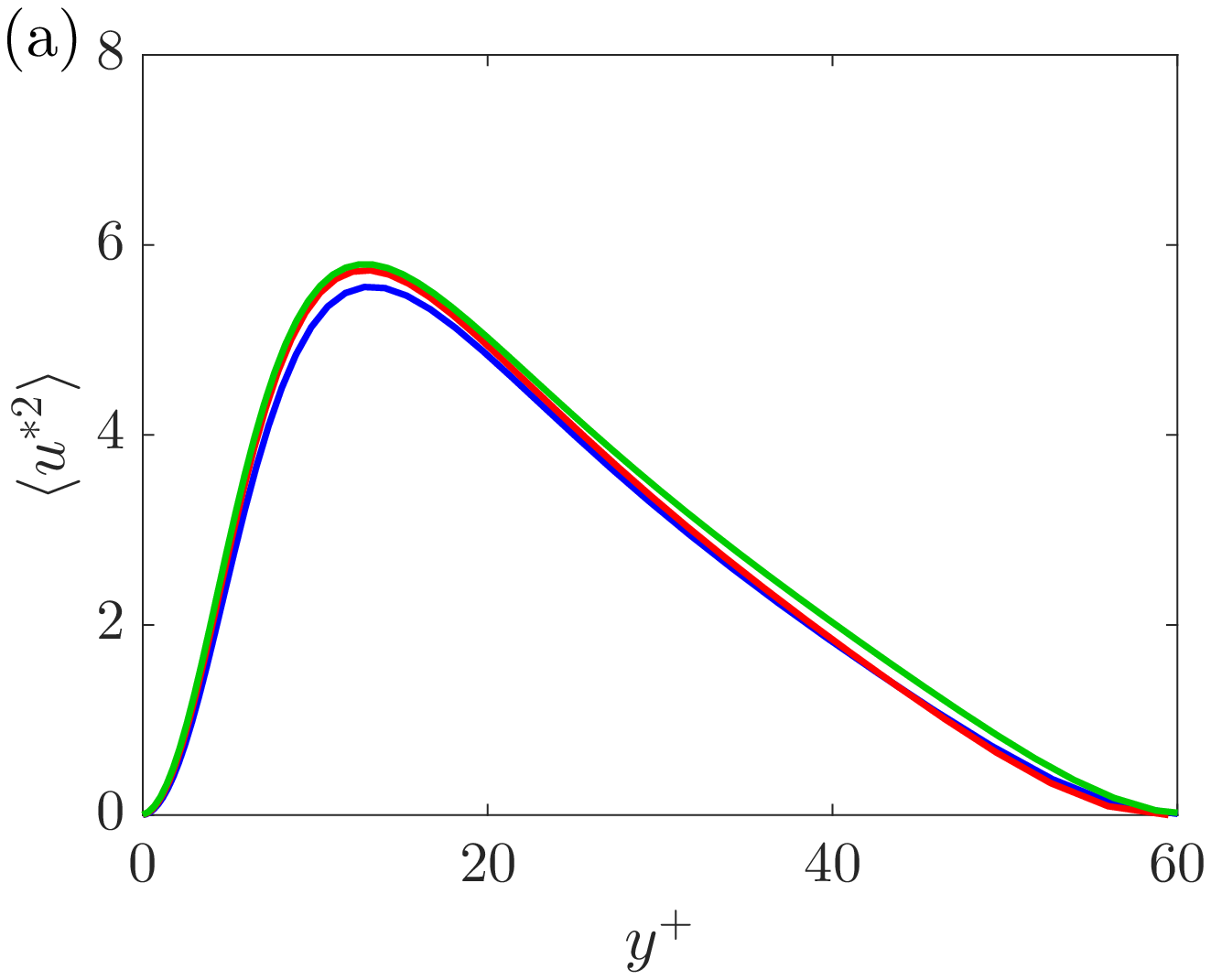}}
\end{minipage}
\hfill
\begin{minipage}{0.49\linewidth}
\centerline{\includegraphics[width=\textwidth]{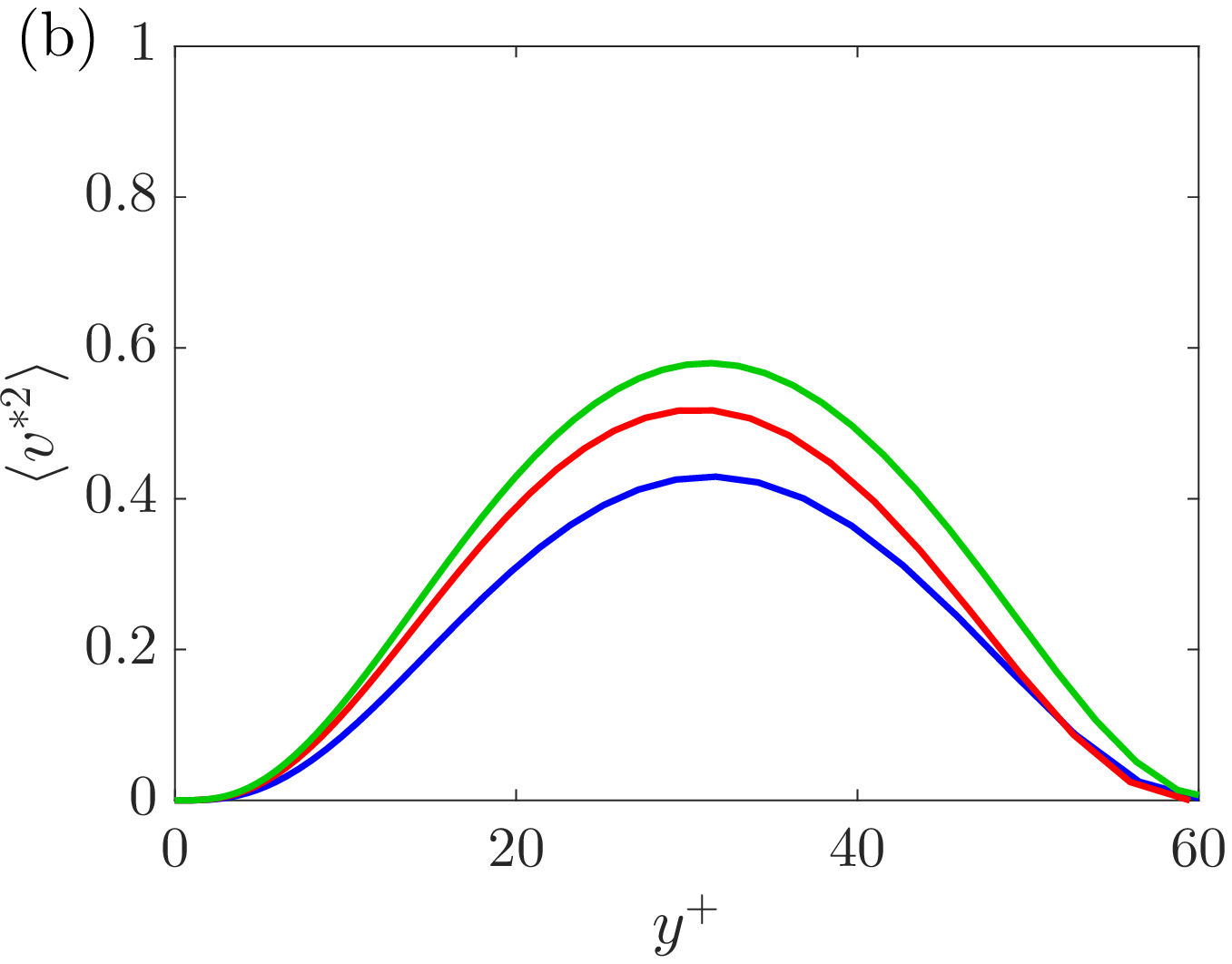}}
\end{minipage}
\vfill
\begin{minipage}{0.49\linewidth}
\centerline{\includegraphics[width=\textwidth]{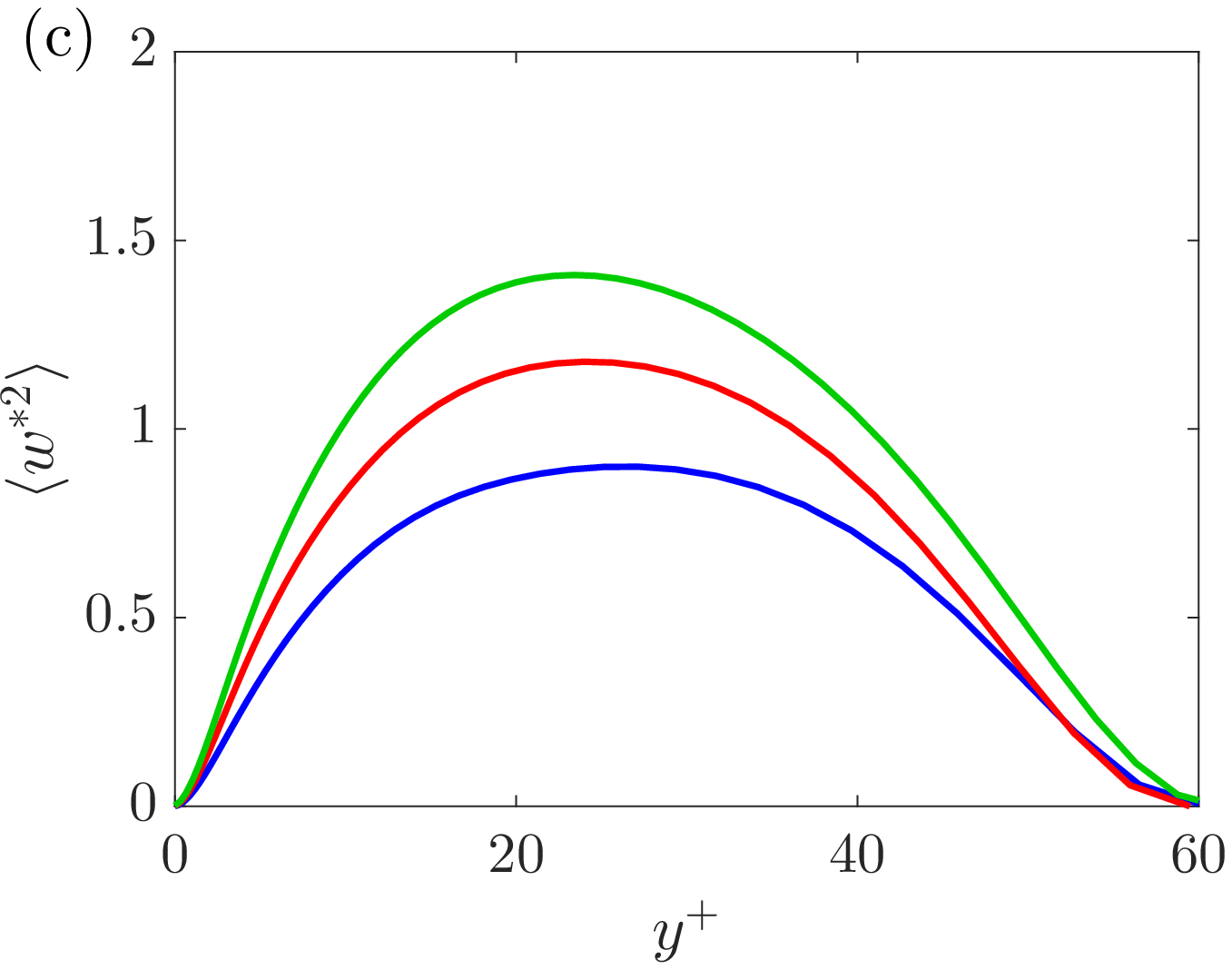}}
\end{minipage}
\caption{Intensities of the extracted (a) $u^*$, (b) $v^*$ and (c) $w^*$ with lowering the outer reference height to $y^+_O$ = 60 at $Re_\tau$ = 180, 310 and 600. Refer to table~\ref{tab:tab1} for the line colors.}
\label{fig:fig20}
\end{figure}

We have shown that excellent viscous scaling of $u^*$, $v^*$ and $w^*$ can be obtained with $y^+_O=100$ at $Re_\tau=1000\sim5200$, while low-Reynolds-number effect exists at $Re_\tau=180\sim600$ using the same $y_O^+$. It may be argued that $y^+=100$ is located in the outer region at the low Reynolds numbers with the outer scaling ($\delta$ and $u_\tau$). Therefore it may be interesting to see what will happen if further lowering $y^+_O$ at the low Reynolds numbers. 
\textcolor{black}{FIG}~\ref{fig:fig20} shows the intensities of the extracted $u^*$, $v^*$ and $w^*$ at the low Reynolds numbers from 180 to 600, with $y^+_O$ = 60. It is seen that the scaling of $u^*$ is improved slightly, compared to \textcolor{black}{FIG}~\ref{fig:fig12} (a), while the low-Reynolds-number effect still exists for $v^*$ and $w^*$. Therefore, it may indicate that using smaller $y_O^+$ can not relieve the anomalous scaling, and the low-Reynolds-number effect may be an objective phenomenon. 

%\subsection{\label{sec:level2}Second-level heading: Formatting}

%\subsubsection{Wide text (A level-3 head)}
% \nocite{*}

\bibliography{ref}% Produces the bibliography via BibTeX.

\end{document}